\documentclass[12pt]{iopart}
\usepackage{iopams}                            
\usepackage{subfigure}
\usepackage{graphicx}
\usepackage{color}
\usepackage{subeqn}                            
\usepackage[en-US]{datetime2}       

\usepackage{cite}
\usepackage{epsfig}
\newcommand{\Schrodinger}{Schr{\"o}dinger}     
\newcommand{\PT}{\mathcal{PT}}                 
\newcommand{\calF}{\mathcal{F}}

\newcommand{\bbV}{\mathbb{V}}
\newcommand{\bbW}{\mathbb{W}}
\newcommand{\ef}[1]{(\ref{#1})}                

\newcommand{\sech}{\,\mathrm{sech}}

\newcommand{\tint}{\!\int\!}                   
\newcommand{\tpsi}{\tilde{\psi}}               
\newcommand{\tPsi}{\tilde{\Psi}}               

\newcommand{\pdv}[2]{\frac{\partial #1}{\partial #2}}
\newcommand{\qc}{\>,\quad}
\newcommand{\qs}{\>,\>\>}

\newcommand{\notag}{\nonumber}                 

\eqnobysec                                     
\begin{document}
\title[]  
{Stability and response of trapped solitary wave solutions of coupled nonlinear \Schrodinger\ 
equations in an external, $\PT$- and supersymmetric potential}

\author{Efstathios G. Charalampidis$^1$, John F. Dawson$^2$, 
        Fred Cooper$^{3,4}$, Avinash Khare$^5$, Avadh Saxena$^4$}
\address{$^1$Mathematics Department, California Polytechnic State University, San Luis Obispo,
CA 93407-0403, United States of America} 
\address{$^2$Department of Physics, University of New Hampshire, Durham, NH 03824, 
United States of America}
\address{$^3$The Santa Fe Institute, 1399 Hyde Park Road, Santa Fe, NM 87501, United States of America}
\address{$^4$Theoretical Division and Center for Nonlinear Studies, Los Alamos National Laboratory, 
Los Alamos, NM 87545, United States of America}
\address{$^5$Physics Department, Savitribai Phule Pune University, Pune 411007, India}
%
%
\ead{echarala@calpoly.edu}
\ead{john.dawson@unh.edu}
\ead{cooper@santafe.edu}
\ead{khare@physics.unipune.ac.in}
\ead{avadh@lanl.gov}
\vspace{10pt}
\begin{indented}
\item[]\DTMnow
\end{indented}
\begin{abstract}
We present trapped solitary wave solutions of a coupled nonlinear \Schrodinger\ system 
in $1$+$1$ dimensions in the presence of an external, supersymmetric and complex $\PT$-symmetric 
potential. The \Schrodinger\ system this work focuses on possesses exact solutions whose 
existence, stability, and spatio-temporal dynamics are investigated by means of analytical 
and numerical methods. Two different variational approximations are considered where the 
stability and dynamics of the solitary waves are explored in terms of eight and twelve 
time-dependent collective coordinates. We find regions of stability for specific potential 
choices as well as analytic expressions for the small oscillation frequencies in the collective
coordinate approximation. Our findings are further supported by performing systematic numerical 
simulations of the nonlinear \Schrodinger\ system. \\
LA-UR-20-22820
%

\end{abstract}

\submitto{\jpa}
\vspace{2pc}
\section{\label{s:Intro}Introduction}

The nonlinear \Schrodinger\ equation (NLSE) arises in many areas of 
physics including Bose-Einstein condensation, plasmas, water waves 
and nonlinear optics~\cite{cNLSreview}. The possibility of experimentally
coupling two component NLSE's in matrix complex potentials has recently
been investigated in nonlinear optics situations in which two wave guides
are locally coupled through an antisymmetric medium~\cite{PhysRevA.99.013823}.  

On the other hand, $\PT$ symmetry was first introduced into physics as an
alternative to Hermiticity in quantum mechanics, yet with real eigenvalues~%
\cite{r:Bender:2007nr,0305-4470-39-32-E01,1751-8121-41-24-240301,1751-8121-45-44-440301}.  
The similarity of the \Schrodinger\ equation with Maxwell's equations in
the paraxial approximation facilitates the realization of $\PT$ invariant 
systems in a variety of contexts such as
optics~\cite{PhysRevLett.100.103904,Makris2011,0305-4470-38-9-L03,PhysRevLett.101.080402,PhysRevLett.103.123601,PhysRevB.80.235102,PhysRevA.81.022102,r:Ruter:2010mz,PhysRevLett.103.093902,r:Regensburger:2012gf},
photonic lattices~\cite{SzameitPRA11}, 
electronic circuits~\cite{PhysRevA.84.040101,1751-8121-45-44-444029}, 
mechanical circuits \cite{r:Bender:2013ly}, whispering-gallery microcavities~\cite{r:Peng:2014ul}, 
among many other physical settings.

Supersymmetry (SUSY) originally considered in high-energy physics to relate 
fermionic and bosonic systems has also been invoked in condensed matter systems
such as fractional quantum Hall states \cite{fqH} and realized in optics~\cite{PhysRevLett.110.233902,r:Heinrich:2014qf}. 
For the \Schrodinger\ equation, supersymmetry relates two potentials which have the 
same spectrum~\cite{r:CooperKhareSukhatmeBOOK, doi:10.1142/S0217751X01004153, Bagchi2000285,Ahmed2001343,PhysRevA.89.032116}.  
Recently in~\cite{1751-8121-50-48-485205}, we studied the stability of exact solutions 
of a single component NLSE in a class of external potentials having SUSY and $\PT$ 
symmetry. 

Our aim in the present work is to extend our considerations to the case 
of two coupled NLSEs in parity-time or $\PT$-symmetric {\it and} supersymmetric
external potentials where the cross interaction between the two components
is dictated by the nonlinear coupling of the equations. In particular, the 
superpotential studied in~\cite{1751-8121-50-48-485205} is generalized to
a matrix form here where we show that it is $\PT$-symmetric. Interestingly,
our potential has a non-trivial coupling between the two components which 
in turn affects the stability of the trapped soliton-like solutions. Our 
numerical investigations on that front are split into two steps. At
first, we will employ a collective coordinate approximation in order to map 
out the domain of stability of the pertinent waveforms of the coupled 
system. Then, we will consider the NLSEs and focus on the existence, stability
and spatio-temporal evolution of the solitary waves. Upon identifying the
steady-state solutions to the NLSEs via fixed-point iterations, we will perform 
parametric continuations over the parameters of the system. This will allow
us to carry out a systematic spectral stability analysis of the solutions and 
identify parametric regions of stability. Those findings will be corroborated
by direct numerical simulations of the NLSEs. Then, we will draw comparisons
between the collective coordinate approximation and numerical simulations
in regimes where the trapped solutions are stable and unstable. In fact, and 
in the unstable parametric regime, we will show that the effect of the coupling 
is responsible for the motion of the solitary waves in opposite directions. 
Also the amplitudes of the two components respond oppositely to small perturbations. 

The structure of the paper is as follows. We discuss the connection to supersymmetry 
in Sec.~\ref{s:supersym}, and give the exact soliton solutions to the coupled NLSEs 
in Sec.~\ref{s:model}. In Sec.~\ref{s:collective} we present the derivation of the 
equations of motion for the collective coordinate approximation using a variational 
method which is based on Rayleigh's dissipation functional. The trial wave functions 
we have chosen together with the respective dynamic equations for the collective 
coordinates derived are discussed in Sec.~\ref{s:trialWF}. We present results for 
the dynamical evolution of the collective coordinates in Sec.~\ref{s:dynamicresults} 
where comparisons of these results with numerical simulations are made. 
In Sec.~\ref{s:CompAnalysis} we present numerical results on the  existence, stability 
and dynamics of the exact solutions to the coupled NLSEs. Finally, 
we state our conclusions in Sec.~\ref{s:Conclusions}.

\section{\label{s:supersym}Supersymmetry}
We consider here a two-component nonlinear \Schrodinger\ (NLS) system 
in $1$+$1$ (one spatial and one temporal) dimensions of the form:
\begin{equation}\label{AntiPTeq}
      \rmi \, \partial_t\Psi(x,t)
      +
      \partial_{x}^2\Psi(x,t) 
      +
      \gamma \, [\, \Psi^{\dag}(x,t) \Psi(x,t) \,]\Psi(x,t)
      -
      \bbV(x)\Psi(x,t)     
      =
      0 \>,
\end{equation}
with
\begin{equation}\label{Psidef}
   \Psi(x,t)
   =
   \Biggl ( \begin{array}{c}
      \psi_1(x,t) \\
      \psi_2(x,t)
   \end{array} \Biggr )\in\mathbb{C}^{2}\>,
\end{equation}
where $\psi_{j}(x,t)$ is the wave function of the first ($j=1$) 
and second components ($j=2$), respectively, and $\gamma$ is the 
nonlinearity strength. The subscripts in Eq.~(\ref{Psidef})
stand for differentiation with respect to $t$ and $x$, respectively,
and $\left(\cdot\right)^\dag$ corresponds to conjugate transpose.

For a superpotential of the form:
\begin{equation}\label{Wsuper}
   \bbW(x) = r \,\sigma_0 \, \tanh(x) + \rmi \, s \, \sigma_3 \, \sech(x) \>,
\end{equation}
where $\sigma_{i}$ are the Pauli matrices, the 
SUSY partner potentials are given by
\begin{eqnarray}\label{Vpm}
   \bbV_{\pm}(x)
   &=
   \bbW^2(x) \pm  \partial_x \bbW(x)
   \\
   &=
   \sigma_0 \, r^2 - \sigma_0 \, b_{\pm}^2 \sech^2(x) 
   + 
   \rmi \, \sigma_3 \, d_{\pm} \sech(x) \tanh(x) \>,
   \notag
\end{eqnarray}
where
\begin{equation}\label{bddefs}
   b_{\pm}^2 = s^2 + r ( r \pm 1)
   \qc
   d_{\pm} =  s \, (\, 2 r \mp 1 ) \>.
\end{equation}
Note that the partner potentials $\bbV_{\pm}(x)$ are $\PT$-symmetric.  
%
\section{\label{s:model}Model potential}
%
%
\begin{figure}[htp]
\centering
\includegraphics[height=.16\textheight, angle =0]{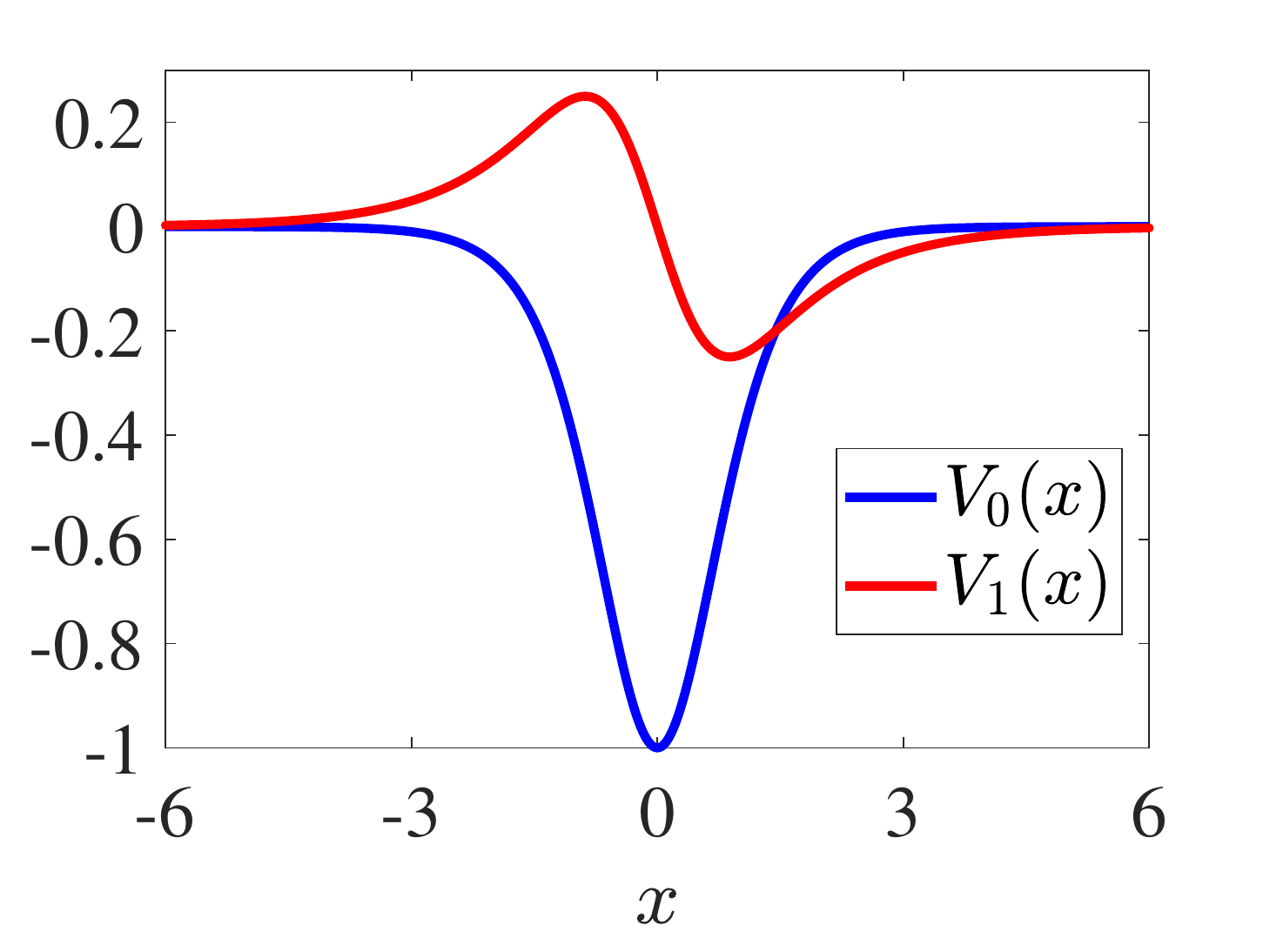}
\includegraphics[height=.16\textheight, angle =0]{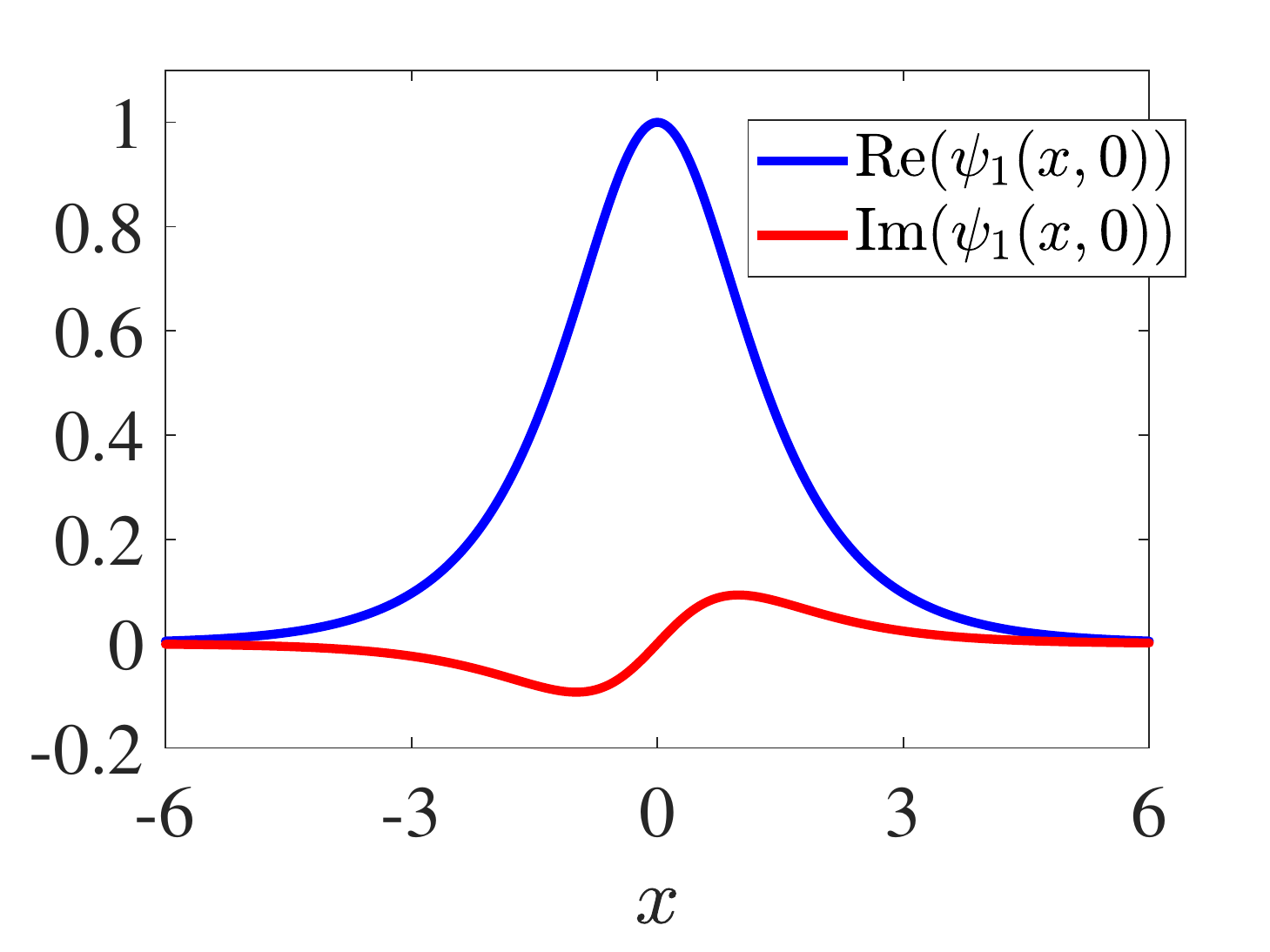}
\includegraphics[height=.16\textheight, angle =0]{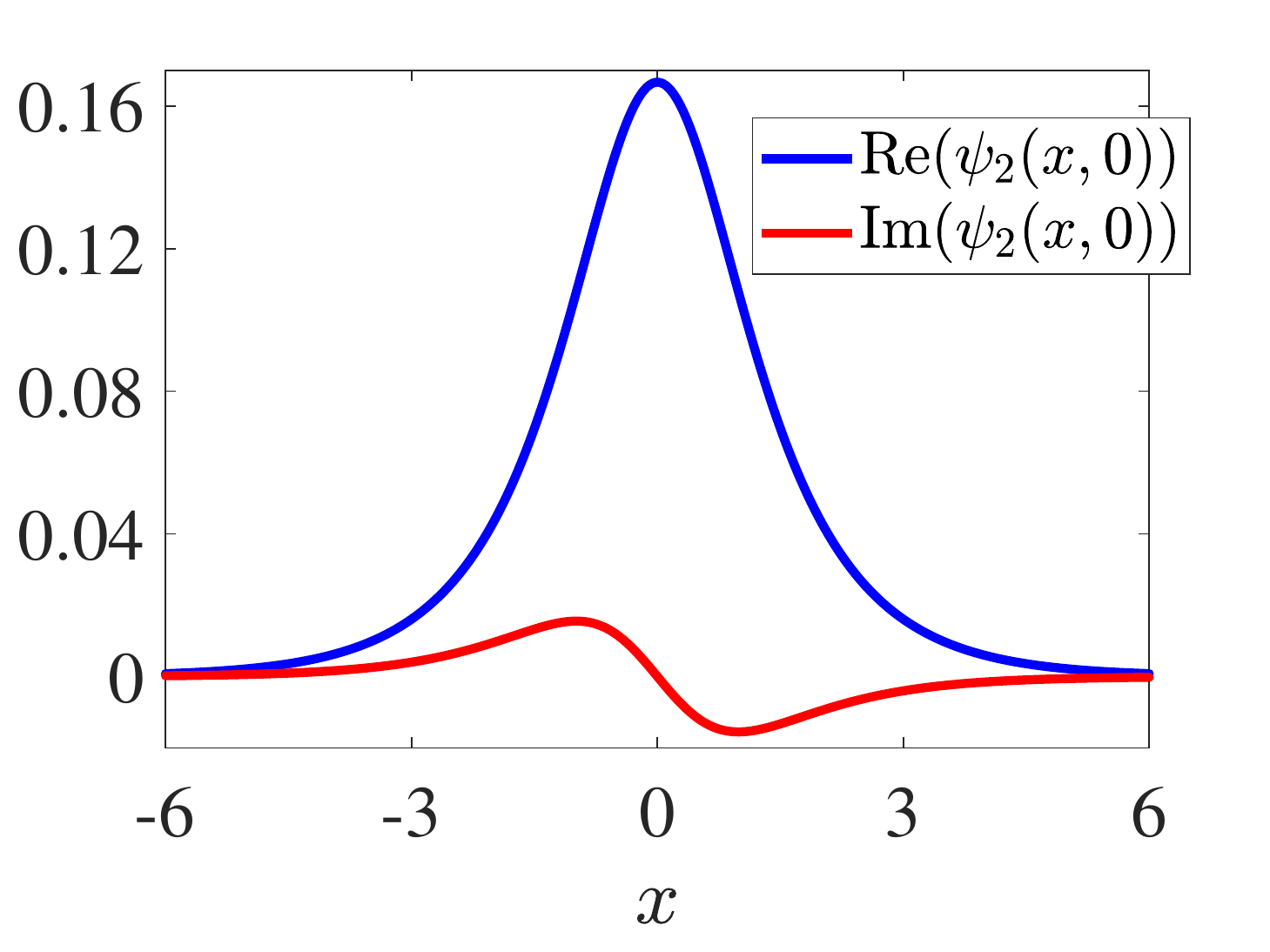}
\caption{\label{f:V01Psi}
Potential functions and exact solutions are shown in the left, middle 
and right panels, respectively, for $\gamma=1$, $b = 1$, $d = 1/2$, 
and $A_{1}=1$. The value of $A_{2}$ is determined by Eq.~(\ref{balphaconditions}).}
\end{figure}

%
%
The equation we want to solve is \ef{AntiPTeq}, where the external potential $\bbV(x)$ is given by
\begin{equation}\label{bbV}
   \bbV(x)
   =
   \sigma_0 \, V_0(x) + \rmi \, \sigma_3 \, V_1(x) \>.
\end{equation}
Since we are interested in variational approximations to the moments of the NLSEs, we now show that these equations can be derived from a modified Euler-Lagrange equation by utilizing a Rayleigh dissipation functional. The usual conservative part of the action is
\begin{equation}\label{eq:gamma-definition}
   \Gamma[\,\Psi^{\dag},\Psi\,] = \tint dt\, \sigma_0 \, L[\,\Psi^{\dag},\Psi\,] ,  
\end{equation}
where the conservative part of the Lagrangian $L$ is given by
\begin{equation}
   L[\,\Psi^{\dag},\Psi\,]
   =
   T[\,\Psi^{\dag},\Psi\,] - H[\,\Psi^{\dag},\Psi\,] \>,
\end{equation}
with
\begin{subeqnarray}\label{eq:THdefs}
   \fl
   T[\,\Psi^{\dag},\Psi\,]
   &=
   \tint dx 
   \frac{\rmi}{2} \,
   \biggl \{\,
      \Psi^{\dag}(x,t) [\, \partial_t \Psi(x,t) \,]
      -
      [\, \partial_t \Psi^{\dag}(x,t) \,] \, \Psi(x,t) \,
   \biggr \} \>,
   \label{eq:Tdef} \\
   \fl
   H[\,\Psi^{\dag},\Psi\,]
   &=
   \tint dx 
   \Bigl \{\,
      | \, \partial_x \Psi(x,t) \, |^2
      -
      \frac{\gamma}{2} \, |\, \Psi^{\dag}(x,t) \Psi(x,t) \,|^2
      +
      V_0(x) \, \Psi^{\dag}(x,t) \Psi(x,t) \,
   \Bigr \} \>.   
   \label{eq:Hdef}
\end{subeqnarray} 
We introduce the dissipation functional $F$ via
\begin{equation}
   \calF[\,\Psi^{\dag},\Psi;\Psi^{\dag}_t,\Psi_t\,] 
   = \tint dt \,F[\,\Psi^{\dag},\Psi;\Psi^{\dag}_t,\Psi_t\,]
\end{equation}
where
\begin{equation}
   \fl
   F[\,\Psi^{\dag},\Psi;\Psi^{\dag}_t,\Psi_t\,]
   =
   - \rmi \tint dx \,V_1(x) \,
   \Bigl \{\,
      [\, \partial_t \Psi^{\dag}(x,t) \,] \, \sigma_3 \, \Psi(x,t) 
      -
      \Psi^{\dag}(x,t) \, \sigma_3 \, [\, \partial_t \Psi(x,t) \,] \,
   \Bigr \} \>.
\end{equation}
The equations of motion for $\Psi(x,t)$ in the presence of a complex potential 
follow from the generalized Euler-Lagrange equations: 
\begin{equation} \label{eq:EulerLagrange}
   \frac{\delta \Gamma[\,\Psi^{\dag},\Psi\,]}
        {\delta \Psi^{\dag}(x,t)} 
   = 
   - \frac{\delta F[\,\Psi^{\dag},\Psi;\Psi^{\dag}_t,\Psi_t\,]}
          {\delta \Psi_{t}^{\dag}(x,t)} \>,
\end{equation}
which lead to the equations of motion
\begin{equation}\label{}
   \frac{\partial L}{\partial \Psi^{\dag}(x,t)}
   -
   \frac{d}{dt} \Bigl ( \pdv{L}{\Psi_t(x,t)} \Bigr )
   =
   -
   \pdv{F}{\Psi_t(x,t)} \>,
\end{equation}
and reproduce Eq.~\ef{AntiPTeq} with the potential \ef{bbV}.  
%
%
%
\subsection{\label{ss:exactsolution}Exact solution}
In component form, Eq.~\ef{AntiPTeq} reads
\begin{subeqnarray}\label{final}
\hskip -2.5cm  
      \rmi \,
      \partial_t\psi_1(x,t)
      +  
      \partial_x^2\psi_1(x,t)
      + 
      \gamma \, \left( | \psi_1(x,t) |^{2}+ | \psi_2(x,t) |^{2}\right)\psi_1(x,t) 
      -
      V(x)^{\phantom{\ast}}\psi_1(x,t)
      &= 
      0 \>,
   \label{final-a} \\
\hskip -2.5cm  
      \rmi \,
      \partial_t\psi_2(x,t)
      +  
      \partial_x^2\psi_2(x,t)
      +
      \gamma \, \left( | \psi_1(x,t) |^{2}+ | \psi_2(x,t) |^{2}\right)\psi_2(x,t)
      - 
      V^{\ast}(x)\psi_2(x,t)
      &= 0 \>,  
   \label{final-b}
\end{subeqnarray}
where $V(x)$ is of the form: $V(x) = V_0(x) + \rmi \, V_1(x)$, and where we have chosen 
\begin{subeqnarray}\label{V01}
   V_0(x)
   &= 
   - b^2 \sech^2(x) \>,
   \label{V0def} \\
   V_1(x)
   &= 
   - d \sech(x) \tanh(x) \>.
   \label{V1def}   
\end{subeqnarray}
The exact solutions of the system [cf. Eq.~\ef{final}] are given by
\begin{subeqnarray}\label{A-Psi}
   \psi_1(x,t)
   &=
   A_1 \sech(x) \, 
   \exp \{\, \rmi \, [\, t + \phi(x) \,] \, \} \>,
   \label{A-Psi-a} \\
   \psi_2(x,t)
   &=
   A_2 \sech(x) \, 
   \exp \{\, \rmi \, [\, t - \phi(x) \,] \, \} \>,
   \label{A-Psi-b}
\end{subeqnarray}
where $\phi(x) = 2 d \tan^{-1}[\, \tanh(x/2) \,]$, provided that
\begin{equation}\label{balphaconditions}
   \gamma \, (\, A_1^2 + A_2^2 \,)
   =
   2 + (d/3)^2 - b^2 \ge 0 \>.
\end{equation}
The left panel of Fig.~\ref{f:V01Psi} showcases the potentials
$V_0(x)$ (solid blue line) and $V_1(x)$ (solid red line) whereas
the middle and right panels of the figure depict the real (solid 
blue line) and imaginary (solid red line) parts of the exact 
solutions $\psi_{1}(x,t)$ and $\psi_{2}(x,t)$ at $t=0$, respectively, 
for values of the parameters $\gamma=1$, $b = 1$, $d = 1/2$, and 
$A_{1}=1$. 

%

\section{\label{s:collective}Collective coordinates (CCs)}
In section, we consider two variational approximations for studying the 
stability and time evolution of the trapped solitary waves. This way, we
will be able to compare our findings with numerical simulations of the 
NLSEs in Sec.~\ref{s:dynamicresults} (see, also Sec.~\ref{s:CompAnalysis}
discussing our computational analysis). We review here the method of collective
coordinates, abbreviated CC hereafter (see for example Ref.~\cite{1751-8121-50-48-485205}) 
applied to our case. 

The time dependent variational approximation relies on introducing a finite 
set of time-dependent real parameters in a trial wave function that hopefully
captures the time evolution of a perturbed solution. By doing this one obtains
a simplified set of ordinary differential equations for the CCs in place of 
solving the full partial differential equations associated with the NLS system.
We begin our discussion by setting
\begin{eqnarray}\label{e:VT-1}
   \Psi(x,t)
   \mapsto
   \tilde{\Psi}[\,x,Q(t)\,]\, , \quad Q(t) 
   = 
   \{\, Q^1(t),Q^2(t),\dots,Q^{2n}(t) \,\} \in \mathbb{R}^{2n} \>,
\end{eqnarray}
where $Q(t)$ corresponds to the CCs. It should be pointed out that the success of 
the method depends greatly on the choice of the trial wave function $\tilde{\Psi}[\,x,Q(t)\,]$. 
The generalized dissipative Euler-Lagrange equations lead to Hamilton's equations 
for $Q(t)$. The Lagrangian in terms of the CCs is given by
\begin{equation}\label{e:VT-2}
   L(Q,\dot{Q})
   =
   T(Q,\dot{Q}) - H(Q) \>,
\end{equation}
where the kinetic term $T(Q,\dot{Q})$ and Hamiltonian $H[\,Q\,]$ are given 
by
\begin{equation}\label{Tdef}
   \fl
   T(Q,\dot{Q})
   =
   \frac{\rmi}{2} \tint dx
   \biggl \{ \, 
      \tPsi^{\dag}(x,Q)\, \tPsi_t(x,Q)
      - 
      \tPsi^{\dag}_t(x,Q) \, \tPsi(x,Q) \,
   \biggr \}
   =
   \pi_\mu(Q) \, \dot{Q}^\mu \>,
\end{equation}
and 
\begin{equation}\label{e:VT-4}
   \fl
   H(Q)
   =
   \tint dx 
   \biggl \{ \,
       |\partial_x \tPsi(x,Q) |^2
       -
       V_0(x) \, |\tPsi(x,Q)|^{2}
       -
       (\gamma/2) \, |\tPsi(x,Q)|^{4} \,
    \biggr \} \>,
\end{equation}
respectively. Note that $\pi_\mu(Q)$ in Eq.~(\ref{Tdef}) is defined by
\begin{equation}\label{e:VT-3}
   \fl
   \pi_\mu(Q)
   =
   \frac{\rmi}{2} \tint dx
   \biggl \{ \, 
      \tPsi^{\dag}(x,Q)\,[\, \partial_\mu \tPsi(x,Q) \,]
      - 
      [\, \partial_\mu \tPsi^{\dag}(x,Q) \,] \, \tPsi(x,Q) \,
   \biggr \} \>,
\end{equation}
where we have introduced the notation $\partial_{\mu} \equiv \partial / \partial Q^{\mu}$.

The dissipation functional in terms of the CCs is given by
\begin{equation}\label{e:VT-4.1}
   \fl
   F(Q,\dot{Q})
   =
   \rmi \tint dx\, V_1(x) \,
   \biggl \{\,
      \tPsi^{\dag}(x,Q)\, \sigma_3 \tPsi_t(x,Q)
      -
      \tPsi_t^{\dag}(x,Q)\, \sigma_3 \tPsi(x,Q) \,
   \biggr \}
   =   
   w_{\mu}(Q) \, \dot{Q}^{\mu} \>,
\end{equation}
where
\begin{equation}\label{e:VT-4.2}
   \fl
   w_{\mu}(Q)
   =
   \rmi \tint dx \,V_1(x) \,
   \biggl\{ \, 
      \tPsi^{\dag}(x,Q)\, \sigma_3 [\, \partial_\mu \tPsi(x,Q) \,]
      - 
      [\, \partial_\mu \tPsi^{\dag}(x,Q) \,] \, \sigma_3 \, \tPsi(x,Q)
   \,\biggr\} \>.
\end{equation}
Upon introducing the antisymmetric $2n \times 2n$ symplectic matrix:
\begin{equation}\label{e:VT-7}
   f_{\mu\nu}(Q)
   =
   \partial_\mu \pi_\nu(Q) - \partial_\nu \pi_\mu(Q)\, ,
\end{equation}
the generalized Euler-Lagrange equations
\begin{equation}\label{e:VT-5}
   \frac{\partial L}{\partial Q^\mu}
   -
   \frac{d}{dt} \Bigl ( \pdv{L}{\dot{Q}^\mu} \Bigr )
   =
   -
   \pdv{F}{\dot{Q}^\mu} \>
\end{equation}
can be written in the form
\begin{equation}\label{e:VT-6}
   f_{\mu\nu}(Q) \, \dot{Q}^\nu
   =
   u_{\mu}(Q)
   =
   v_{\mu}(Q) - w_{\mu}(Q) \>, 
\end{equation}
by setting $v_{\mu}(Q) = \partial_\mu H(Q)$. If $\det\left[{f(Q)}\right]\ne 0$, 
we can define an inverse as the contra-variant matrix with upper 
indices:
\begin{equation}\label{e:VT-8}
   f^{\mu\nu}(Q) \, f_{\nu\sigma}(Q) = \delta^\mu_\sigma \>,
\end{equation}
in which case the equations of motion \ef{e:VT-6} can be formulated in the
symplectic form:
\begin{equation}\label{e:VT-9}
   \dot{Q}^\mu
   =
   f^{\mu\nu}(Q) \, u_{\nu}(Q) \>.
\end{equation}
We solve this set of equations for our choice of CCs.  

%
\section{\label{s:trialWF}Trial wave function}

We will choose trial wave functions similar to that used for the single-component 
NLSE in a $\PT$ symmetric complex external potential \cite{1751-8121-50-48-485205}:
\begin{subeqnarray}\label{t-psi12}
   \tpsi_1[x,Q_1(t)]
   &=
   A_1(t) \sech[\, \beta_1(t) (x - q_1(t)) \,] \, \rme^{\rmi\, \phi_1[x,Q_1(t)] } \>,
   \label{t-psi1} \\
   \tpsi_2[x,Q_2(t)]
   &=
   A_2(t) \sech[\, \beta_2(t) (x - q_2(t)) \,] \, \rme^{\rmi\, \phi_2[x,Q_2(t)] } \>,   
\end{subeqnarray}
where  
\begin{subeqnarray}\label{phi12def}
   \fl
   \phi_1[x,Q_1(t)]
   &=
   - \theta_1(t) + p_1(t) \, (x - q_1(t)) + \Lambda_1(t) \, (x - q_1(t))^2 + \phi(x) \>,
   \label{phi12def-a} \\
   \fl
   \phi_2[x,Q_2(t)]
   &=
   - \theta_2(t) + p_2(t) \, (x - q_2(t)) + \Lambda_2(t) \, (x - q_2(t))^2 - \phi(x) \>,
   \label{phi12def-b}
\end{subeqnarray}
together with $\phi(x) = (2 d/3) \tan^{-1}[\, \tanh(x/2) \,]$. Let us now define
\begin{equation}\label{Midefs}
   M_i(t)
   =
   \tint dx\,|\psi_i(x,t)|^2
   =
  \frac{| A_i(t)|^2}{\beta_i(t)} \tint dz \sech^2(z)
  =
  \frac{2 \, | A_i(t)|^2}{\beta_i(t)}  \>,
\end{equation}
where the integral in the right hand side is calculated in~\ref{s:Integrals}
(alongside with other integrals useful for the present work).

We consider two sets of variational parameters:
\begin{subeqnarray}\label{Qparms}
   Q_1(t)
   &=
   \biggl\{\, M_1(t), \theta_1(t), q_1(t), p_1(t), \beta_1(t), \Lambda_1(t) \, \biggr\} \>,
   \label{Q1parms} \\
   Q_2(t)
   &=
   \biggl\{\, M_2(t), \theta_2(t), q_2(t), p_2(t), \beta_2(t), \Lambda_2(t) \, \biggr\} \>,   
\end{subeqnarray}
with initial conditions
\begin{subeqnarray}\label{Q12initial}
   p_1(0) = 0
   \qc
   \beta_1(0) = 1
   \qc
   \Lambda_1(0) = 0 \>,
   \label{Q1initial} \\
   p_2(0) = 0
   \qc
   \beta_2(0) = 1
   \qc
   \Lambda_2(0) = 0 \>,
   \label{Q2initial}   
\end{subeqnarray}
and with $q_1(0) = q_2(0) = \delta q_0$.  We make this perturbation to study the response of the exact solution which has $\delta q_0 = 0$ to small initial perturbations.  We also require that
\begin{equation}\label{A1plusA2}
   \frac{\gamma}{2} \, [\, M_1(0) + M_2(0) \,]
   =
   \gamma \, [\, A_1^2(0) + A_2^2(0) \,] = 2 + (d/3)^2 - b^2 \>.
\end{equation}
This way, the set of variational trial wave functions~(\ref{t-psi12}) 
satisfies the exact solution [cf. Eq.~\ef{A-Psi}] at $t=0$. We also 
set $\theta_i(0) = 0$, and require that $\theta'_i(0) = -1$. 

%
\subsection{\label{ss:Dynamic}Dynamic term}

From Eq.\ef{Tdef}, the dynamic term splits into the sum of two 
independent parts:
\begin{equation}\label{Ttot1t2}
   T(Q,\dot{Q})
   =
   t(Q_1,\dot{Q}_1) + t(Q_2,\dot{Q}_2) \>,
   \notag
\end{equation}
where
\begin{eqnarray}\label{Tcalc}
   t(Q,\dot{Q})
   &=
   \frac{\rmi}{2} \tint dx\,
   \biggl \{ \, 
      \tpsi^{\ast}(x,Q)\, \tpsi_{t}(x,Q)
      - 
      \tpsi^{\ast}_{t}(x,Q) \, \tpsi(x,Q) \,
   \biggr \}
   \\
   &=
   M \,
   \Biggl \{\,
      \dot{\theta} + p \, \dot{q} -  \frac{\pi^2}{12\,\beta^2} \, \dot{\Lambda} \,
   \Biggr \} 
   =
   \pi_{\mu}(Q) \, \dot{Q}^{\mu} \>.
   \notag
\end{eqnarray}   
From this expression one easily determines the symplectic matrix 
\begin{equation}
f_{\mu \nu} = \partial_\mu \pi_\nu- \partial_\nu \pi_\mu \,,
\end{equation}
from which we obtain its inverse $f^{\mu \nu}$:
\begin{equation}\label{invfmunu}
   f^{\mu\nu}(Q)
   =
   \Bigl ( \begin{array}{cc} 
      g^{\mu\nu}(Q_1) & 0 \\
      0 & g^{\mu\nu}(Q_2)
   \end{array} \Bigr ) \>,
\end{equation}
where
\begin{equation}\label{gmat}
   \fl
   g^{\mu\nu}(Q)
   =
   \frac{1}{M}
   \left ( \begin{array}{cccccc}
      0 & -M & 0 & 0 & 0 & 0 \\
      M & 0 & 0 & -p & c & 0 \\
      0 & 0 & 0 & 1 & 0 & 0 \\
      0 & p & -1 & 0 & 0 & 0 \\
      0 & -c & 0 & 0 & 0 & -d \\
      0 & 0 & 0 & 0 & d & 0
   \end{array} \right )
   \qs
   c = \frac{\beta}{2} 
   \qc
   d = \frac{6 \, \beta^3}{\pi^2} \>.
\end{equation}

%
\subsubsection{Hamiltonian and its decomposition}

Based on Eq.~\ef{e:VT-4}, the Hamiltonian can be written as the sum of 
three parts:
\begin{equation}\label{HamThreeParts}
   H(Q) = H_{\mathrm{kin}}(Q) + H_{\mathrm{pot}}(Q) + H_{\mathrm{nl}}(Q) \>,
\end{equation}
where $H_{\mathrm{kin}}$, $H_{\mathrm{pot}}$, $H_{\mathrm{nl}}$ stand for 
the kinetic, potential, and nonlinear terms, respectively. 

Let us consider the kinetic term first which itself splits into two parts:
\begin{eqnarray}\label{Hkin}
   H_{\mathrm{kin}}(Q)
   =
   h_{\mathrm{kin}}(Q_1) + h_{\mathrm{kin}}(Q_2)
   \qc
   h_{\mathrm{kin}}(Q)
   =
   \tint dx \,|\partial_x \tpsi(x,Q) |^2 \>.   
\end{eqnarray}
Using the integral definitions of \ref{s:Integrals}, we find:
{\small
\begin{eqnarray}\label{gradpsiII}
\hskip -2.3cm     
   h_{\mathrm{kin}}(Q)
   &=
   M \!
   \Biggl \{
      \frac{1}{3} \,\beta^2
      +
      p^2
      +
      \frac{\pi^2}{3} \, \frac{\Lambda^2}{\beta^2}
      +
      \frac{\beta \, \alpha^2}{8} \, I_{3}(\beta,q)
      +
      \kappa \, \frac{\alpha \, \beta \, p}{2} \, I_{1}(\beta,q)
      +
      \kappa \, \alpha \, \beta \, \Lambda \, I_{2}(\beta,q)\!
   \Biggr \} \>. 
\end{eqnarray}}
In a similar fashion, the potential term also splits into two parts:
\begin{equation}\label{Hpot}
   H_{\mathrm{pot}}(Q)
   =
   h_{\mathrm{pot}}(Q_1) + h_{\mathrm{pot}}(Q_2) \>,
\end{equation}
where
\begin{eqnarray}\notag
\hskip -1cm
   h_{\mathrm{pot}}(Q)
   &=
   \tint dx \,V_0(x) \, |\psi(x,t)|^2
   \\
   &=
   -
   \frac{ \beta M}{2} \, b^2 \tint dx
   \sech^2[\beta (x - q)] \, \sech^2(x)
   =
   -
   \frac{\beta M}{2} \,  b^2 \, I_3(\beta,q) \>.
   \label{hpot} 
\end{eqnarray}
Finally, we consider the nonlinear term. Unlike the kinetic and 
potential terms, the nonlinear term does \emph{not} split into 
two parts. Here we have
\begin{eqnarray}
   H_{\mathrm{nl}}(Q) \notag
   &=
   - \frac{\gamma}{2}
   \tint dx \,|\tPsi(x,Q)|^{4}
   \\
   &=
   - \frac{\gamma}{2}
   \tint dx\, 
   \Bigl [\,
      |\psi_1(x,t)|^4 + 2 \, |\psi_1(x,t)|^2 |\psi_2(x,t)|^2 + |\psi_2(x,t)|^4 \,
   \Bigr ] \>
   \notag \\
   &=
   h_{\mathrm{nl}}(Q_1) + c(Q_1,Q_2) + h_{\mathrm{nl}}(Q_2) \>,
   \label{HnonlinearI}
\end{eqnarray}
where
\begin{equation}\label{hnl}
   h_{\mathrm{nl}}(Q)
   =
   - \frac{\gamma}{2} \, \Bigl ( \frac{\beta \, M}{2} \Bigr )^2
   \tint dx \sech^4[\beta(x-q)]
   =
   - \frac{\gamma}{6} \, \beta \, M^2 \>.
\end{equation}
The cross term, i.e., $c(Q_1,Q_2)$ in Eq.~(\ref{HnonlinearI}) is given by
\begin{equation}
c(Q_1,Q_2)
   =
   - \frac{\gamma}{4} \, \beta_1 M_1 \, \beta_2 M_2 \, C(\, \beta_1,q_1,\beta_2,q_2 \,) \>,
   \label{cterm}
\end{equation}
which involves the mixing integral (see, \ref{ss:mixed})
\begin{equation}
C(\, \beta_1,q_1,\beta_2,q_2 \,)
   =
   \tint dx \sech^2[\beta_1(x-q_1)] \sech^2[\beta_2(x-q_2)] \>.
\end{equation}
Note that $ C(\, \beta_1,q_1,\beta_2,q_2 \,)$ is invariant under 
$ \{ \beta_1,q_1 \} \leftrightarrow  \{\beta_2,q_2 \}   $, and  
$ c(Q_1,Q_2)$ is invariant under $ \{M_1, \beta_1,q_1 \} \leftrightarrow  \{M_2,\beta_2,q_2 \}   $. 

%
%

%
%
\subsection{\label{ss:Derivatives}Derivatives of the Hamiltonian}

The Hamiltonian is made up of three terms: 
\begin{equation}\label{Hfinal}
   H(Q_1,Q_2)
   =
   h(Q_1) + c(Q_1,Q_2) + h(Q_2) \>,
\end{equation}
where
\begin{eqnarray}\notag
   h(Q)
   &=
   M \,
   \Biggl \{\,
      p^2
      +
      \frac{1}{3} \,\beta^2
      +
      \frac{\pi^2}{3} \, \frac{\Lambda^2}{\beta^2}
      +
      \beta \, \frac{d^2 - 9 \, b^2}{18} \, I_{3}(\beta,q)
      \\
      & \hspace{2em}
      +
      \kappa \, \frac{d \, \beta}{3} \, 
      \bigl [\, 
         p \, I_{1}(\beta,q) + 2 \Lambda \, I_{2}(\beta,q) \,
      \bigr ] \,
   \Biggr \}
   -
   \frac{\gamma}{6} \, \beta M^2 \>,
\label{hq}
\end{eqnarray}
and the coupling term $c(Q_1,Q_2)$ is given by Eq.~\ef{cterm}. From 
Eqs.~\ef{cterm}, \ef{Hfinal}, and~\ef{hq}, we can then determine $v_{\mu}$
\begin{equation} \label{vmu2}
v_{\mu} = \partial_{\mu} H(Q_1,Q_2)
\end{equation} 
needed to obtain the first order equations of motion~\ef{e:VT-6}. For 
the derivatives of the coupling term~\ef{cterm} we explicitly have:
\begin{subeqnarray}\label{Cderivatives}
   \fl
   c_{M_1}(Q_1,Q_2)
   &\equiv
   \partial_{M_1} c(Q_1,Q_2)
   =
   - \frac{\gamma}{4} \, \beta_1 \beta_2 M_2 \,
   C(\, \beta_1,q_1,\beta_2,q_2 \,) \>,
   \\
   \fl
   c_{M_2}(Q_1,Q_2)
   &\equiv
   \partial_{M_2} c(Q_1,Q_2)
   =
   - \frac{\gamma}{4} \, \beta_1 \beta_2 M_1 \,
   C(\, \beta_1,q_1,\beta_2,q_2 \,) \>,
   \\
   \fl
   c_{q_1}(Q_1,Q_2)
   &\equiv
   \partial_{q_1} c(Q_1,Q_2)
   =
   - \frac{\gamma}{4} \, \beta_1 M_1 \beta_2 M_2 \,
   C_{q_1}(\, \beta_1,q_1,\beta_2,q_2 \,) \>,
   \\
   \fl
   c_{q_2}(Q_1,Q_2)
   &\equiv
   \partial_{q_2} c(Q_1,Q_2)
   =
   - \frac{\gamma}{4} \, \beta_1 M_1 \beta_2 M_2 \,
   C_{q_2}(\, \beta_1,q_1,\beta_2,q_2 \,) \>,
   \\
   \fl
   c_{\beta_1}(Q_1,Q_2)
   &\equiv
   \partial_{\beta_1} c(Q_1,Q_2)
   \\
   \fl
   &=
   - \frac{\gamma}{4} \, \beta_2 M_1 M_2 \,
   \bigl [\,
      C(\, \beta_1,q_1,\beta_2,q_2 \,)
      +
      \beta_1 \, C_{\beta_1}(\, \beta_1,q_1,\beta_2,q_2 \,) \,
   \bigr ] \>,
   \notag \\
   \fl
   c_{\beta_2}(Q_1,Q_2)
   &\equiv
   \partial_{\beta_2} c(Q_1,Q_2)
   \\
   \fl
   &=
   - \frac{\gamma}{4} \, \beta_1 M_1 M_2 \,
   \bigl [\,
      C(\, \beta_1,q_1,\beta_2,q_2 \,)
      +
      \beta_2 \, C_{\beta_2}(\, \beta_1,q_1,\beta_2,q_2 \,) \,
   \bigr ] \>.
   \notag
\end{subeqnarray} 
%

\subsection{\label{ss:disterm}Dissipative term}

The dissipative term \ef{e:VT-4.1} also splits into two parts:
\begin{equation}\label{Fsplit}
   F(Q,\dot{Q})
   =
   f(Q_1,\dot{Q}_1) - f(Q_2,\dot{Q}_2) \>,
\end{equation}
where
\begin{equation}\label{fdef}
   f(Q,\dot{Q})
   =
   \rmi \tint dx\,V_1(x) \,
   \biggl \{\,
      \tpsi^{\ast}(x,Q)\, \tpsi_t(x,Q) - \tpsi^{\ast}_t(x,Q)\, \tpsi(x,Q) \,
   \biggr \} \>.
\end{equation}
Again using the integral definitions in \ref{s:Integrals}, we find:
\begin{equation}\label{fcompute}
   \fl
   f(Q,\dot{Q})
   =
   -
   \beta \, M \, d \,
   \biggl \{\,
      (\, \dot{\theta} + p \, \dot{q} \,) \, f_{1}(\beta,q)
      -
      (\, \dot{p} - 2\, \Lambda \, \dot{q} \,) \, f_{2}(\beta,q)
      -
      \dot{\Lambda} \, f_{3}(\beta,q) \,
   \biggr \} \>,
\end{equation}
where the derivatives of $f(Q,\dot{Q})$ with respect to $\dot{Q}^{\mu}$ 
are given by
\begin{subeqnarray}\label{wmu-values}
   w_{M} 
   &= 
   0 \>,
   \\
   w_{\theta}
   &=
   -
   \kappa \, \beta \, M \, d \, f_{1}(\beta,q) \,, 
   \\
   w_{q}
   &=
   -
   \kappa \, \beta \, M \, d \,
   [\, p \, f_{1}(\beta,q) + 2 \, \Lambda \, f_{2}(\beta,q) \,] \,, 
   \\
   w_{p}
   &=
   \kappa \, \beta \, M \, d \, f_{2}(\beta,q) \,, 
   \\
   w_{\beta}
   &=
   0 \>,
   \\
   w_{\Lambda}
   &=
   \kappa \, \beta \, M \, d \, f_{3}(\beta,q) \>.
\end{subeqnarray}

%
\subsection{\label{ss:varEOM}Equations of motion}

From Eqs.~\ef{vmu2} and \ef{wmu-values} we can now obtain the equations 
$u_{\mu}(Q) = v_{\mu}(Q) - w_{\mu}(Q)$. The latter read
{\small
\begin{subeqnarray}\label{umu-values}
   \fl
   u_{M}\!
   &=\!
   p^2
   +\!
   \frac{1}{3} \,\beta^2
   +\!
   \frac{\pi^2}{3} \, \frac{\Lambda^2}{\beta^2}
   +
   \beta \, \frac{d^2 - 9 \, b^2}{18} \, I_{3}(\beta,q)
   +
   \kappa \, \frac{d \, \beta}{3} \, 
   \bigl [\, 
      p \, I_{1}(\beta,q) + 2 \Lambda \, I_{2}(\beta,q) \,
   \bigr ]\!
   -\!
   \frac{\gamma}{3} \, \beta M \>,
   \\
   \fl
   u_{\theta}\!
   &=
   \kappa \, \beta \, M \, d \, f_{1}(\beta,q) \>,
   \\
   \fl
   u_{q}\!
   &=
   M \,
   \Biggl \{\,
      \beta \, \frac{d^2 - 9 \, b^2}{18} \, I_{3,q}(\beta,q)
      +
      \kappa \, \frac{2}{3} \, d \, \beta \, 
      [\, p \, f_{1}(\beta,q) + 2 \, \Lambda \, f_{2}(\beta,q) \,] \,
   \Biggr \} \>,
   \\
   \fl
   u_{p}\!
   &=
   M \, 
   \{\, 
      2 \, p
      +
      \kappa \, \frac{d \, \beta }{3} \, I_{1}(\beta,q)
      -
      \kappa \, \beta \, d \, f_{2}(\beta,q) \,
   \} \>,
   \\
   \fl
   u_{\beta}\!
   &=
   M \,
   \Biggl \{\,
      \frac{2}{3} \,\beta
      -
      \frac{2\,\pi^2}{3} \, \frac{\Lambda^2}{\beta^3}
      +
      \frac{d^2 - 9 \, b^2}{18} \, 
      [\, I_{3}(\beta,q) + \beta \, I_{3,\beta}(\beta,q) \,]
      \\
      \fl
      & \hspace{1em}
      +
      \kappa \, \frac{d}{3} \, 
      \Bigl [\, 
         p \, [\, I_{1}(\beta,q) + \beta I_{1,\beta}(\beta,q) ]
         + 
         2 \Lambda \, [\, I_{2}(\beta,q) + \beta I_{2,\beta}(\beta,q) \,] \,
      \Bigr ] \,
   \Biggr \}
   -
   \frac{\gamma}{6} \, M^2 \>,
   \notag \\
   \fl
   u_{\Lambda}\!
   &=
   M \,
   \Biggl \{\,
      \frac{2\,\pi^2}{3} \, \frac{\Lambda}{\beta^2}
      +
      \kappa \, \frac{2 \, d \, \beta}{3} \, I_{2}(\beta,q)
      -
      \kappa \, \beta \, d \, f_{3}(\beta,q) \,
   \Biggr \} \>.
\end{subeqnarray}
}
The net current splits into two blocks. We use the indices $\mu_j$ ($j=1,2$) to 
refer to the $j$th parameter, i.e., $Q_j$, and reserve $\mu$ for the index for 
both sets. Let us define
\begin{subeqnarray}\label{U12defs}
   U_{\mu_1}^{(1)}(Q_1,Q_2)
   &=
   u_{\mu_1}(Q_1) +  c_{\mu_1}(Q_1,Q_2) \>,
   \\
   U_{\mu_2}^{(2)}(Q_1,Q_2)
   &=
   u_{\mu_2}(Q_2) + c_{\mu_2}(Q_1,Q_2) \>, 
\end{subeqnarray}
where $u_{\mu}(Q)$ is given by \ef{umu-values} and the mixed currents 
$c_{\mu}(Q_1,Q_2)$ by \ef{Cderivatives}. The only terms that involve 
the mixed currents are for $\mu = \{\, M,q,\beta \,\}$. The $Q^{\mu}$ 
and $U_{\mu}(Q_1,Q_2)$ vectors are then defined by
\begin{equation}\label{QUvectors}
   Q^{\mu}
   =
   \Biggl ( \begin{array}{c}
      Q_1^{\mu} \\
      Q_2^{\mu}
   \end{array} \Biggr )
   \qc   
   U_{\mu}(Q_1,Q_2) 
   = 
   \Biggl ( \begin{array}{c}
      U_{\mu_1}^{(1)}(Q_1,Q_2) \\
      U_{\mu_2}^{(1)}(Q_1,Q_2)
   \end{array} \Biggr ) \>, 
\end{equation}
whence the equations of motion \ef{e:VT-9} with $f^{\mu\nu}(Q)$ 
given in \ef{invfmunu} become
\begin{subeqnarray}\label{Xeom}
   \dot{Q}_1^{\mu}
   &=
   g^{\mu\nu}(Q_1) \, U_{\nu}^{(1)}(Q_1,Q_2) \>,
   \label{Xeom-a} \\
   \dot{Q}_2^{\mu}
   &=
   g^{\mu\nu}(Q_2) \, U_{\nu}^{(2)}(Q_1,Q_2) \>.  
   \label{Xeom-b}
\end{subeqnarray}
This way, the associated rates are given by
\begin{subeqnarray}\label{QdotsU}
   \fl
   \dot{M}
   &=
   - \kappa \, d \, \beta \, M \, f_{1}(\beta,q) \>,
   \\
   \fl
   \dot{\theta}
   &=
   -
   p^2
   +
   \frac{2}{3} \,\beta^2
   +
   \beta \, \frac{d^2 - 9 \, b^2}{36} \, 
   [\, 3 I_{3}(\beta,q) + \beta \, I_{3,\beta}(\beta,q) \,]
   +
   \kappa \, d \, p \, \beta \, f_{2}(\beta,q)
   \\
   \fl
   & \hspace{0em}
   +
   \kappa \, \frac{d}{3} \, 
   \Bigl [\, 
      p \, [\, I_{1}(\beta,q) + \beta I_{1,\beta}(\beta,q) ]
      + 
      2 \Lambda \, [\, 3 \,I_{2}(\beta,q) + \beta I_{2,\beta}(\beta,q) \,] \,
   \Bigr ] \,
   -
   \gamma \, \frac{5}{12} \, \beta \, M \>, 
   \notag \\
   \fl
   \dot{q}
   &=
   2 \, p
   +
   \kappa \, \frac{d \, \beta }{3} \, 
   [\, I_{1}(\beta,q) - 3 \, f_{2}(\beta,q) \,] \>,
   \\
   \fl
   \dot{p}
   &=
   -
   \beta \, \frac{d^2 - 9 \, b^2}{18} \, I_{3,q}(\beta,q)
   +
   \kappa \, d \, \frac{1}{3} \, \beta  \, 
   \bigl [\, 
      p \, f_{1}(\beta,q) - 4 \Lambda \, f_{2}(\beta,q) \,
   \bigr ] \>,
   \\
   \fl
   \dot{\beta}
   &=
   - 
   \kappa \, d \, \frac{\beta^2}{2} \, f_{1}(\beta,q)
   -
   4 \, \beta \, \Lambda
   -
   \kappa \, d \, \frac{2}{\pi^2} \, \beta^4 \,
   [\, 2 \, I_{2}(\beta,q) - 3 \, f_{3}(\beta,q) \, ] \>, 
   \\
   \fl
   \dot{\Lambda}
   &=
   \frac{4}{\pi^2} \, \beta^4 
   -
   4 \, \Lambda^2
   +
   \frac{\beta^3(d^2 - 9 \, b^2)}{3 \pi^2} \, 
   [\, I_{3}(\beta,q) + \beta \, I_{3,\beta}(\beta,q) \,]
   \\
   \fl
   & \hspace{-1em}
   +
   \kappa \, d \, \frac{2 \, \beta^3}{\pi^2} \, 
   \Bigl \{\, 
      p \, [\, I_{1}(\beta,q) + \beta I_{1,\beta}(\beta,q) ]
      + 
      2 \Lambda \, [\, I_{2}(\beta,q) + \beta I_{2,\beta}(\beta,q) \,] \,
   \Bigr \} \,
   -
   \gamma \, \frac{\beta^3}{\pi^2} \, M \>,
\end{subeqnarray}
where we add the mixed derivative terms $R_{\mu}$ to them 
\begin{subeqnarray}\label{Crates1}
   \fl
   R_{M_1}
   &=
   0 \>,
   \\
   \fl
   R_{\theta_1}
   &=
   c_{M_1}(q_1,\beta_1,q_2,\beta_2)
   +
   \frac{\beta_1}{2 M_1} \, c_{\beta_1}(q_1,\beta_1,q_2,\beta_2) \> 
   \\
   \fl
   &=
   - 
   \frac{3\,\gamma}{8} \, \beta_1 \beta_2 M_2 \, C(\, \beta_1,q_1,\beta_2,q_2 \,)
   -
   \frac{\gamma}{8} \, \beta_1^2 \beta_2 M_2 \, C_{\beta_1}(\, \beta_1,q_1,\beta_2,q_2 \,) \,, 
   \notag \\
   \fl   
   R_{q_1}
   &=
   0 \>,
   \\
   \fl
   R_{p_1}
   &=
   - \frac{1}{M_1} \, c_{q_1}(q_1,\beta_1,q_2,\beta_2)
   =
   \frac{\gamma}{4} \, \beta_1 \beta_2 M_2 \,
   C_{q_1}(\, \beta_1,q_1,\beta_2,q_2 \,) \>,
   \\
   \fl
   R_{\beta_1}
   &=
   0 \>,
   \\
   \fl
   R_{\Lambda_1}
   &=
   \frac{1}{M_1} \frac{6 \beta_1^3}{\pi^2} \, 
   c_{\beta_1}(q_1,\beta_1,q_2,\beta_2)
   \\
   \fl
   &=
   -
   \gamma \, \frac{3}{2\pi^2} \, \beta_1^3 \beta_2 M_2 \,
   \bigl [\,
      C(\, \beta_1,q_1,\beta_2,q_2 \,)
      +
      \beta_1 \, C_{\beta_1}(\, \beta_1,q_1,\beta_2,q_2 \,) \,
   \bigr ] \>,
   \notag   
\end{subeqnarray}
and
\begin{subeqnarray}\label{Crates2}
   \fl
   R_{M_2}
   &=
   0 \>,
   \\
   \fl
   R_{\theta_2}
   &=
   c_{M_2}(q_1,\beta_1,q_2,\beta_2)
   +
   \frac{\beta_2}{2 M_2} \, c_{\beta_2}(q_1,\beta_1,q_2,\beta_2) \> 
   \\
   \fl
   &=
   - 
   \frac{3\,\gamma}{8} \, \beta_1 \beta_2 M_1 \, C(\, \beta_1,q_1,\beta_2,q_2 \,)
   -
   \frac{\gamma}{8} \, \beta_1 \beta_2^2 M_1 \, C_{\beta_2}(\, \beta_1,q_1,\beta_2,q_2 \,) \,, 
   \notag \\
   \fl   
   R_{q_2}
   &=
   0 \>,
   \\
   \fl
   R_{p_2}
   &=
   - \frac{1}{M_2} \, c_{q_2}(q_1,\beta_1,q_2,\beta_2)
   =
   \frac{\gamma}{4} \, \beta_1 \beta_2 M_1 \,
   C_{q_2}(\, \beta_1,q_1,\beta_2,q_2 \,) \>,
   \\
   \fl
   R_{\beta_2}
   &=
   0 \>,
   \\
   \fl
   R_{\Lambda_2}
   &=
   \frac{1}{M_2} \frac{6 \beta_2^3}{\pi^2} \, 
   c_{\beta_2}(q_1,\beta_1,q_2,\beta_2)
   \\
   \fl
   &=
   -
   \gamma \, \frac{3}{2\pi^2} \, \beta_1 \beta_2^3 M_1 \,
   \bigl [\,
      C(\, \beta_1,q_1,\beta_2,q_2 \,)
      +
      \beta_2 \, C_{\beta_2}(\, \beta_1,q_1,\beta_2,q_2 \,) \,
   \bigr ] \>,
   \notag   
\end{subeqnarray}
for the sets $Q_1$ and $Q_2$, respectively. The rate equations for
the $Q_2$ parameters are identical to the above with $Q_1$ and $Q_2$ 
interchanged. Recall that $\kappa_1 = +1$ for the $Q_1$ and $\kappa_2 = -1$ 
for the $Q_2$ parameters. The full equations for the CCs are obtained 
by adding either \ef{Crates1} or \ef{Crates2} to  \ef{QdotsU}.
This completes the derivation of the rate equations for the 12 variational 
parameters.

%
\subsection{\label{ss:reduction}Reduction to 8 CCs}

A reasonable approximation in the stable regime is to consider that the two 
components do {\it not} separate over their time evolution and that their 
widths are similar. Thus, we can can assume that $q_1(t)\equiv q_2(t)= q(t) $, 
$p_1(t)\equiv p_2(t)=p(t)$, $\beta_1(t)\equiv\beta_2(t)= \beta(t) $, and 
$\Lambda_1(t)\equiv\Lambda_2(t)=\Lambda(t)$ in our analysis. 
Using the formalism of Sec.~\ref{s:collective} we can directly obtain the equations 
of motion for these 8 CCs:
\begin{equation}\label{eightparams}
   Q^{\mu}
   =
   \{\, M_1(t),\theta_1(t),M_2(t),\theta_2(t),q(t),p(t),\beta(t),\Lambda(t) \,\} \>,
\end{equation}
from Eq.~\ef{e:VT-9}.  The results we obtain directly can also be obtained by a reduction process from the 12 collective coordinates equation by setting $q_1=q_2$, $p_1=p_2$, $\beta_1=\beta_2$, and $\Lambda_1=\Lambda_2$, and then defining the time derivatives as follows:
%
%
\begin{equation}\label{qavedef}
   \dot{q}
   =
   \frac{M_1 \, \dot{q}_1 + M_2, \dot{q}_2}{M_1 + M_2} \,
   \Big |_{q_1=q_2}
\end{equation}
with similar relations for the average values of $\dot{p}$, $\dot{\beta}$, and $\dot{\Lambda}$.  

Upon following the steps described above, the equations of motion for the case of 8 CCs
are given by
\begin{subeqnarray}\label{dotQFred}
   \fl
   \dot{M}_1
   &=
   - d \, \beta \, M_1 \, f_{1}(\beta,q) \>,
   \\
   \fl
   \dot{\theta}_1
   &=
   -
   p^2
   +
   \frac{2}{3} \,\beta^2
   +
   \beta \, \frac{d^2 - 9 \, b^2}{36} \, 
   [\, 3 I_{3}(\beta,q) + \beta \, I_{3,\beta}(\beta,q) \,]
   +
   d \, p \, \beta \, f_{2}(\beta,q)
   \\
   \fl
   & \hspace{0em}
   +
   \frac{d}{3} \, 
   \Bigl [\, 
      p \, [\, I_{1}(\beta,q) + \beta I_{1,\beta}(\beta,q) ]
      + 
      2 \Lambda \, [\, 3 \,I_{2}(\beta,q) + \beta I_{2,\beta}(\beta,q) \,] \,
   \Bigr ] \,
   -
   \gamma \, \frac{5}{12} \, \beta \, (M_1 + M_2) \>, 
   \notag \\
   \fl
   \dot{M}_2
   &=
   d \, \beta \, M_2 \, f_{1}(\beta,q) \>,
   \\
   \fl
   \dot{\theta}_2
   &=
   -
   p^2
   +
   \frac{2}{3} \,\beta^2
   +
   \beta \, \frac{d^2 - 9 \, b^2}{36} \, 
   [\, 3 I_{3}(\beta,q) + \beta \, I_{3,\beta}(\beta,q) \,]
   -
   d \, p \, \beta \, f_{2}(\beta,q)
   \\
   \fl
   & \hspace{0em}
   -
   \frac{d}{3} \, 
   \Bigl [\, 
      p \, [\, I_{1}(\beta,q) + \beta I_{1,\beta}(\beta,q) ]
      + 
      2 \Lambda \, [\, 3 \,I_{2}(\beta,q) + \beta I_{2,\beta}(\beta,q) \,] \,
   \Bigr ] \,
   -
   \gamma \, \frac{5}{12} \, \beta \, (M_1 + M_2) \>, 
   \notag \\
   \fl
   \dot{q}
   &=
   2 \, p
   +
   \frac{M_1 - M_2}{M_1 + M_2}  \, \frac{d \, \beta }{3} \, 
   [\, I_{1}(\beta,q) - 3 \, f_{2}(\beta,q) \,] \>,
   \\
   \fl
   \dot{p}
   &=
   -
   \beta \, \frac{d^2 - 9 \, b^2}{18} \, I_{3,q}(\beta,q)
   +
   \frac{M_1 - M_2}{M_1 + M_2}  \, d \, \frac{1}{3} \, \beta  \, 
   \bigl [\, 
      p \, f_{1}(\beta,q) - 4 \Lambda \, f_{2}(\beta,q) \,
   \bigr ] \>,
   \\
   \fl
   \dot{\beta}
   &=
   -
   4 \, \beta \, \Lambda
   - 
   \frac{M_1 - M_2}{M_1 + M_2} \,
   \Biggl \{\,
      d \, \frac{\beta^2}{2} \, f_{1}(\beta,q)
      +
      d \, \frac{2}{\pi^2} \, \beta^4 \,
      [\, 2 \, I_{2}(\beta,q) - 3 \, f_{3}(\beta,q) \, ] \,
   \Biggr \} \>, 
   \\
   \fl
   \dot{\Lambda}
   &=
   \frac{4}{\pi^2} \, \beta^4 
   -
   4 \, \Lambda^2
   +
   \frac{\beta^3(d^2 - 9 \, b^2)}{3 \pi^2} \, 
   [\, I_{3}(\beta,q) + \beta \, I_{3,\beta}(\beta,q) \,]
   -
   \gamma \, \frac{\beta^3}{\pi^2} \, (M_1 + M_2)
   \\
   \fl
   & \hspace{-1em}
   +
   \frac{M_1 - M_2}{M_1 + M_2}  \, d \, \frac{2 \, \beta^3}{\pi^2} \, 
   \Biggl \{\, 
      p \, [\, I_{1}(\beta,q) + \beta I_{1,\beta}(\beta,q) ]
      + 
      2 \Lambda \, [\, I_{2}(\beta,q) + \beta I_{2,\beta}(\beta,q) \,] \,
   \Biggr \} \>. 
\end{subeqnarray}
%
%
The results of the reduction agree with the direct determination of \ef{dotQFred} which is a consistency check.

%
\subsection{\label{ss:smallamps}Small amplitude approximation}

%
%
\begin{figure}[t]
\centering
\includegraphics[height=.18\textheight, angle =0]{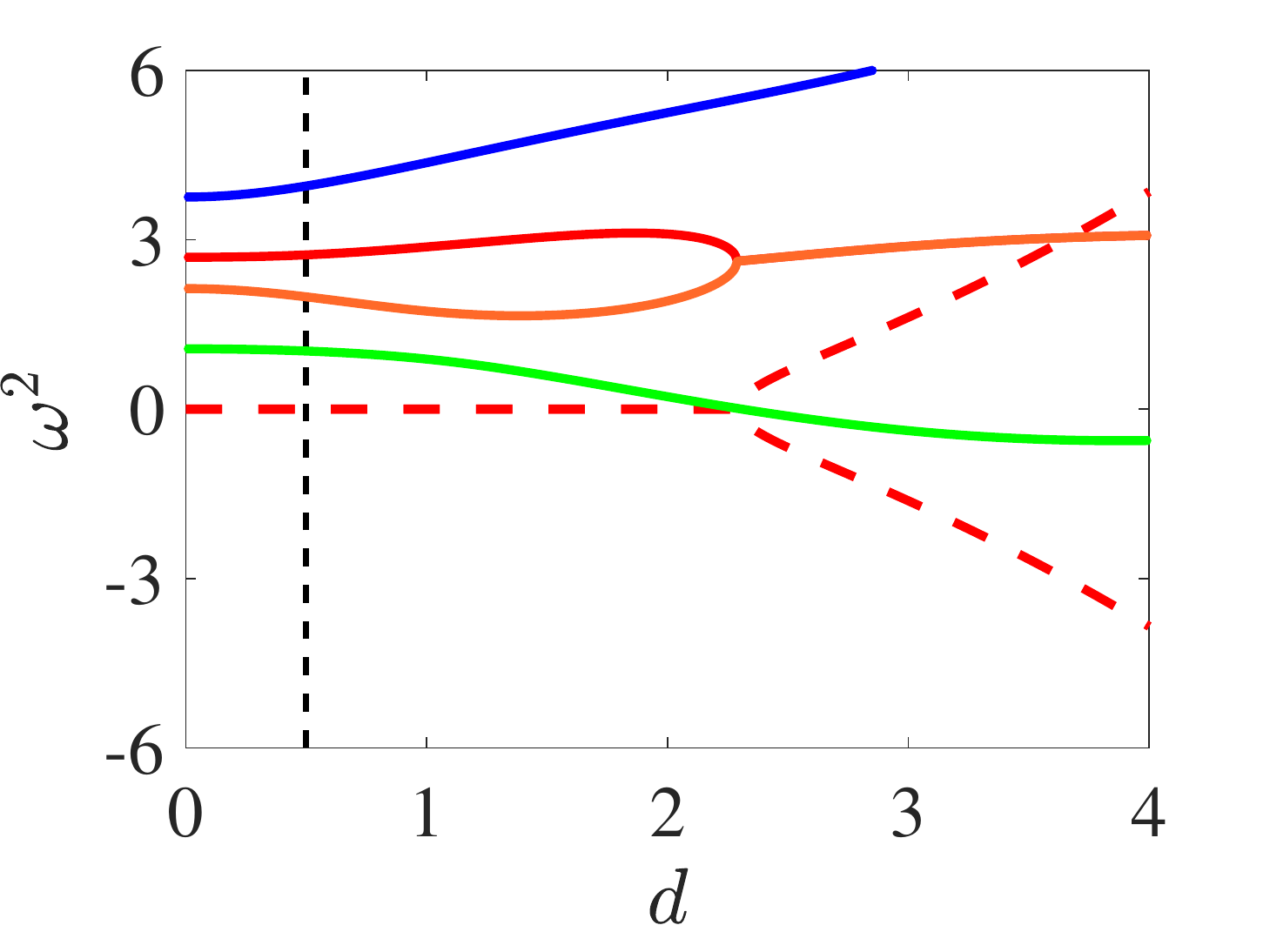}
\includegraphics[height=.18\textheight, angle =0]{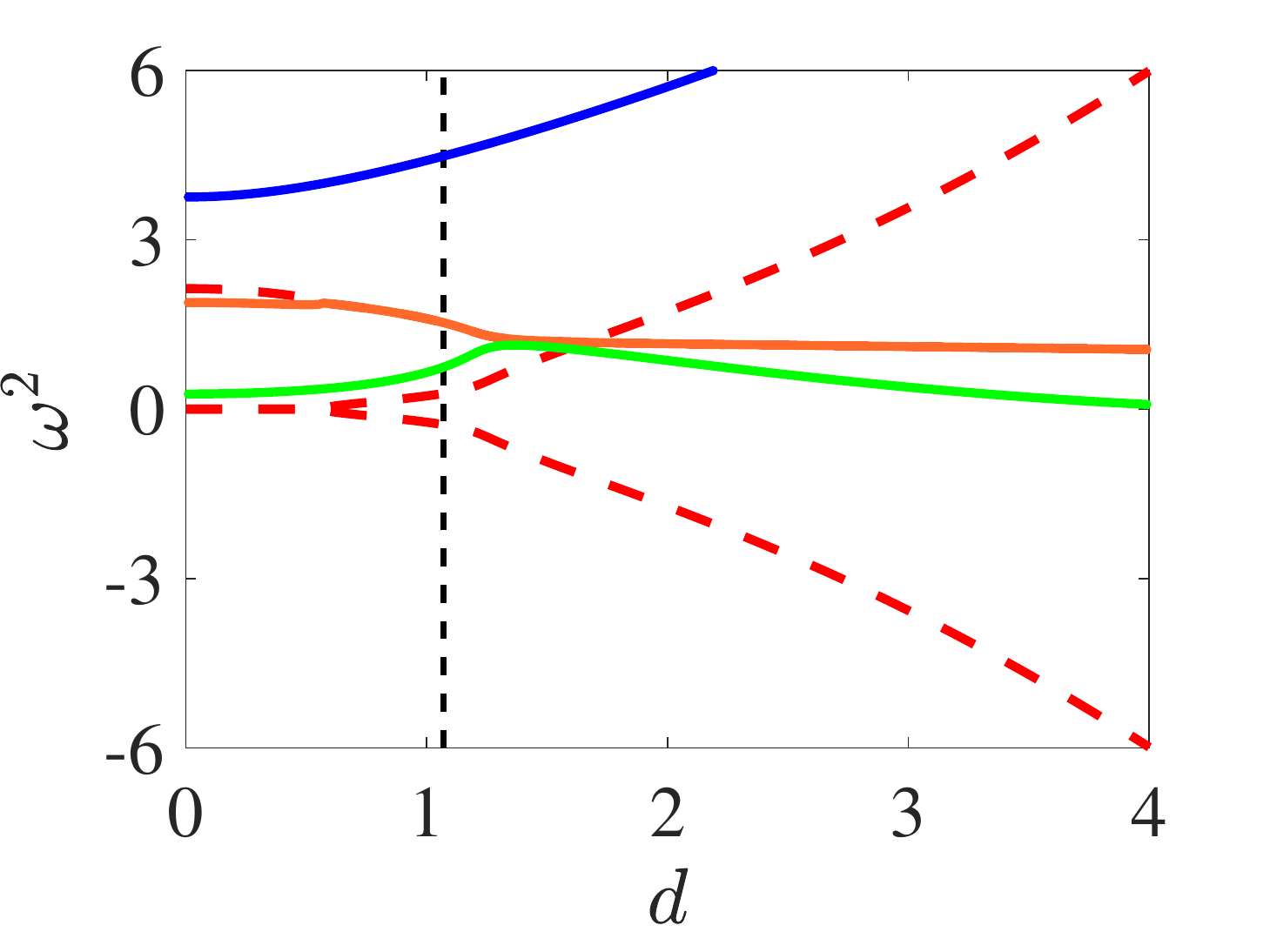}
\caption{\label{f:eigenvalues}Eigenvalues of $W^{\mu}{}_{\nu}[Q_{0}]$ as 
functions of $d$ for the case of $b = 1$, $A_1 = 1$ (left panel) and for 
$b = 1/2$, $A_1 = 1/4$ (right panel). The solid red and orange lines are 
the real parts of two of the eigenvalues, and the red dotted lines are 
the imaginary parts. The vertical lines correspond to values of $d$ used 
for illustrating a stable (left panel) and unstable (right panel) solution, 
respectively.}
\end{figure}
%
%
The parametric regions of stability can be determined by performing a
small amplitude approximation to the full CC equations. This way, we
will be able to obtain the eigenfrequencies of the linearized system 
which is derived by expanding the rate equations to first order in the
parameters using the expansions of integrals of the~\ref{ss:expand}. 
Indeed, for Eqs.~\ef{QdotsU}, \ef{Crates1}, and \ef{Crates2}, we find
\begin{subeqnarray}\label{SO-Qrates}
   \fl
   \delta \dot{M}
   &=
   -\kappa d \, \frac{\pi}{4} \, M_0 \,\delta q \>,
   \\
   \fl
   \delta \dot{\theta}
   &=
   -
   1
   +
   \frac{5}{6} \,
   \Biggl [\, 2 + \frac{d^2}{9} - b^2 - \frac{\gamma \, M_0}{2} \, \Biggr ]
   -
   \frac{5}{12} \, \gamma \,\delta M
   +
   \kappa \, d \, \frac{2 \, \pi}{9} \, \delta p
   \\
   \fl
   & \hspace{2em}
   +
   \Biggl \{\,
      -
      \frac{\pi^2}{45} \, \Biggl [\, \frac{d^2}{9} - b^2 \, \Biggr ]
      +
      \frac{1}{2} \,
      \Biggl [\,
         \frac{8}{3} + \frac{d^2}{9} - b^2 - \frac{5 \, \gamma M_0}{6} \,
      \Biggr ] \,
   \Biggr \} \, \delta \beta \>,
   \notag \\
   \fl
   \delta \dot{q}
   &=
   2 \, \delta p 
   - 
   \kappa \, d \,  
   \Biggl ( \frac{4 \pi}{9} - \frac{\pi^3}{16} \Biggr ) \, \delta \beta \>,
   \\
   \fl
   \delta \dot{p}
   &=
   \frac{8}{15} \, \Biggl ( \frac{d^2}{9} - b^2 \Biggr ) \, \delta q 
   - 
   \kappa \, d \, \frac{2 \, \pi}{9} \, \delta \Lambda \>,
   \\
   \fl
   \delta \dot{\beta}
   &=
   \kappa \, d \, \Biggl ( \frac{\pi}{4} - \frac{10}{3 \pi} \Biggr ) \, \delta q
   -
   4 \, \delta \Lambda \>,
   \\
   \fl
   \delta \dot{\Lambda}
   &=
   \frac{2}{\pi^2} \, \Biggl [\, 2 + \frac{d^2}{9} - b^2 - \gamma \, \frac{M_0}{2} \, \Biggr ]
   -
   \frac{\gamma}{\pi^2} \, \delta M 
   +
   \frac{\kappa \, d}{3 \, \pi} \, \delta p
   \\
   \fl
   & \hspace{2em}
   +
   \Biggl \{\,
      \frac{4}{15} \, 
      \Biggl [\, b^2 - \frac{d^2}{9} \, \Biggr ]
      +
      \frac{6}{\pi^2} \,
      \Bigl [\,
         \frac{8}{3} + \frac{d^2}{9} - b^2 - \gamma \frac{M_0}{2} \,
      \Bigr ] \,
   \Biggr \} \, \delta \beta \>.
   \notag
\end{subeqnarray}
For the mixing rates, we find
\begin{subeqnarray}\label{SO-Crates}
   \fl
   \delta R_{\theta_1}
   &=
   -
   \frac{5}{12} \, \gamma \, M_{2,0}
   -
   \Biggl [\,
      \frac{1}{4} - \frac{\pi^2}{90} \,
   \Biggr ] \, \gamma \, M_{2,0} \, \delta \beta_1
   -
   \frac{5}{12} \gamma \, \delta M_{2}
   -
   \Biggl [\,
      \frac{1}{6} + \frac{\pi^2}{90} \,
   \Biggr ] \, \gamma \, M_{2,0} \, \delta \beta_2 \>,
   \\
   \fl
   \delta R_{p_1}
   &=
   - \frac{4}{15} \, \gamma \, M_{2,0} \,
   (\, \delta q_1 - \delta q_2 \, ) \>,
   \\
   \fl
   \delta R_{\Lambda_1}
   &=
   - 
   \frac{1}{\pi^2} \, \gamma \, M_{2,0}
   +
   \Biggl [\,
      \frac{2}{15} - \frac{3}{\pi^2} \,
   \Biggr ] \, \gamma \, M_{2,0} \, \delta \beta_1
   -
   \frac{1}{\pi^2} \, \gamma \, \delta M_{2}
   -
   \frac{2}{15} \, \gamma M_{2,0} \, \delta \beta_2 \>,
   \\
   \fl
   \delta R_{\theta_2}
   &=
   -
   \frac{5}{12} \, \gamma \, M_{1,0}
   -
   \frac{5}{12} \, \gamma \, \delta M_{1}
   -
   \Biggl [\,
      \frac{1}{6} + \frac{\pi^2}{90} \,
   \Biggr ] \, \gamma \, M_{1,0} \, \delta \beta_1
   -
   \Biggl [\,
      \frac{1}{4} - \frac{\pi^2}{90} \,
   \Biggr ] \, \gamma \, M_{1,0} \, \delta \beta_2 \>,
   \\
   \fl
   \delta R_{p_2}
   &=
   - \frac{4}{15} \, \gamma \, M_{1,0} \, (\, \delta q_2 - \delta q_1 \, ) \>,
   \\
   \fl
   \delta R_{\Lambda_2}
   &=
   -
   \frac{1}{\pi^2} \, \gamma \, M_{1,0}
   -
   \frac{1}{\pi^2} \, \gamma \, \delta M_{1}
   -
   \frac{2}{15} \, \gamma \, M_{1,0} \, \delta \beta_1
   +
   \Biggl [\,
      \frac{2}{15} - \frac{3}{\pi^2} \,
   \Biggr ] \, M_{1,0} \, \delta \beta_2 \>,
\end{subeqnarray}
and by using Eqs.~\ef{SO-Qrates} and \ef{SO-Crates}, we find 
the rate equations for $Q_1$
\begin{subeqnarray}\label{12paramSO}
   \fl
   \delta \dot{M}_1
   &=
   -\kappa_1 \, d \, \frac{\pi}{4} \, M_{1,0} \,\delta q_1 \>,
   \\
   \fl
   \delta \dot{\theta}_1
   &=
   -
   1   
   -
   \frac{5}{12} \, \gamma \, (\, \delta M_1 + \delta M_2 \, )
   +
   \kappa_1 \, d \, \frac{2 \, \pi}{9} \, \delta p_1
   \\
   \fl
   & \hspace{1em}
   +
   \Biggl \{\,
      -
      \frac{\pi^2}{45} \, \Biggl [\, \frac{d^2}{9} - b^2 - \frac{\gamma \, M_{2,0} }{2} \, \Biggr ]
      +
      \frac{1}{2} \,
      \Biggl [\,
         \frac{8}{3} + \frac{d^2}{9} - b^2 - \frac{5 \, \gamma M_{1,0}}{6} \,
      \Biggr ] \,
   \Biggr \} \, \delta \beta_1 \>,
   \notag \\
   \fl
   \delta \dot{q}_1
   &=
   2 \, \delta p_1 
   - 
   \kappa_1 \, d \,  
   \Biggl ( \frac{4 \pi}{9} - \frac{\pi^3}{16} \Biggr ) \, \delta \beta_1 \>,
   \\
   \fl
   \delta \dot{p}_1
   &=
   \frac{8}{15} \, \Biggl [\,  \frac{d^2}{9} - b^2 \, \Biggr ] \, \delta q_1
   -
   \frac{4}{15} \, \gamma \, M_{2,0} \, ( \delta q_1 - \delta q_2 )
   - 
   \kappa_1 \, d \, \frac{2 \, \pi}{9} \, \delta \Lambda_1 \>,
   \\
   \fl
   \delta \dot{\beta}_1
   &=
   \kappa_1 \, d \, \Biggl ( \frac{\pi}{4} - \frac{10}{3 \pi} \Biggr ) \, \delta q_1
   -
   4 \, \delta \Lambda_1 \>,
   \\
   \fl
   \delta \dot{\Lambda}_1
   &=
   -
   \frac{1}{\pi^2} \, \gamma \, ( \delta M_1 + \delta M_2 )
   +
   \frac{\kappa_1 \, d}{3 \, \pi} \, \delta p_1
   \\
   \fl
   & \hspace{1em}
   +
   \Biggl \{\,
      \frac{4}{\pi^2}
      +
      \frac{4}{15} \, 
      \Biggl [\, b^2 - \frac{d^2}{9} \, \Biggr ] \,
   \Biggr \} \, \delta \beta_1
   +
   \frac{2}{15} \, \gamma \, M_{2,0} \, ( \, \delta \beta_1 -\delta \beta_2 ) \>,
   \notag
\end{subeqnarray}
where we have used Eq.~\ef{A1plusA2} as well. The equations for the $Q_2$ 
variables are obtained from the above by interchanging $1 \leftrightarrow 2$.
Notice also that the $\delta$-rate equations [cf. Eqs.~\ef{12paramSO}] vanish
when all $\delta Q^{\mu}$ are set to zero. For the $\delta \dot{\theta}_1$ term
we obtain $\delta \dot{\theta}_1 = -1$, as required. Also, when we set $Q_1 = Q_2$, 
we obtain the respective equations for the 8-parameter case.

We turn our focus now on Eqs.~\ef{12paramSO} and the ones corresponding to $Q_2$.
Those could be written in the following form
\begin{equation}\label{deltaQform}
   \delta \dot{Q}^{\mu}
   =
   M^{\mu}{}_{\nu}[Q_0] \, \delta Q^{\nu} \>,
\end{equation}
from which we find:
\begin{equation}\label{QeigenvalueEq}
   \delta \ddot{Q}^{\mu} 
   + 
   W^{\mu}{}_{\nu}[Q_0] \, \delta Q^{\nu}
   =
   0
   \>, \qquad
   W^{\mu}{}_{\nu}[Q_0]
   =
   -
   M^{\mu}{}_{\sigma}[Q_0] \, M^{\sigma}{}_{\nu}[Q_0] \>,
\end{equation}
where $W^{\mu}{}_{\nu}(Q_0)$ is Hermitian with 
\begin{equation}\label{Wform}
   M^{\mu}{}_{\nu}[Q_{0}]
   =
   \Biggl ( \begin{array}{cc}
      A[Q_{1,0},Q_{2,0}] & B[Q_{1,0},Q_{2,0}] \\
      B[Q_{2,0},Q_{1,0}] & A[Q_{2,0},Q_{1,0}]
   \end{array} \Biggr ) \>. 
\end{equation}
We can ignore the $\delta \dot{\theta}_i$ equations since the rest of 
the equations do not couple with them.  The $\mu$ and $\nu$ indices then run over the ten values:
\begin{equation}\label{eq:munuindices}
   \{\, \delta M_1,\delta q_1,\delta p_1,\delta \beta_1,\delta \Lambda_1,
        \delta M_2,\delta q_2,\delta p_2,\delta \beta_2,\delta \Lambda_2 \, \} \>.
\end{equation}
The matrices $A$ and $B$ are then $5 \times 5$ matrices given by:
\begin{equation}\label{Amat}
   \fl
   A[Q_{1,0},Q_{2,0}]
   =
   \left ( \begin{array}{ccccc}
      0 & - \kappa_1 d \pi \gamma M_{1,0}/4 & 0 & 0 & 0 \\
      0 & 0 & 2 & - \kappa_1 d c & 0 \\
      0 & -2 a_2 & 0 & 0 & - \kappa_1 d 2\pi/9 \\
      0 & \kappa_1 d e & 0 & 0 & -4 \\
      - \gamma/\pi^2 & 0 & \kappa_1 d /(3 \pi) & 4/\pi^2 + a_2 & 0
   \end{array} \right ) \>,   
\end{equation}
and
\begin{equation}\label{Bmat}
   \fl
   B[Q_{1,0},Q_{2,0}]
   =
   \left ( \begin{array}{ccccc}
      0 & 0 & 0 & 0 & 0 \\
      0 & 0 & 0 & 0 & 0 \\
      0 & 4 \gamma M_{2,0}/15 & 0 & 0 & 0 \\ 
      0 & 0 & 0 & 0 & 0 \\
      - \gamma/\pi^2 & 0 & 0 & -2 \gamma M_{2,0}/15 & 0
   \end{array} \right ) \>,
\end{equation}
where we have set
\begin{equation}\label{acedefs}
   c = \frac{4 \pi}{9} - \frac{\pi^3}{16}
   \qc
   e = \frac{\pi}{4} - \frac{10}{3 \pi}
   \qc
   a_i
   =
   \frac{4}{15} \,
   \Biggl [\, b^2 - \frac{d^2}{9} + \frac{\gamma M_{i,0}}{2} \, \Biggr ] \>.
\end{equation}

Plots of the eigenvalues $\omega_i^2$ of the matrix $W^{\mu}{}_{\nu}[Q_{0}]$ 
as functions of $d$ for fixed values of $A_1(:= A_{1}(0))$ and $b$ are shown
in the panels of Fig.~\ref{f:eigenvalues}. In particular, there are two cases 
shown with five eigenvalues, one of which is a zero eigenvalue. It can be discerned
from the panels that as $d$ is increased, some of the eigenvalues become complex, 
indicating the emergence of instability and blow-up of the wave functions. For the
left panel of Fig.~\ref{f:eigenvalues}, and at the intersection of the vertical 
dotted line at $d=1/2$, we find the following five doubly degenerate eigenvalues:
\begin{subeqnarray}\label{12CCeigenvaluesCaseI}
   \omega^2_{\mathrm{12CC}} 
   &= 
   \{\, 3.95008, 2.73053, 1.98677, 1.02733, 0 \, \} \>,
   \label{12CCeigenvaluesCaseI-a} \\
   \tau_{\mathrm{12CC}}
   =
   (2\pi/\omega_{\mathrm{12CC}})
   &=
   \{\, 3.16138, 3.80239, 4.45765, 6.19906 \,\} \>,
   \label{12CCeigenvaluesCaseI-b}
\end{subeqnarray}
which indicates a stable solution. On the other hand, the 8-parameter 
CC approximation gives two doubly degenerate eigenvalues for this case:
\begin{equation}\label{8CCeigenvaluesCaseI}
   \omega^2_{\mathrm{8CC}} = \{\,2.71099, 1.03419 \,\} \>,
\end{equation}
which are very close to two of the eigenvalues found in Eq.~\ef{12CCeigenvaluesCaseI-a}.  
For the right panel of Fig.~\ref{f:eigenvalues}, and at the intersection of the vertical 
dotted line now at $d=1.07$, we find the following five doubly degenerate eigenvalues:
\begin{subeqnarray}\label{12CCeigenvaluesCaseII}
   \fl
   \omega^2_{\mathrm{12CC}} 
   &= 
   \{\, 4.48071, 1.52926 + 0.276596 \,\rmi, 1.52926 - 0.276596 \,\rmi, 0.743116, 0 \, \} \>,
   \label{12CCeigenvaluesCaseII-a} \\
   \fl
   \tau_{\mathrm{12CC}}
   &=
   (2\pi/\omega_{\mathrm{12CC}})
   \label{12CCeigenvaluesCaseII-b} \\
   \fl
   &=
   \{\, 2.96829, 5.02 + 0.450327 \,\rmi, 5.02 - 0.450327 \,\rmi, 7.28872 \, \} \>.
   \notag
\end{subeqnarray}

Because of the decoupling of the small oscillation equations in the 8-CC approximation 
one can write down an analytic expression for the two oscillation frequencies in terms
of $b$ and $d$, and so determine regimes of instability analytically. Indeed, we find
\begin{eqnarray}
 \delta \ddot{q} - [\, A \, \delta q + B \, \delta \Lambda \,] &&= 0 \>,
   \label{e:eigen-1a} \\
   \delta \ddot{\Lambda} - [\, D \, \delta q + E \, \delta \Lambda \,] &&= 0 \>,
   \label{e:eigen-1}
\end{eqnarray}
where
\begin{eqnarray} \label{e:eigen-2}
   A
  & &=
 \frac{1024 \left(d^2-9 b^2\right)+5 \left(3 \pi ^2-40\right) \left(9 \pi ^2-64\right) d^2
   (1-2 r)^2}{8640} \>,
   \label{e:eigen-2a} \\
   B
  &&=
-\frac{1}{12} \pi  \left(3 \pi ^2-16\right) d (2 r-1)\>,
   \label{e:eigen-2b} \\
   D
   &&=
-\frac{d (2 r-1) \left(\pi ^2 \left(423 b^2-47 d^2-540\right)+2 \pi ^4 \left(d^2-9
   b^2\right)+3600\right)}{270 \pi ^3}\>,
   \label{e:eigen-2c} \\
   E
   &&=
\frac{2}{135} \left(-72 b^2+d^2 \left(-20 r^2+20 r+3\right)-\frac{1080}{\pi ^2}\right).
\end{eqnarray}
The two eigenfrequencies are
\begin{equation}\label{e:eigen-3}
   \omega_\pm^2
   = \frac{1}{2} 
    \left(- (A + E) \pm \sqrt{ (A - E)^2 + 4 \, B D }  \right) \>.
\end{equation}
Note that $\omega^2_{\pm}$ can be written entirely in terms of $A_1, b$ 
and $d$,  since $r= M_1/(M_1+M_2)= 2 A_1^2/(M_1+M_2)$, and 
$(M_1+M_2)=2 (2 + d^2/9-b^2)$ with $M_{1,2}:= M_{1,2}(0)$. 

%
\section{\label{s:dynamicresults}Dynamical Results}

\begin{figure}[t]
\centering
\includegraphics[width=0.45\columnwidth]{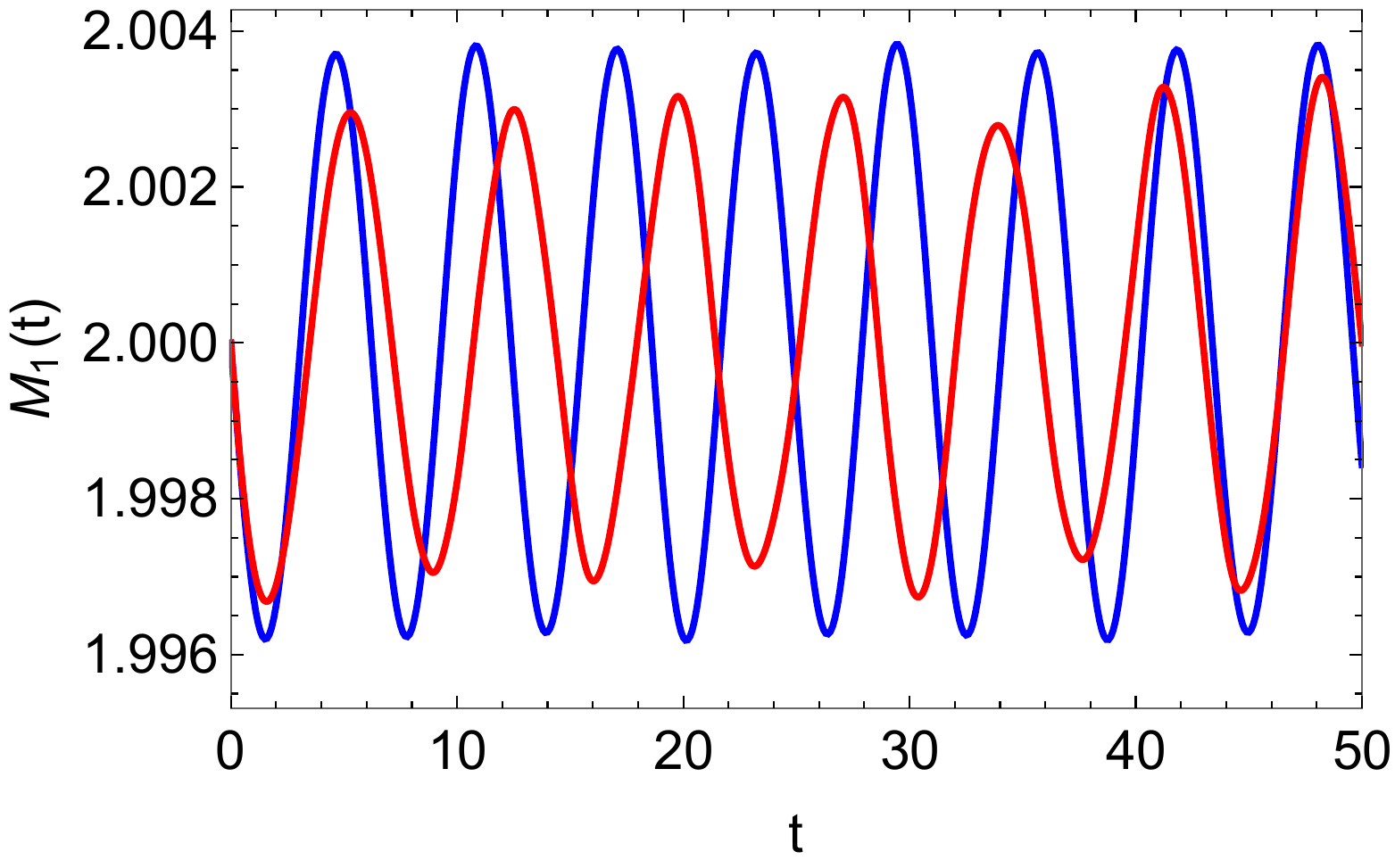}
\includegraphics[width=0.45\columnwidth]{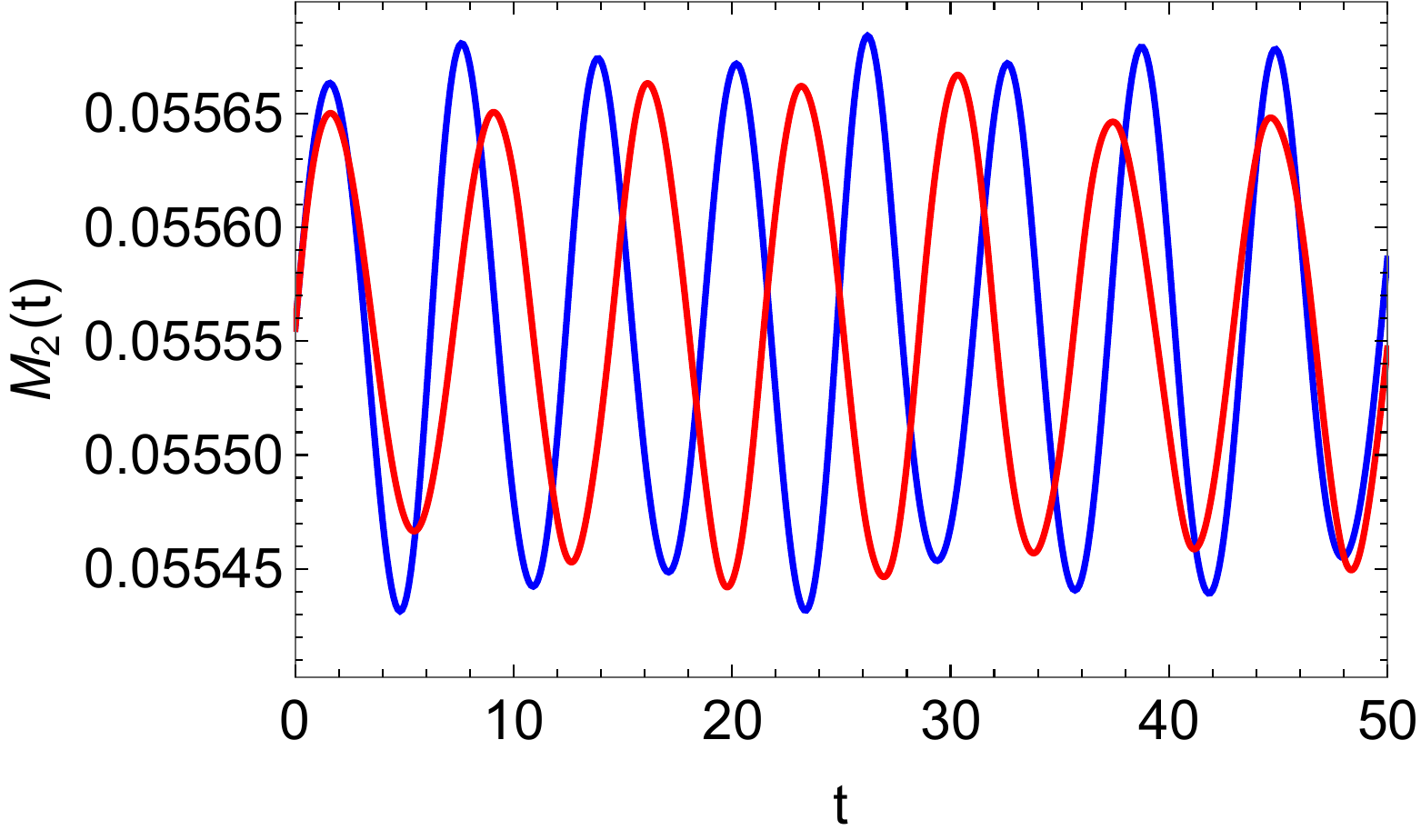}
\includegraphics[width=0.45\columnwidth]{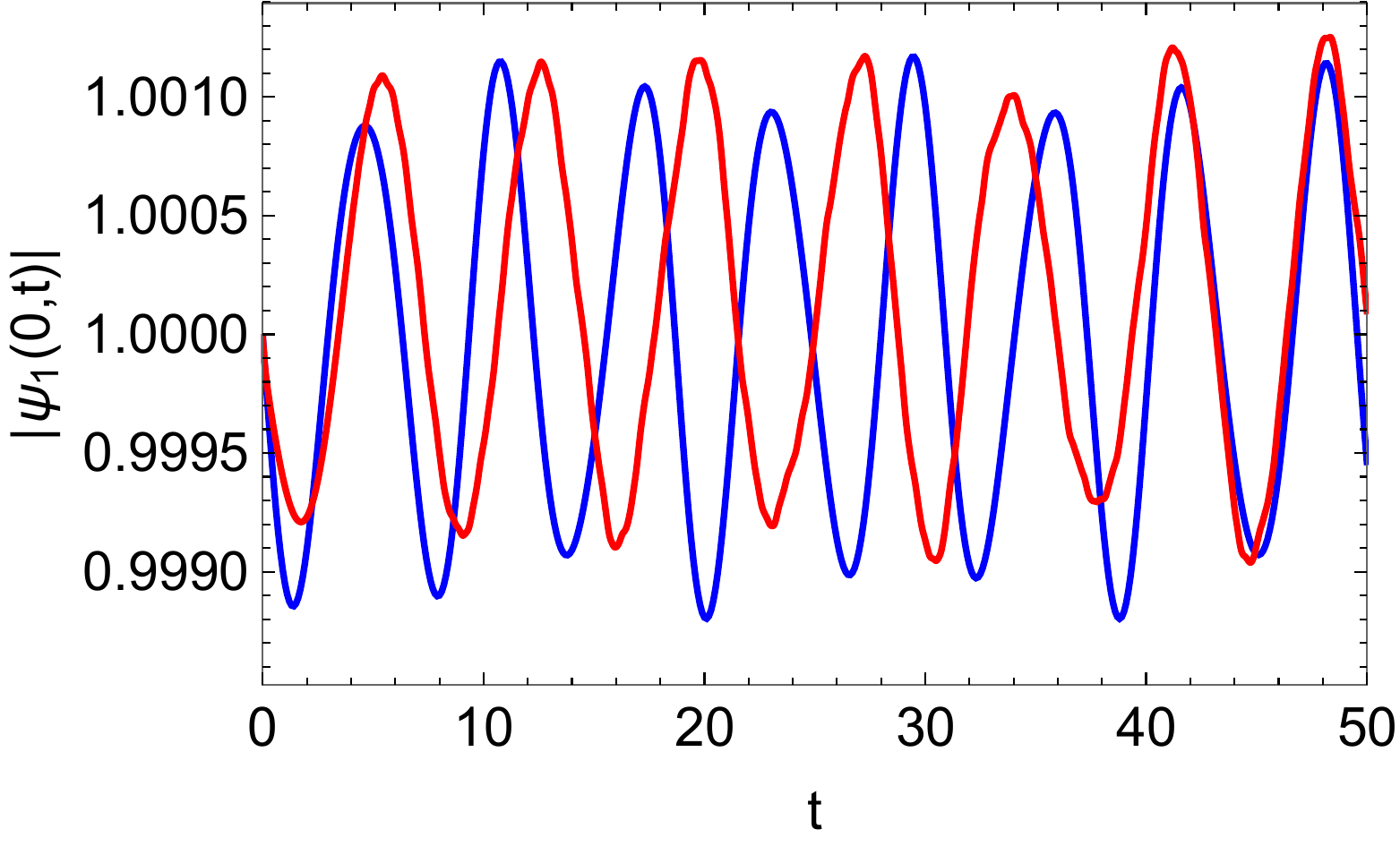}
\includegraphics[width=0.45\columnwidth]{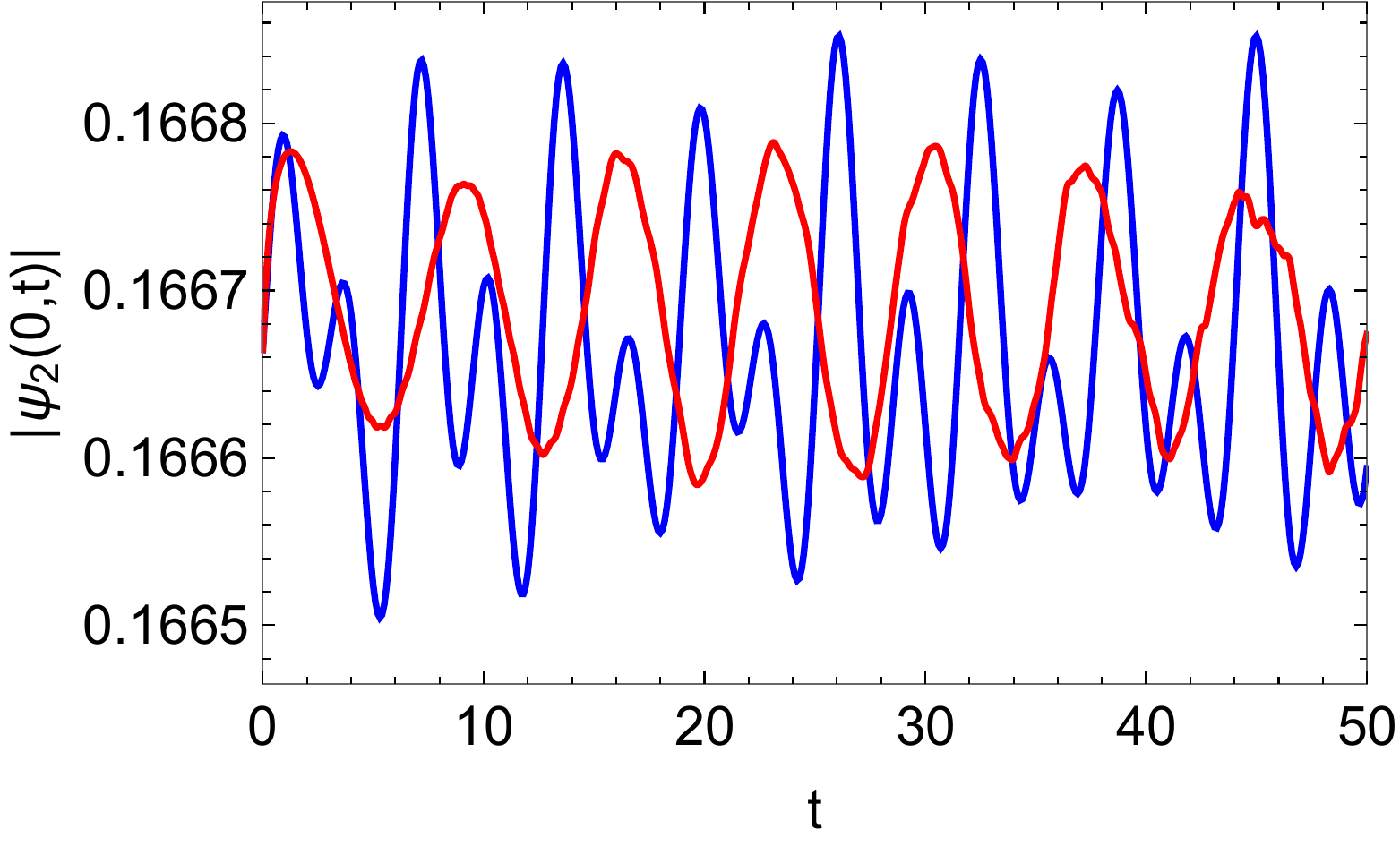}
\includegraphics[width=0.45\columnwidth]{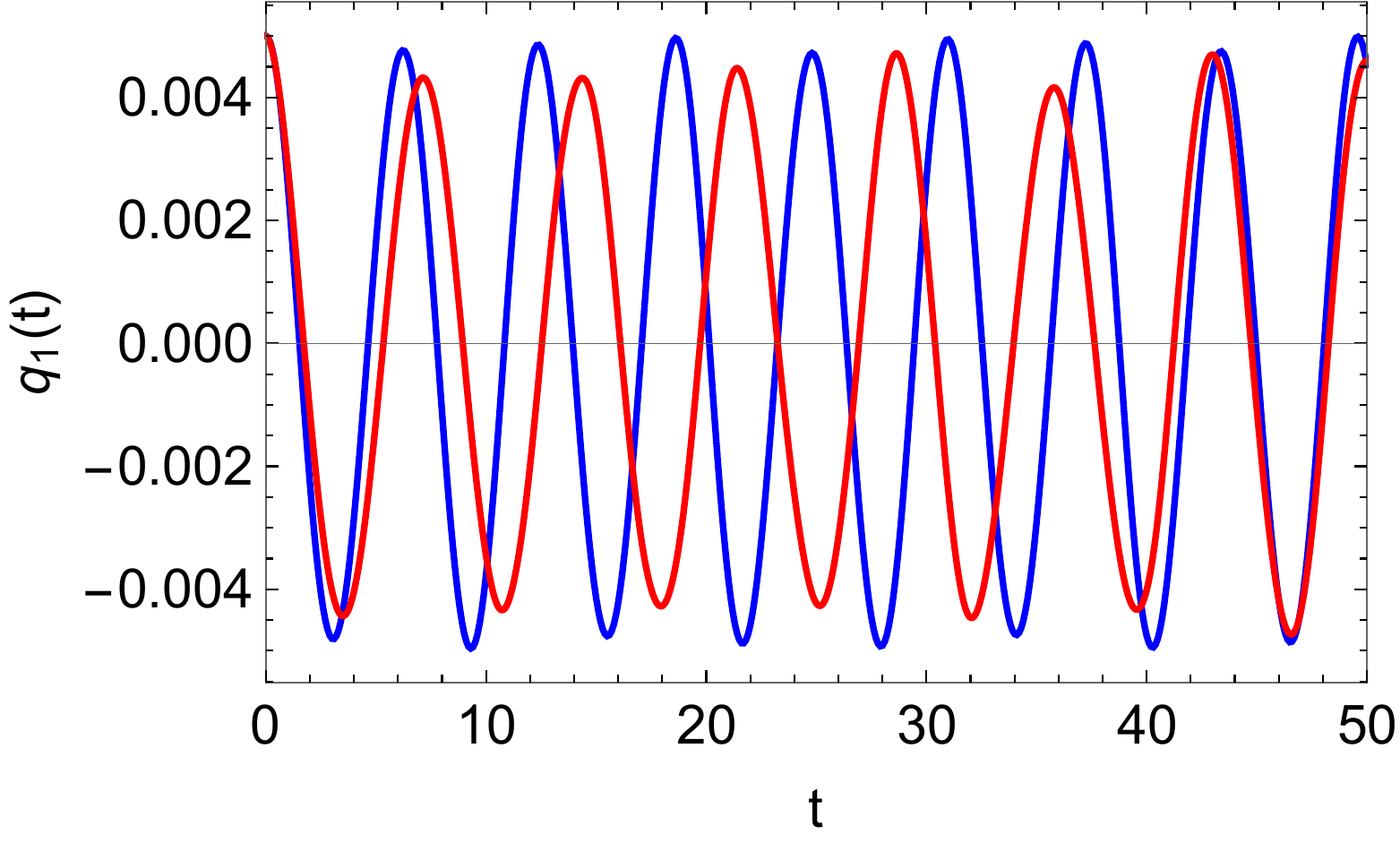}
\includegraphics[width=0.45\columnwidth]{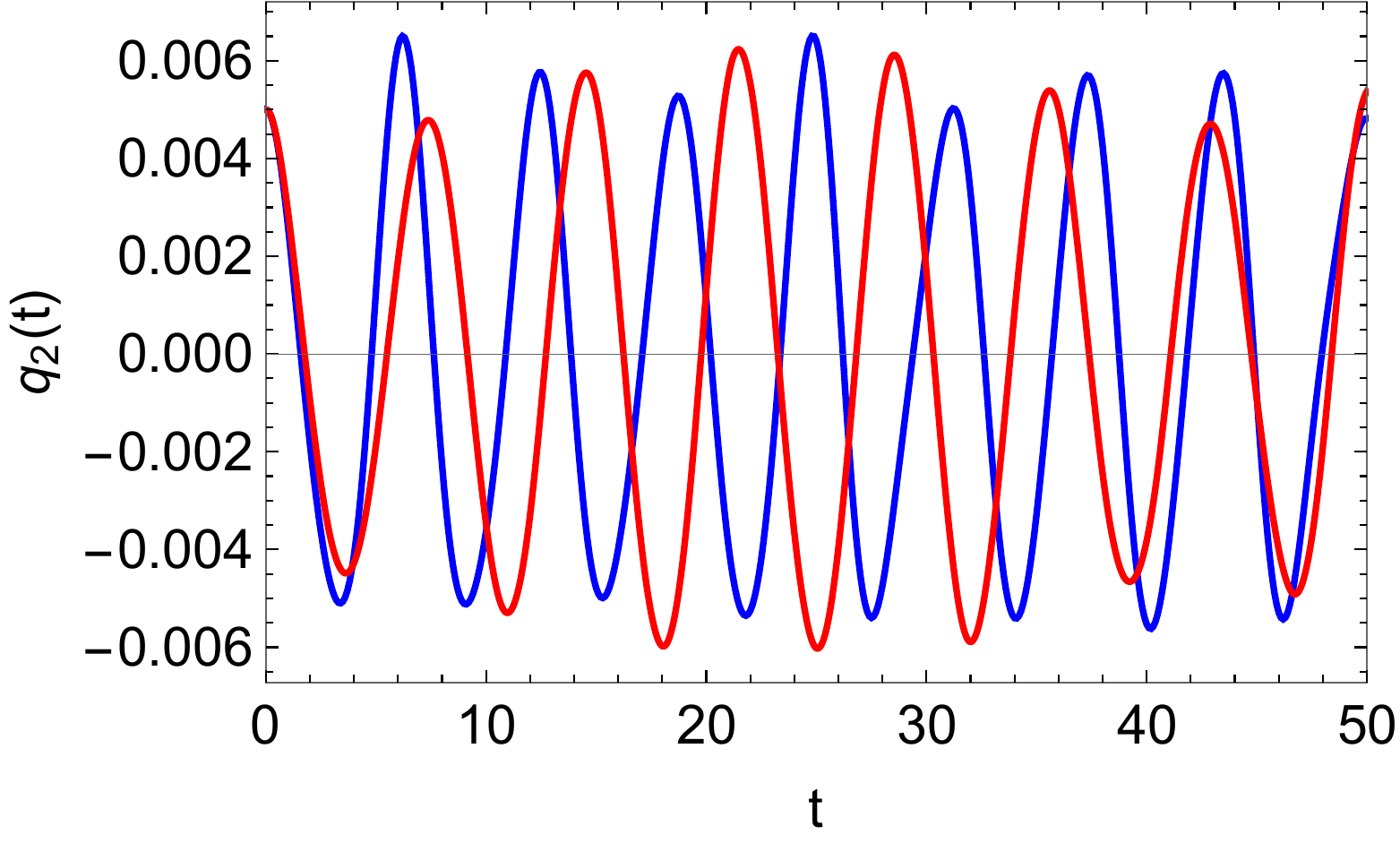}
\caption{\label{f:12parms}Results of the 12-parameter 
variational calculation for the stable case corresponding to
$A_1 = 1$, $b = 1$, and $d = 1/2$ (in blue) compared with 
a numerical calculation of the coupled NLSEs (in red). The 
left column depicts the temporal evolution of $M_{1}(t)$ 
(top), $|\psi_{1}(0,t)|$ (middle), and $q_{1}$ (bottom), 
whereas the right column presents the same quantities but
for the second component.}
\end{figure}

In this section, we present our results on the variational approximations. We begin
our discussion here by considering the stable regime as this was illustrated in the 
left panel of Figure~\ref{f:eigenvalues} corresponding to parameter values of $A_1 = 1$, 
$b = 1$ and $d = 1/2$. The initial conditions were chosen to be the values given in 
Eq.~\ef{Q12initial} but perturbed with $\delta q_{0} = 0.005$. Our numerical 
results on the 12-parameter variational calculation are shown in the panels of 
Fig.~\ref{f:12parms} with solid blue lines where we compare them to the respective 
numerical calculation of the coupled NLSEs represented by solid red lines (see, 
Sec.~\ref{s:CompAnalysis} for details). It can be discerned from the panels of 
Fig.~\ref{f:12parms} that the numerical solutions are reproduced reasonably well. 
However, the variational calculation seems to predict several additional frequency
modes which are not observed in the numerical simulation of the NLSEs.  

\begin{figure}[t]
\centering
\includegraphics[width=0.45\columnwidth]{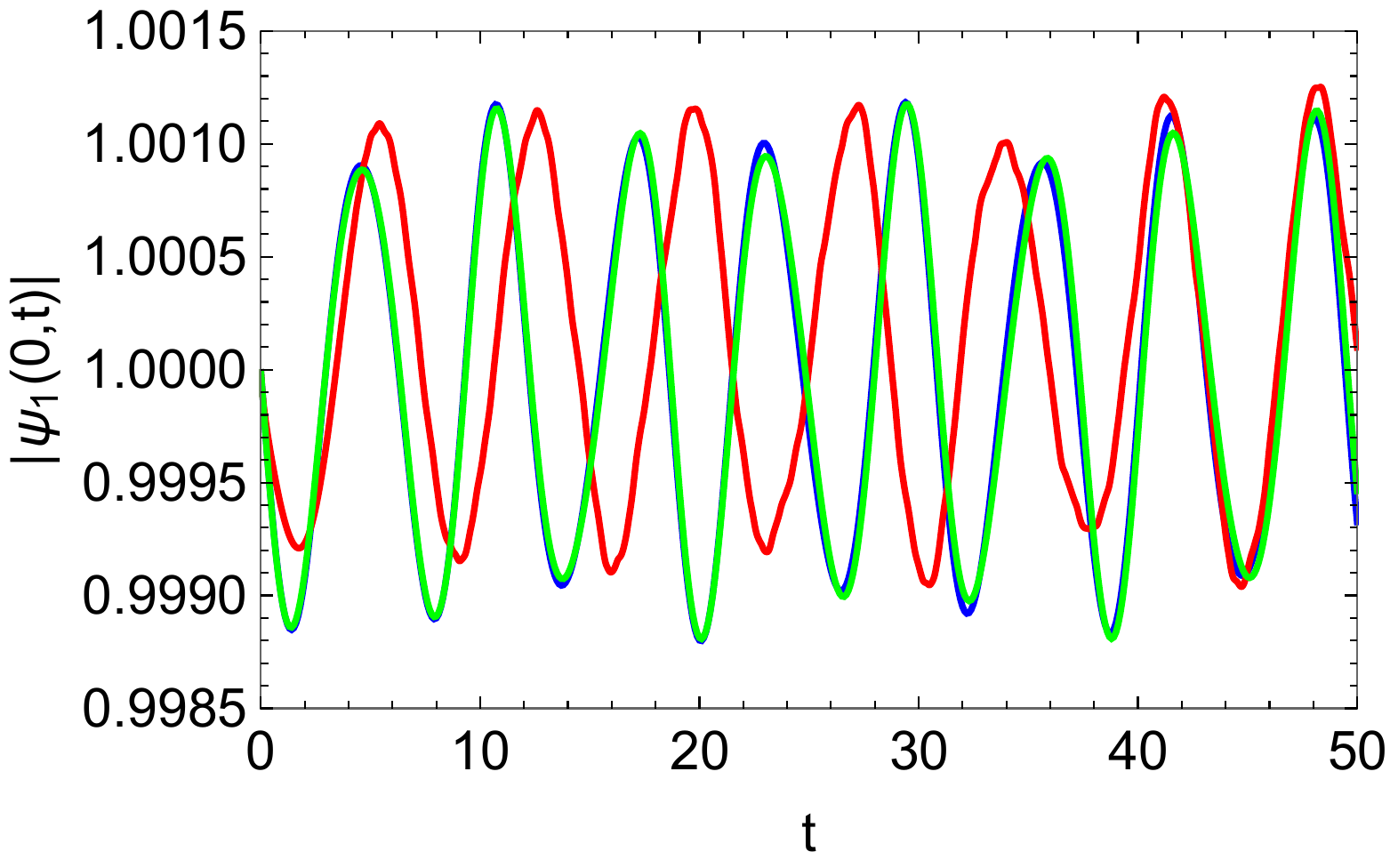}
\includegraphics[width=0.45\columnwidth]{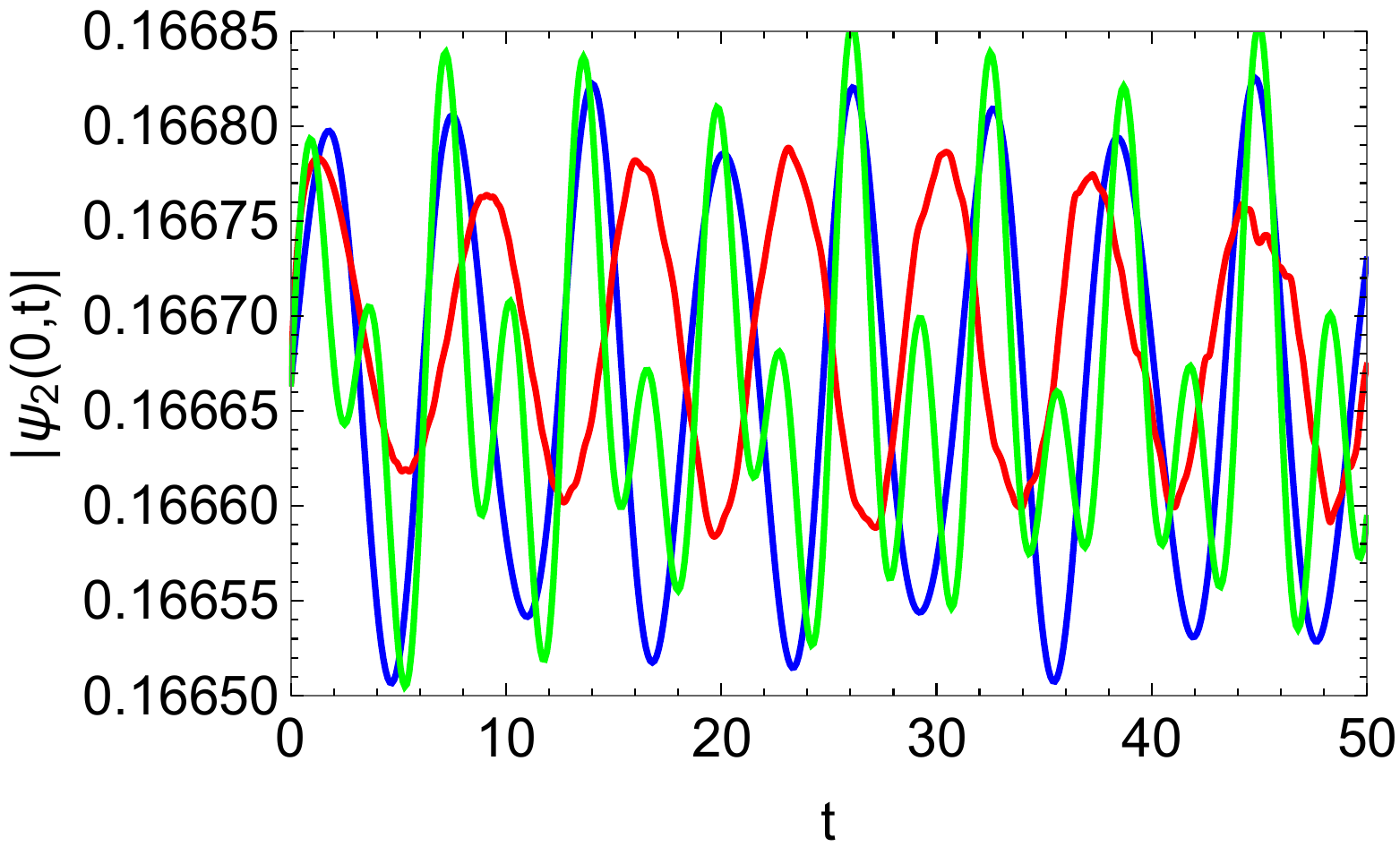}
\includegraphics[width=0.45\columnwidth]{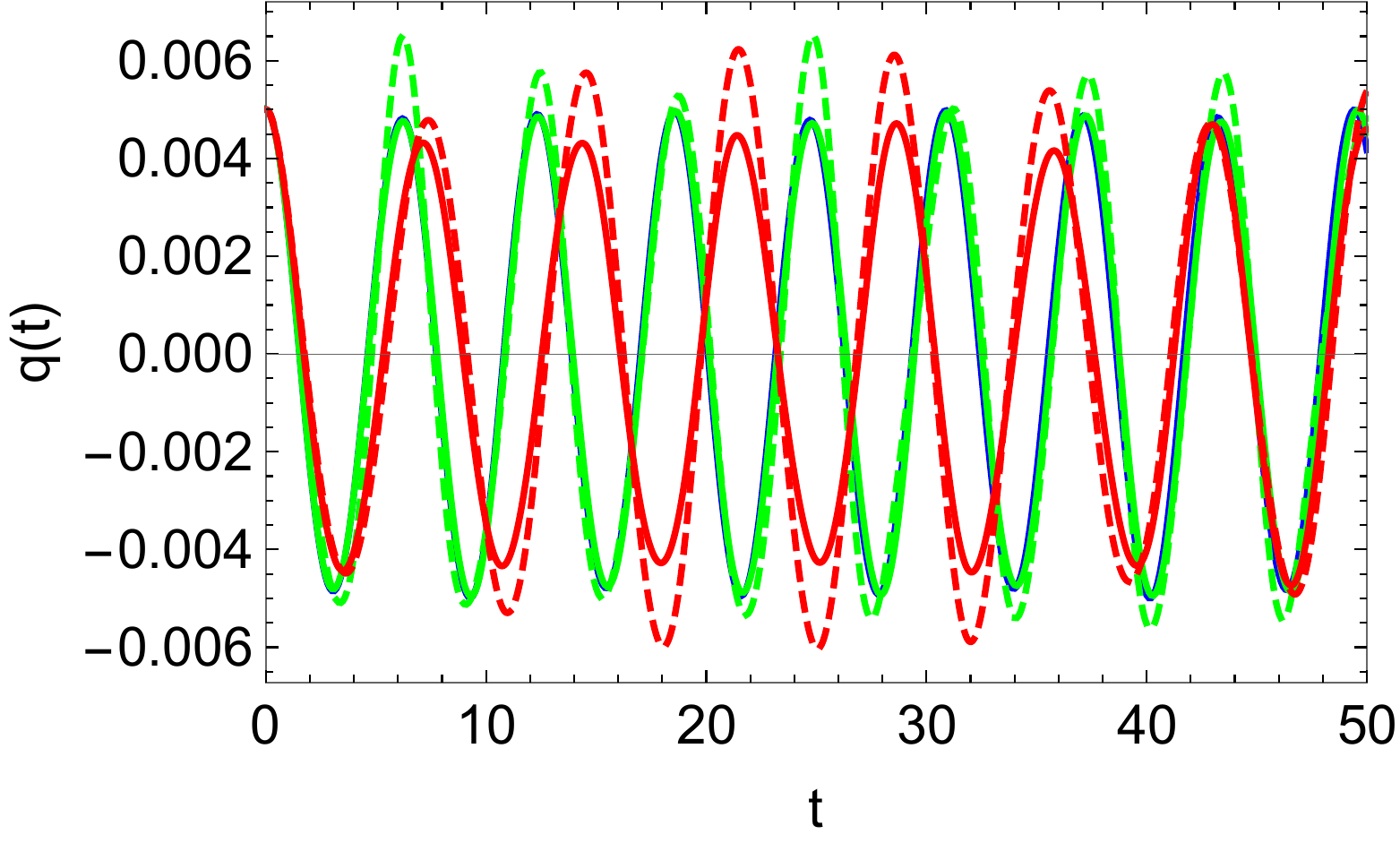}
\includegraphics[width=0.45\columnwidth]{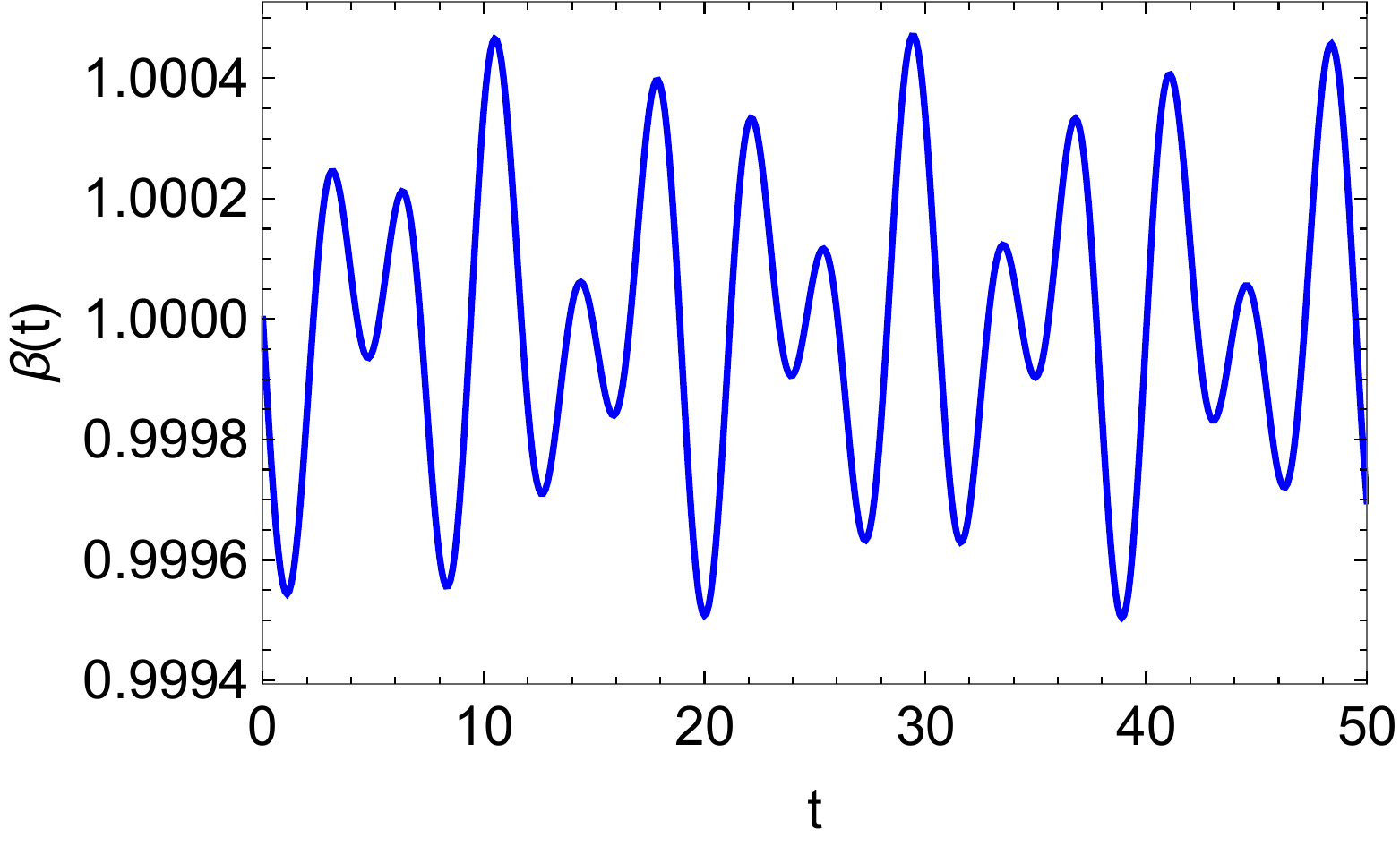}
\caption{\label{f:AllResults} 
Comparison of the 8-parameter variational calculation (in blue) 
with the 12-parameter variational calculation results (in green), 
and the numerical calculation of the coupled NLSEs (in red) for 
the stable solution is shown in the top and bottom left panels.
In particular, the top left and right panels showcase the temporal
evolution of the amplitudes of the first and second components,
respectively, whereas the bottom left depicts the evolution of 
$q(t)$. The bottom right panel showcases the average value of 
$\beta(t)$ from the 12-parameter variational calculation.}
\end{figure}

Furthermore, and in the stable regime, we compare the different variational calculations 
in Fig.~\ref{f:AllResults}. There is very little difference between the 8- and 12-parameter
calculations for $|\psi_1(0,t)|$, however the results for $|\psi_2(0,t)|$ differ 
significantly as a result of additional frequency components in the 12-parameter 
ansatz. For the $q(t)$ results, the 8-parameter case only calculates an average value
whereas the 12-parameter ansatz produced two different results for $q_1(t)$ and $q_2(t)$
as does the numerical calculation. Solutions of the linearized 12-parameter equations
\ef{12paramSO} are indistinguishable from the full 12-parameter calculation. This is 
because the amplitudes of the variational parameters in this case are quite small. 

%
%
%
%

%
%
%
%

\begin{figure}[t]
\centering
\includegraphics[width=0.45\columnwidth]{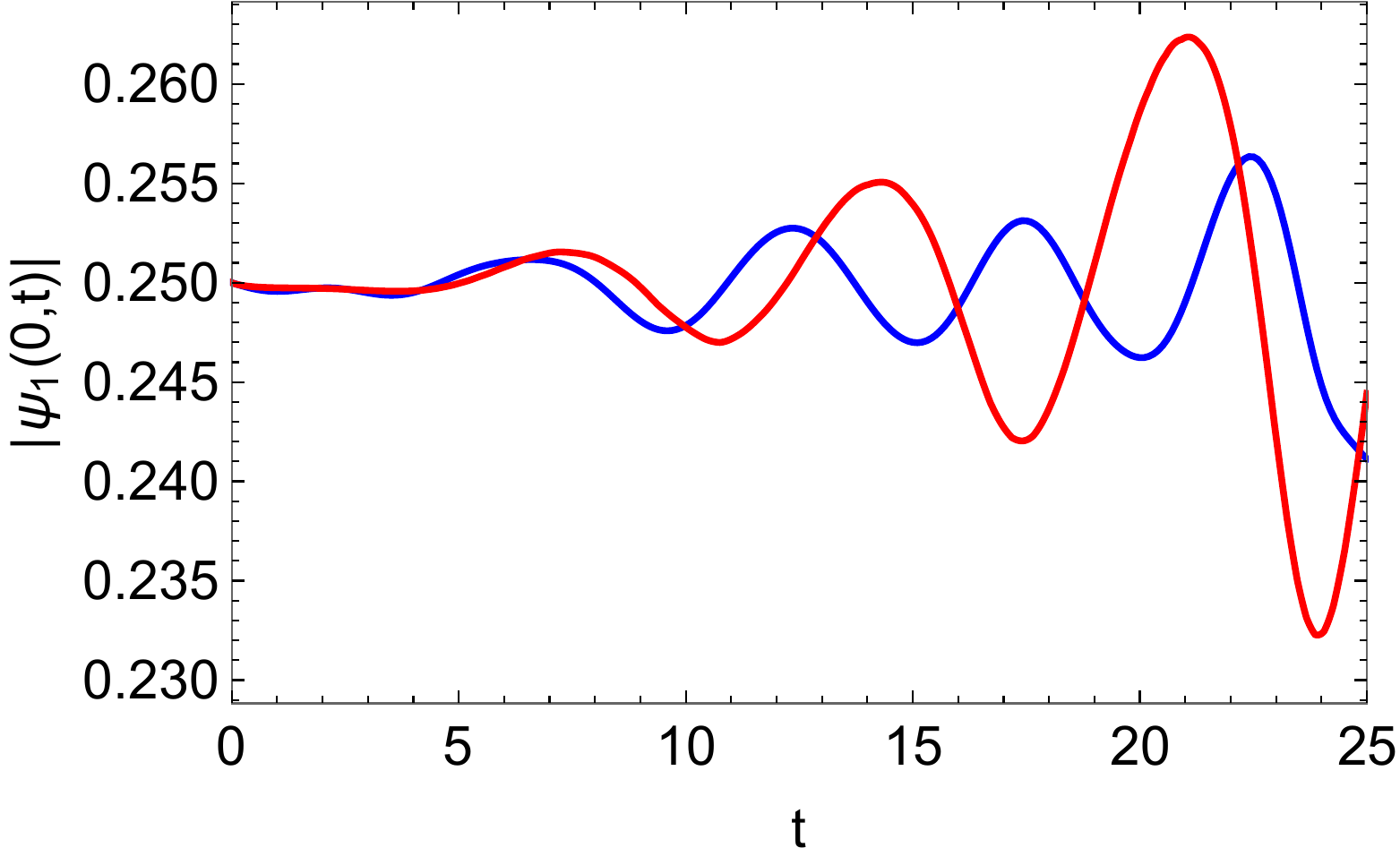}
\includegraphics[width=0.45\columnwidth]{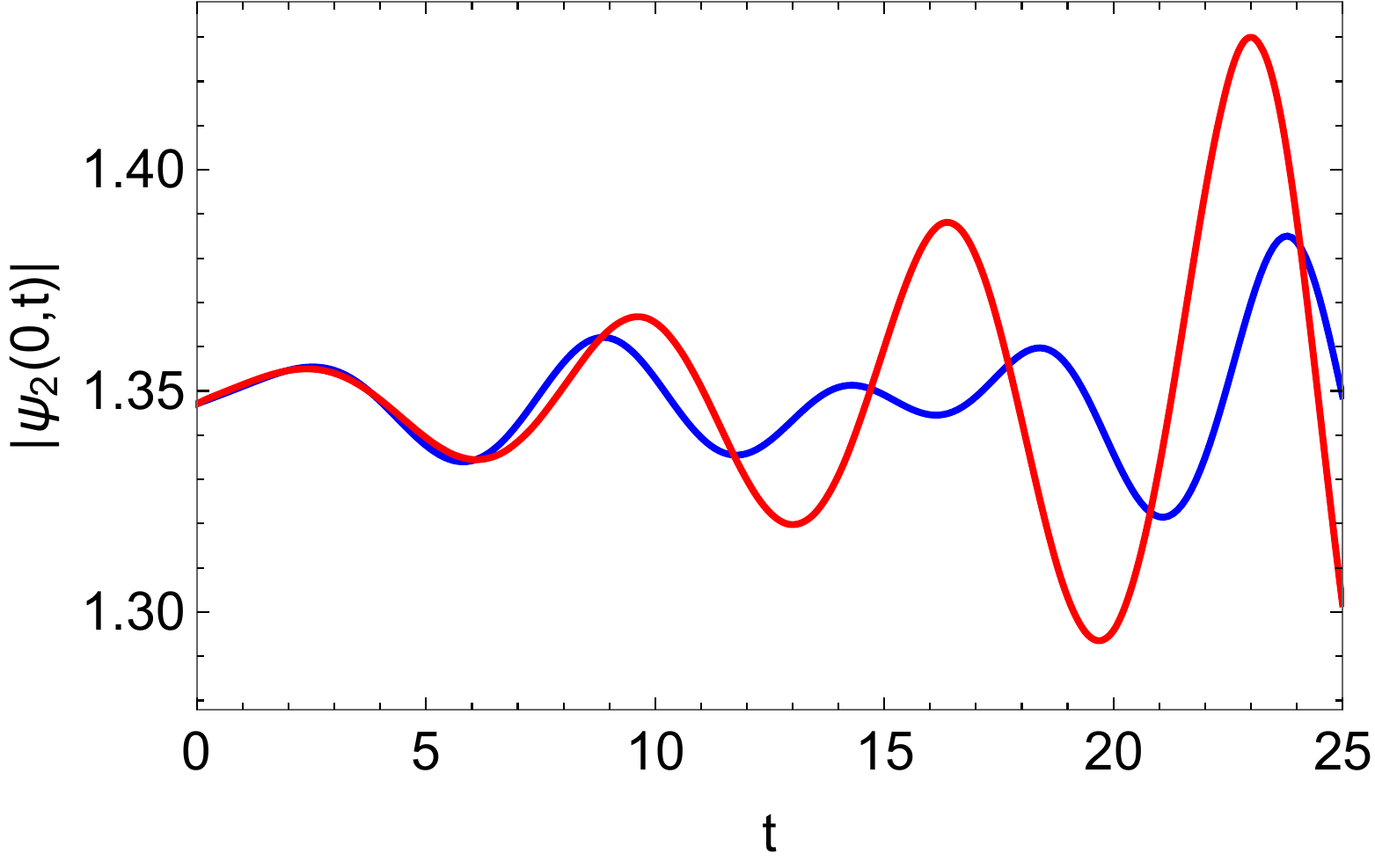}
\includegraphics[width=0.45\columnwidth]{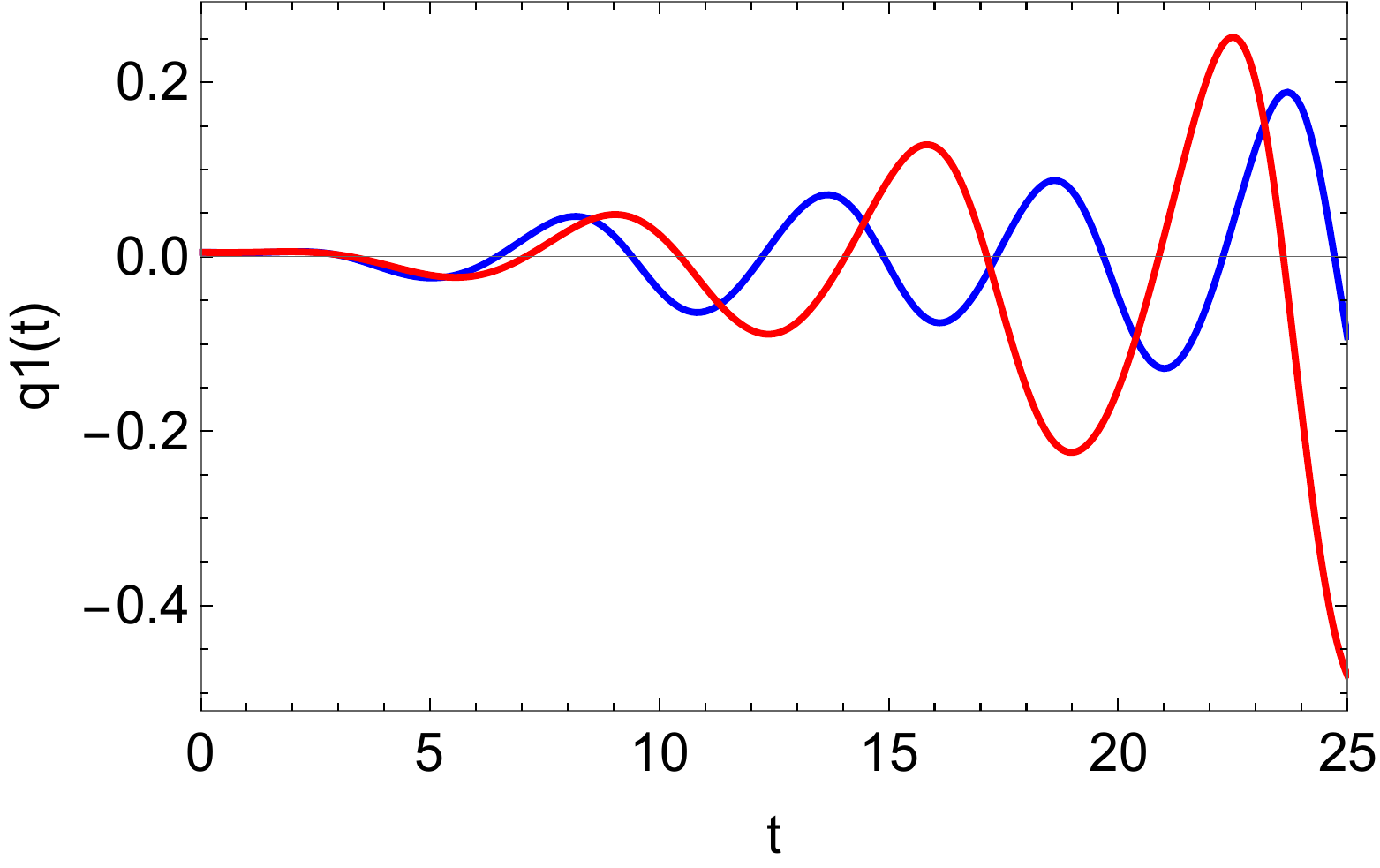}
\includegraphics[width=0.45\columnwidth]{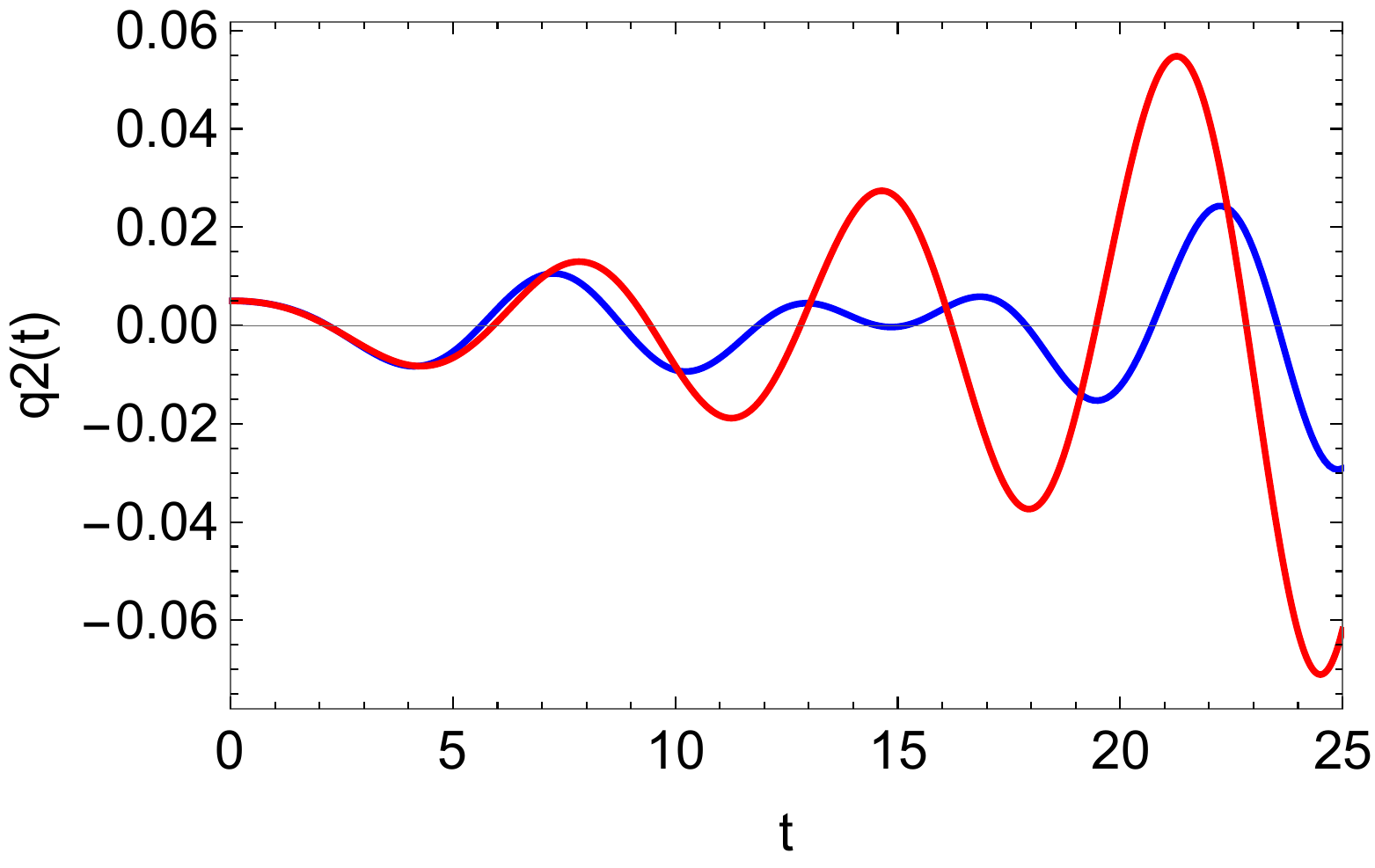}
\caption{\label{f:12parms-unstable}Results of the 12-parameter 
variational calculation for the unstable case corresponding to
parameter values of $b = 1/2$, $d = 1.07$, and $A_1 = 1/4$. Red 
and blue solid lines correspond to the 12-CC approximation and 
numerical calculation of the coupled NLSEs, respectively.}
\end{figure}

Finally, results on dynamics for the unstable case for the 12-parameter 
ansatz corresponding to parameter values of $A_1 = 1/4$, $b = 1/2$, 
and $d = 1.07$, are shown in Fig.~\ref{f:12parms-unstable}. We use the 
same initial conditions as with the stable case. Both the variational 
calculation and the numerical simulations of the NLSEs predict a blow-up 
and instability of the soliton, in agreement with the linear analysis.
Note that in this case $q_1(t)$ diverges from $q_2(t)$ and the 8-parameter 
approximation would not be adequate.  

%
%
%
%

%
%
\section{\label{s:CompAnalysis}Computational analysis and numerical results of the full NLS system}

In this last section, we present our numerical results on the{ \it existence},
{\it stability}, and {\it spatio-temporal evolution} of the solutions~\ef{A-Psi} 
to the coupled NLS system of Eq.~\ef{final}. The existence of solutions is 
investigated by introducing the ansatz
\begin{equation}
\psi_{j}(x,t)=\psi_{j}^{(0)}(x)e^{-i\omega_{j}t}, \quad \psi_{j}^{(0)}(x)\in\mathbb{C},%
\quad j=1,2.
\label{steady_state_ans}
\end{equation}
Upon plugging Eq.~\ef{steady_state_ans} into Eqs.~\ef{final}, we obtain the
system of steady-state equations
\vskip -0.5cm
\begin{subeqnarray}\label{steady_state_bvp}
&&\frac{d^{2}\psi_{1}^{(0)}}{dx^{2}}+%
\gamma\left[\big|\psi_{1}^{(0)}\big|^{2}+\big|\psi_{2}^{(0)}\big|^{2}\right]\psi_{1}^{(0)}%
-V(x)\psi_{1}^{(0)}+\omega_{1}\psi_{1}^{(0)}=0, \label{steady_state_bvp_1} \\
&&\frac{d^{2}\psi_{2}^{(0)}}{dx^{2}}+%
\gamma\left[\big|\psi_{1}^{(0)}\big|^{2}+\big|\psi_{2}^{(0)}\big|^{2}\right]\psi_{2}^{(0)}%
-V^{\ast}(x)\psi_{2}^{(0)}+\omega_{2}\psi_{2}^{(0)}=0.  \label{steady_state_bvp_2}
\end{subeqnarray}
In this work, we solve the boundary-value-problem (BVP) consisting of 
Eqs.~\ef{steady_state_bvp} and zero Dirichlet boundary conditions numerically. 
To that effect, we consider a uniform one-dimensional grid of points on 
$[-20,20]$ with lattice spacing $\Delta x=0.04$. The second-order derivatives
in Eqs.~\ef{steady_state_bvp} are replaced with a second-order accurate, 
central finite difference approximation. It should be noted that we further 
corroborated our results on the existence (and stability) of solutions by 
employing Chebyshev collocation on the unit interval $[-1,1]$ with $N=701$ 
Chebyshev nodes. In that case, the affine transformation $x_{k}=\frac{a+b}{2}+\frac{b-a}{2}\xi_{k}$ 
was employed which maps $[a,b]$ into $[-1,1]$ with $x_{k}\in[a,b]$, $\xi_{k}\in[-1,1]$, 
and $k=1,2,\dots, N$ (here, $a=-20$ and $b=20$). Regardless of the spatial 
discretization, we employed Newton's method to solve the underlying system 
of coupled nonlinear equations. The initial guess that was fed to the solver 
was the steady part of the exact solutions of Eqs.~\ef{A-Psi}, and thus 
Newton's method converged rapidly (typically in two iterations with an error
of $\approx 10^{-12}$ on the residuals). We also used a Newton-Krylov method~\cite{Kelley_nsoli}
to validate our findings. Both methods produced exactly the same results 
and matched perfectly with the exact solutions (up to local truncation error).

Having a steady-state solution $\psi_{j}^{(0)}$ ($j=1,2$) at hand, we perform 
a spectral stability analysis around them. To do so, we introduce the perturbation 
Ans\"atze 
\begin{subeqnarray}\label{lin_ansatz_1}
\widetilde{\psi}_{1}(x,t)&=&e^{-i\omega_{1}t}\left[\psi_{1}^{(0)}+\varepsilon
\left( a(x)e^{\lambda t}+b^{\ast}(x)e^{\lambda^{\ast }t}\right) \right],\quad a(x),b(x)\in\mathbb{C},\\
\label{lin_ansatz_2}
\widetilde{\psi}_{2}(x,t)&=&e^{-i\omega_{2}t}\left[\psi_{2}^{(0)}+\varepsilon
\left(c(x)e^{\lambda t}+d^{\ast}(x)e^{\lambda^{\ast}t}\right) \right], \quad c(x),d(x)\in\mathbb{C},
\end{subeqnarray}
where $\lambda\in\mathbb{C}$ is the eigenvalue and $\varepsilon\ll 1$ is a small 
parameter. Then, we insert Eqs.~\ef{lin_ansatz_1} into Eqs.~\ef{final} and obtain, 
at order $\mathcal{O}(\varepsilon )$, the eigenvalue problem:
\begin{equation}\label{eig_prob}
\left ( 
\begin{array}{cccc}
  A_{11}        &  A_{12}        &  A_{13}        &  A_{14} \\
 -A_{12}^{\ast} & -A_{11}^{\ast} & -A_{14}^{\ast} & -A_{13}^{\ast} \\
  A_{13}^{\ast} &  A_{14}        &  A_{33}        &  A_{34} \\
 -A_{14}^{\ast} & -A_{13}        & -A_{34}^{\ast} & -A_{33}^{\ast}%
\end{array}
\right )
\left (
\begin{array}{c}
a \\
b \\
c \\
d%
\end{array}
\right )
=
\tilde{\lambda}
\left (
\begin{array}{c}
a \\
b \\
c \\
d%
\end{array}
\right )
\end{equation}
with eigenvalues $\tilde{\lambda}=-i \,\lambda$, eigenvectors 
$\mathcal{V}=\left[a \,b \, c \,d\right]^{T}$, and matrix elements 
given by
\begin{subeqnarray}
A_{11}&=&\frac{d^{2}}{dx^{2}}+\gamma\left(2|\psi_{1}^{(0)}|^{2}+|\psi_{2}^{(0)}|^{2}\right)-V(x)+\omega_{1} \,,\\
A_{12}&=&\gamma\left(\psi_{1}^{(0)}\right)^{2} \,,\\
A_{13}&=&\gamma\psi_{1}^{(0)}\left(\psi_{2}^{(0)}\right)^{\ast} \,,\\
A_{14}&=&\gamma\psi_{1}^{(0)}\psi_{2}^{(0)} \,,\\
A_{33}&=&\frac{d^{2}}{dx^{2}}+\gamma\left(|\psi_{1}^{(0)}|^{2}+2|\psi_{2}^{(0)}|^{2}\right)-V^{\ast}(x)+\omega_{2} \,,\\
A_{34}&=&\gamma\left(\psi_{2}^{(0)}\right)^{2} \>.
\end{subeqnarray}
Then, a solution is deemed stable if the eigenvalues $\lambda=\lambda_{r}+i\lambda_{i}$ 
have a nonvanishing (negative) real part, i.e., $\lambda_{r}<0$. On the other hand, if 
$\lambda_{r}>0$, this would indicate the presence of an unstable mode. We compute the 
eigenvalues of the linearization (sparse) matrix $A$ associated with Eq.~\ef{eig_prob}
in MATLAB. The spectra we obtained were further corroborated by the highly accurate FEAST
eigenvalue solver~\cite{kestyn_eric_tang} (and references therein) which considers contour 
integration and involves density-matrix representation techniques from quantum mechanics. 
In the eigenvalue computations using FEAST, an elliptical contour was chosen in such a way
that $\approx 150$ eigenvalues were computed. FEAST converged rapidly (within two iterations
in most of the cases considered in this work) with relative tolerance $10^{-10}$ on the 
residuals of eigenvectors, and the spectra obtained via \verb|eig| and FEAST are identical 
by using both spatial discretizations as well. 

For the numerical computations presented below, we use the parameter fixing mentioned 
in the previous section together with $\gamma=1$ and $A_{1}=1$. We will focus on different
cases in the parameter $b$, and in particular on values of $b=0.2$, $0.4$, $0.6$, $0.8$ 
and $b=1$ while $d$ is treated as a {\it bifurcation} parameter. Then, we will employ a 
{\it sequential} continuation over $d$ (with $\Delta d=0.01$ as our continuation step) 
by using as an initial guess the solution previously found for the new value of $d$.

\begin{figure}[htp]
\begin{center}
\includegraphics[height=.21\textheight, angle =0]{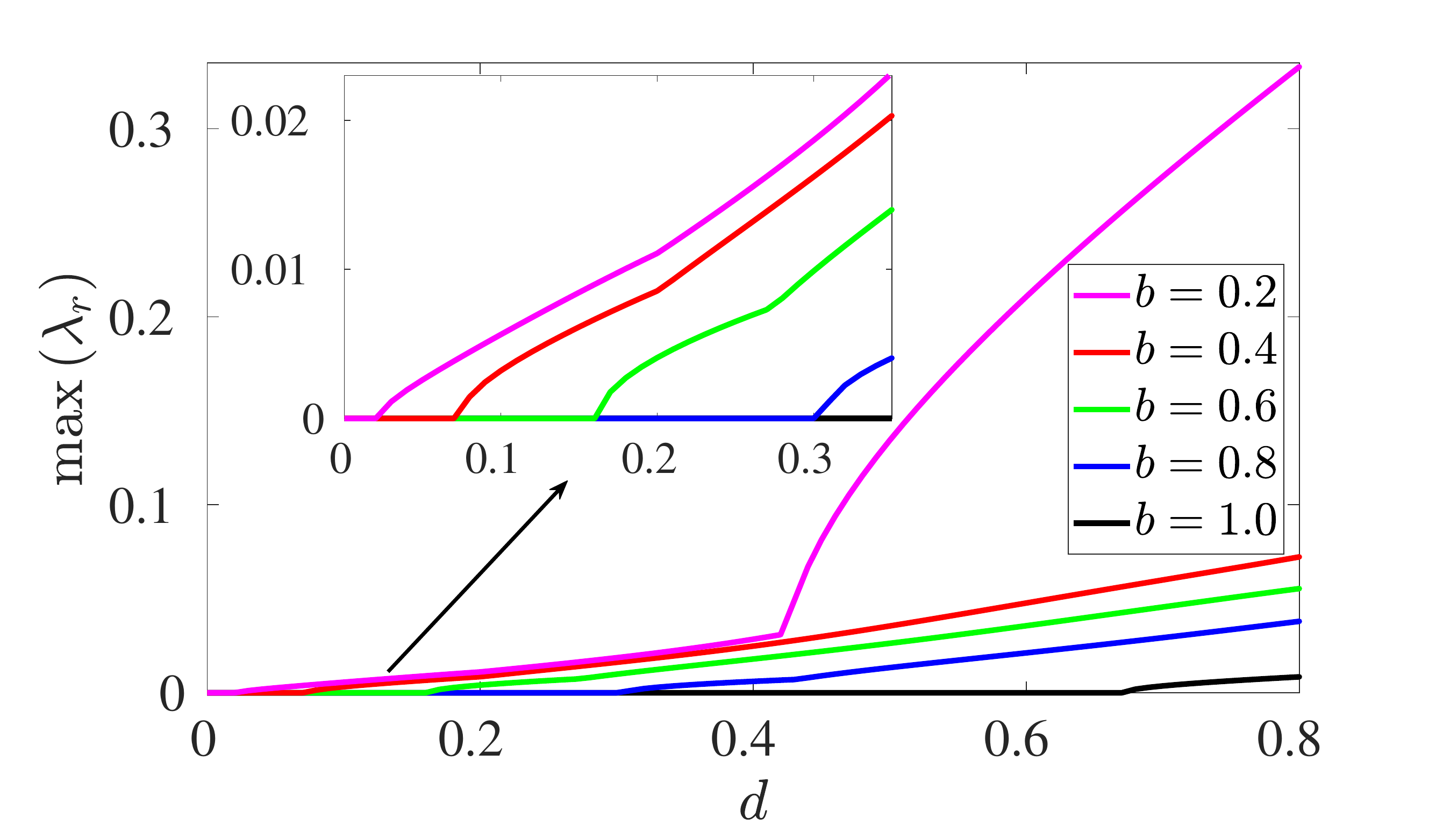}
\includegraphics[height=.13\textheight, angle =0]{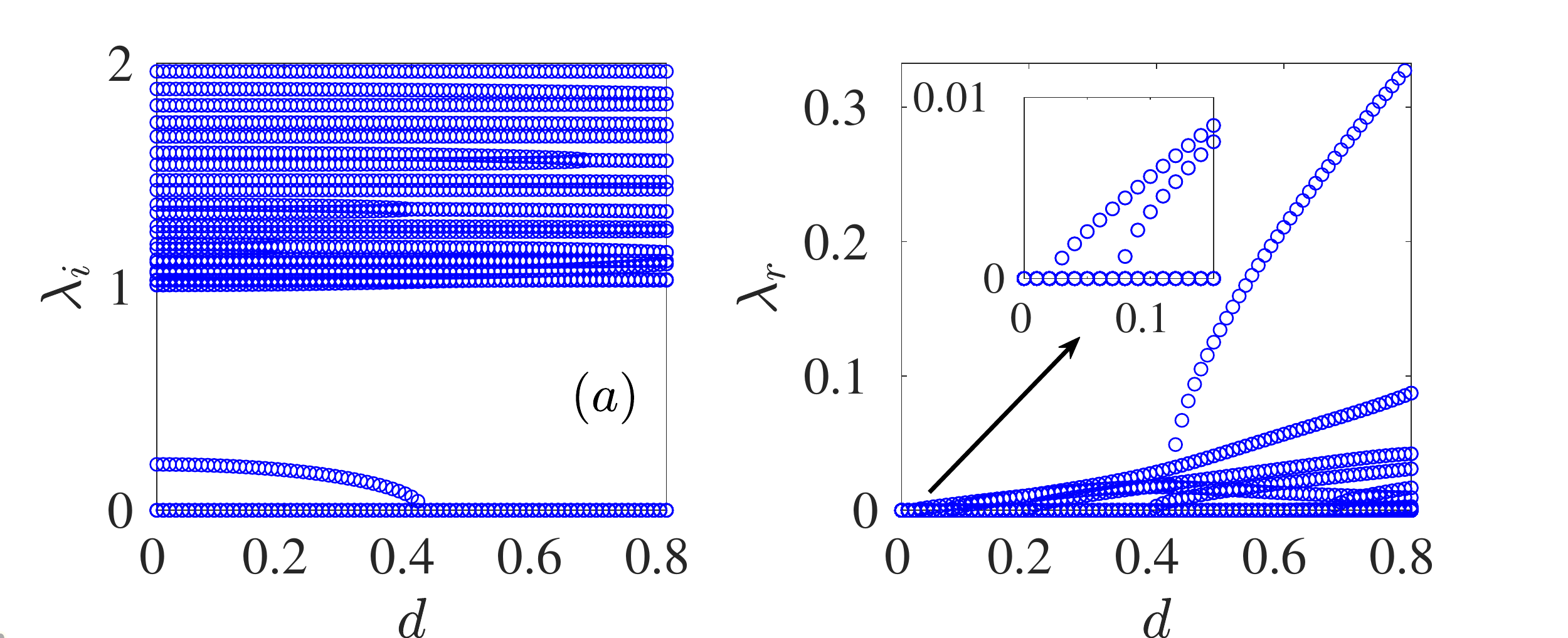}
\includegraphics[height=.13\textheight, angle =0]{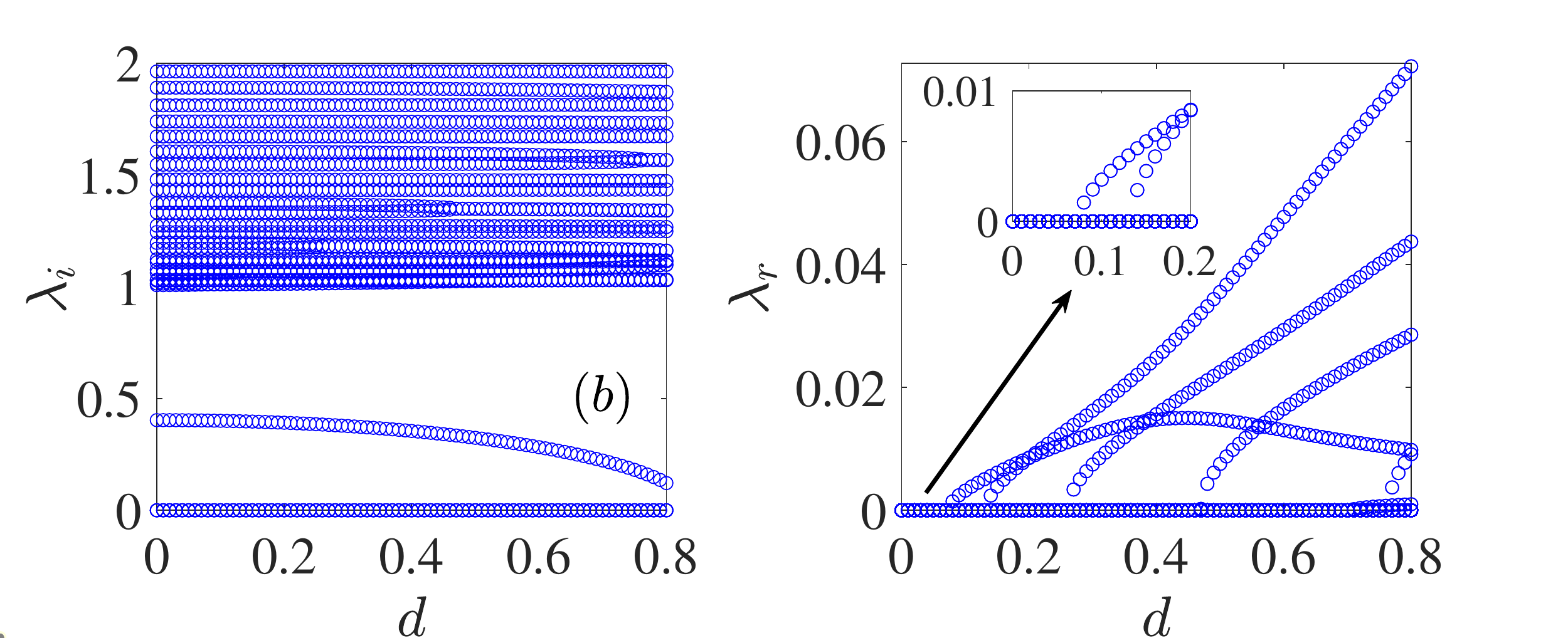}
\includegraphics[height=.13\textheight, angle =0]{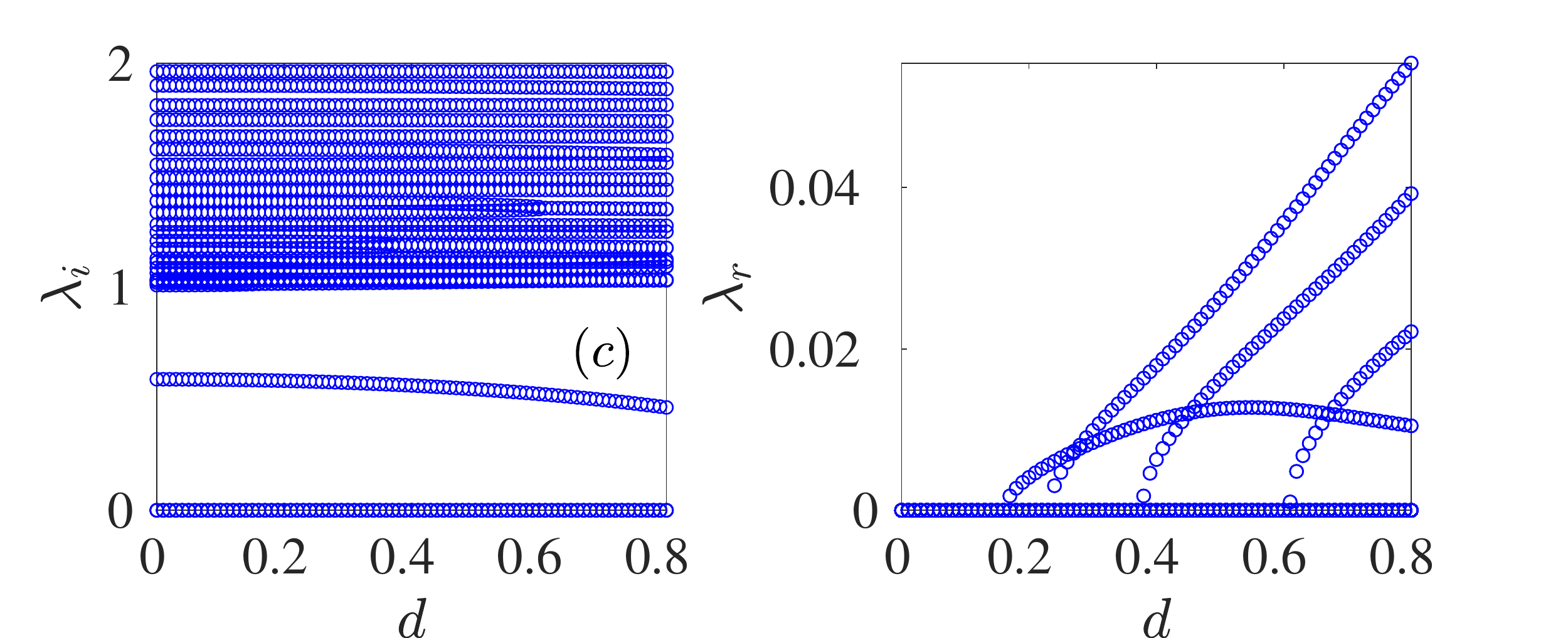}
\includegraphics[height=.13\textheight, angle =0]{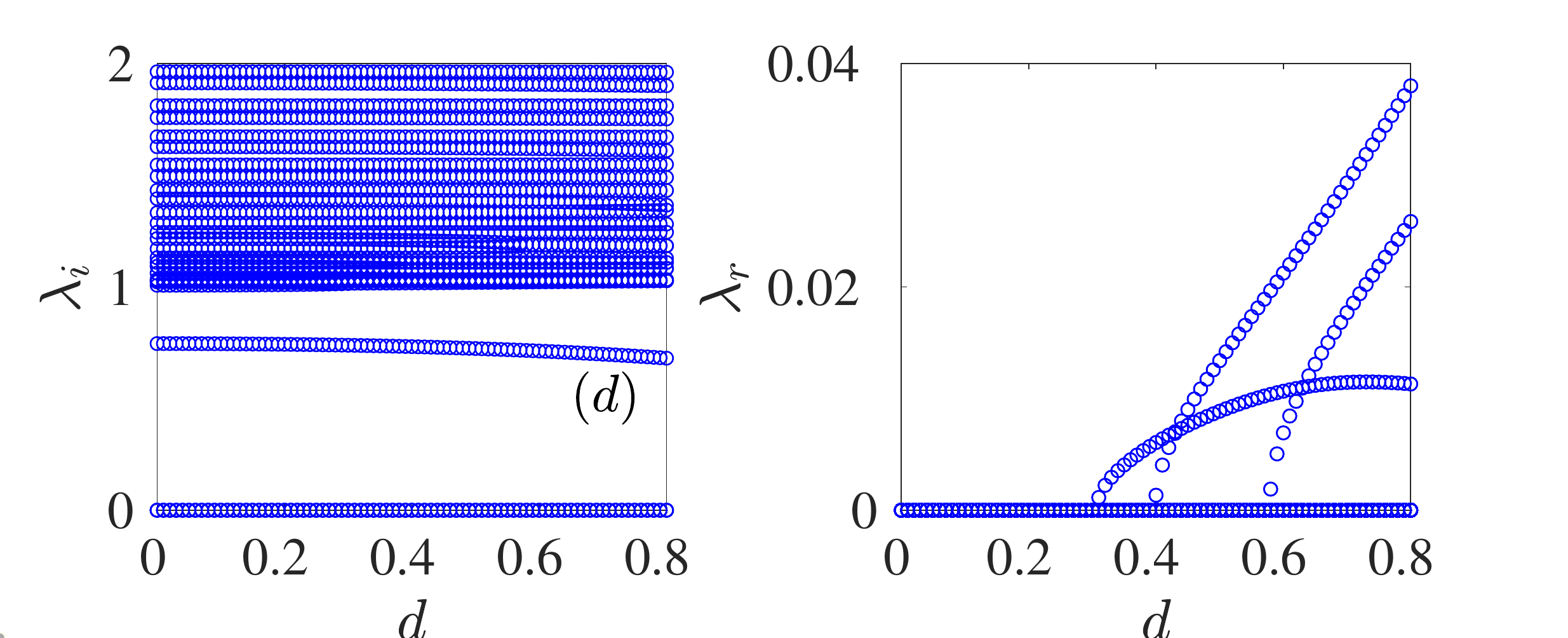}
\includegraphics[height=.13\textheight, angle =0]{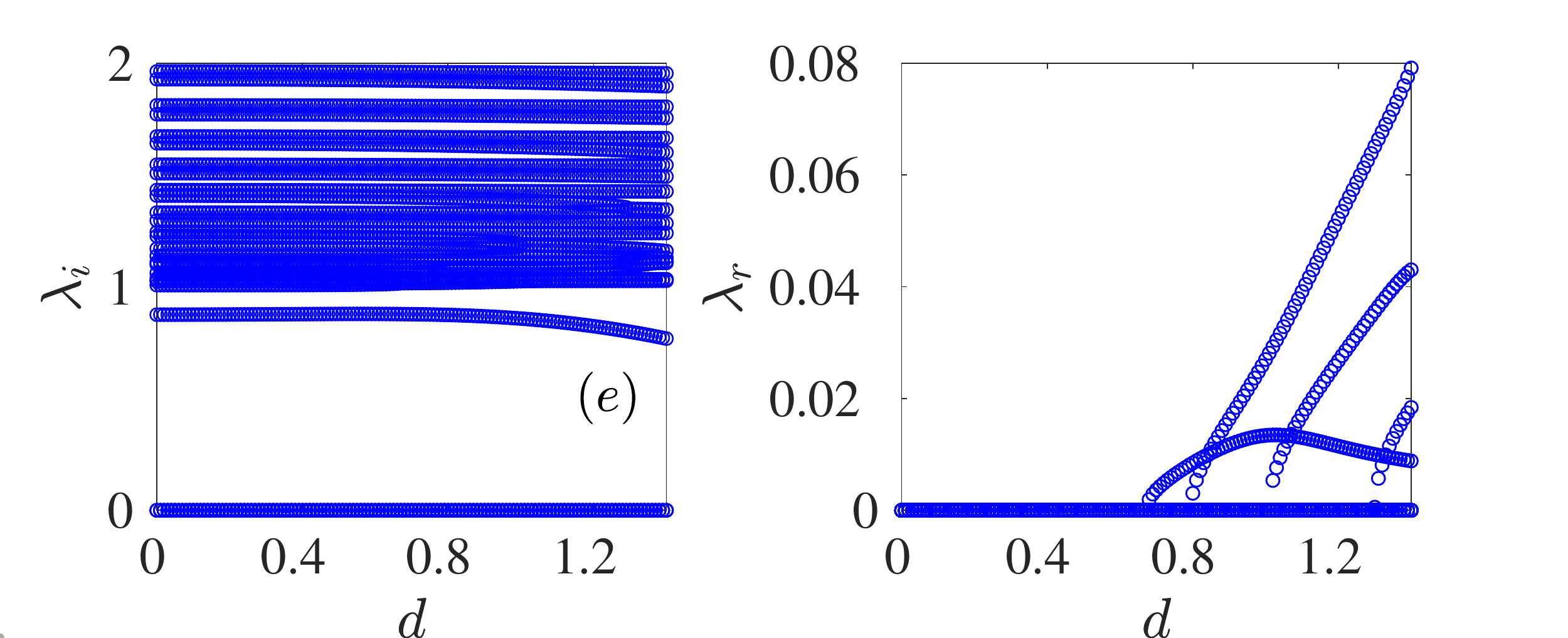}
\end{center}
\caption{
{\it Top row:} The maximal real eigenvalue $\lambda_{r}$, 
i.e., $\max{(\lambda_{r})}$ is shown as a function of $d$, and 
for different values of $b=0.2$, $0.4$, $0.6$, $0.8$ and 
$b=1$, respectively (see, the legend therein).
{\it Second, third and fourth rows:} The imaginary $\lambda_{i}$ 
and real $\lambda_{r}$ parts of the eigenvalues as functions of the 
parameter $d$, for the cases with (a) $b=0.2$, (b) $b=0.4$, (c) $b=0.6$,
(d) $b=0.8$, and (e) $b=1.0$, respectively.
}
\label{fig1}
\end{figure}
\begin{figure}[htp]
\begin{center}
\includegraphics[height=.17\textheight, angle =0]{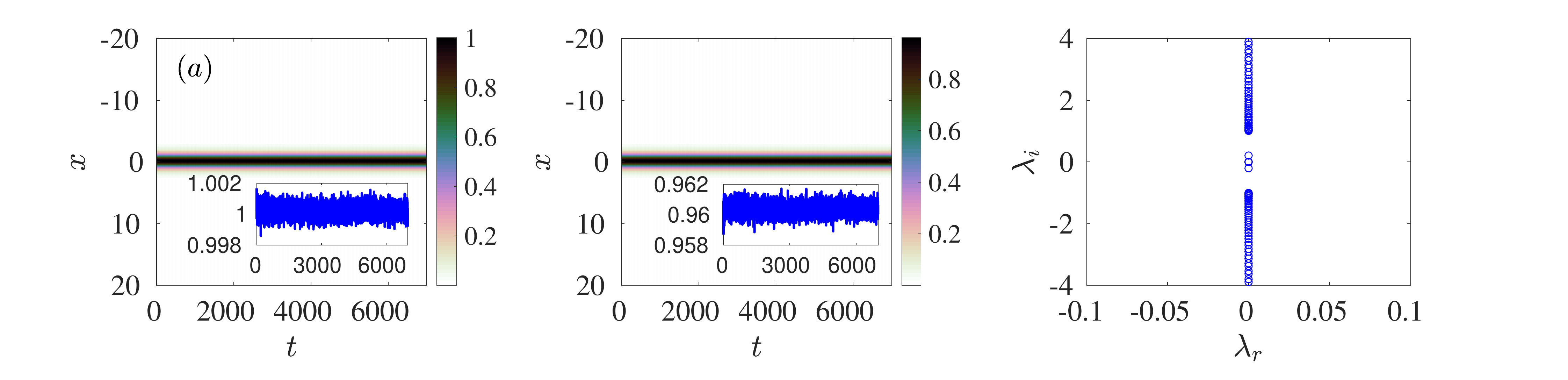}
\includegraphics[height=.17\textheight, angle =0]{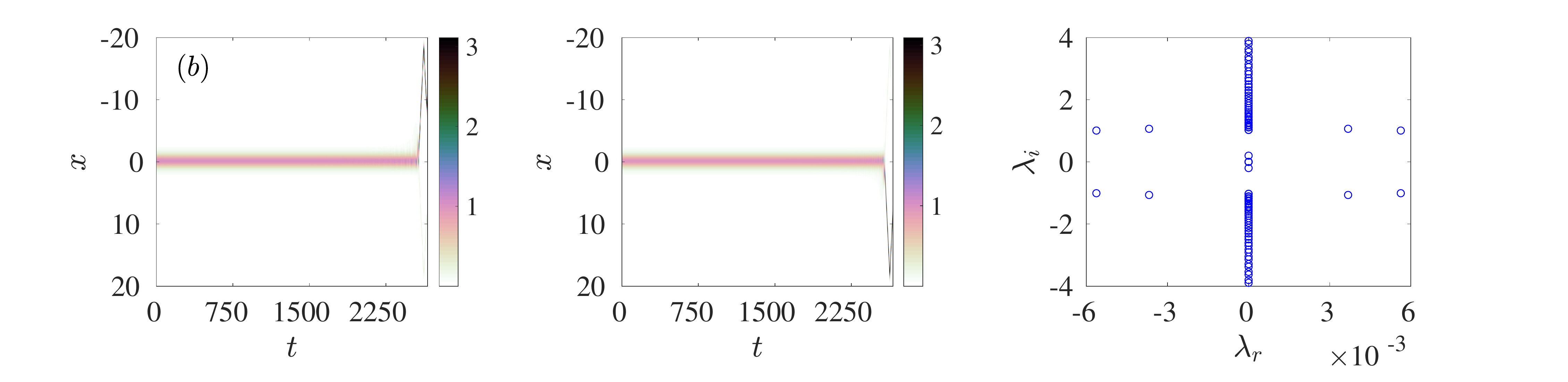}
\includegraphics[height=.17\textheight, angle =0]{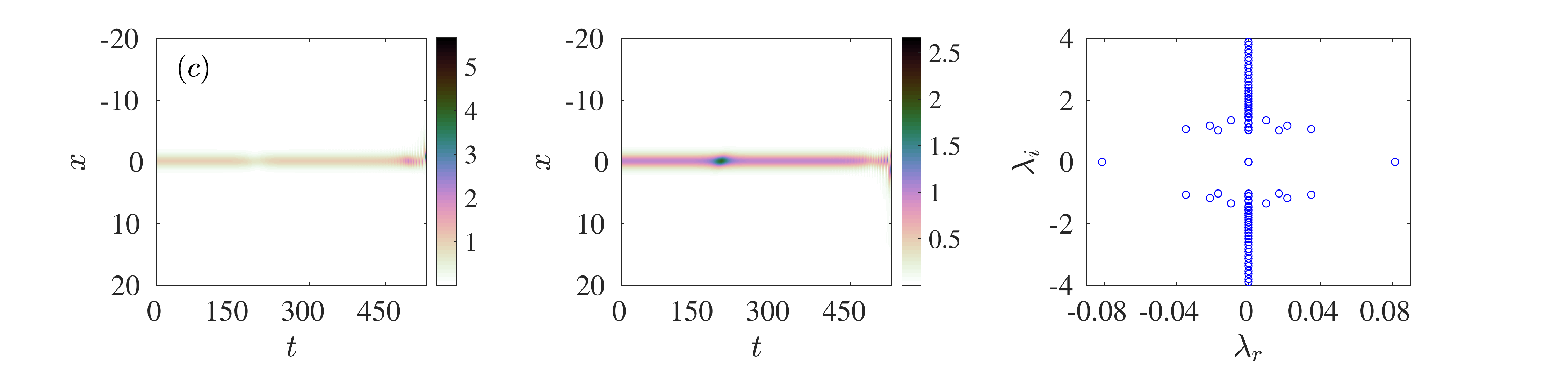}
\includegraphics[height=.17\textheight, angle =0]{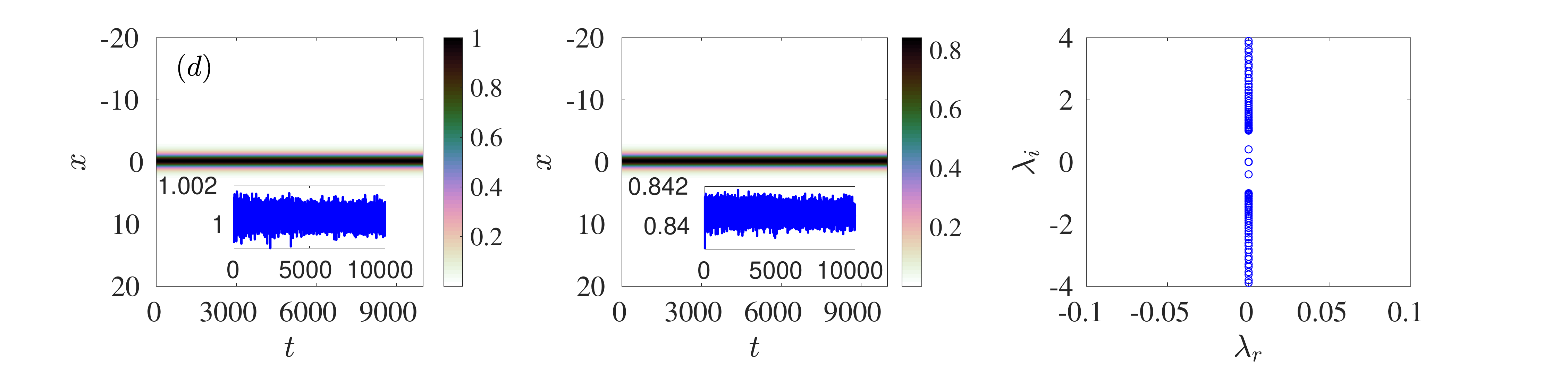}
\includegraphics[height=.17\textheight, angle =0]{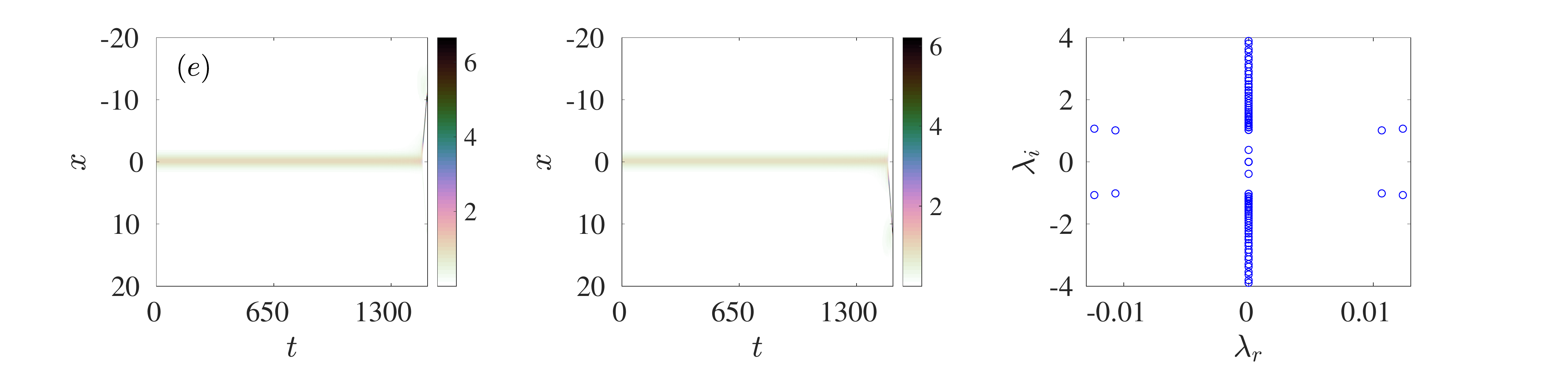}
\end{center}
\caption{
Spatio-temporal evolution of the densities $|\psi_{1}(x,t)|^{2}$
and $|\psi_{2}(x,t)|^{2}$ and associated spectra are shown in 
the left, middle and right columns, respectively. In particular,
the panels (a)-(c) correspond to the cases with $b=0.2$ and 
(a) $d=0.02$, (b) $d=0.1$, and (c) $d=0.45$, respectively. The 
panels (d) and (e) present results for the cases with $b=0.4$ 
and (d) $d=0.05$ and (e) $d=0.25$, respectively. The insets shown 
in the left and middle panels (a) and (d) correspond to $|\psi_{1}(x=0,t)|^{2}$ 
(left panel) and $|\psi_{2}(x=0,t)|^{2}$ (middle panel) as functions 
of time $t$.
}
\label{fig2}
\end{figure}
\begin{figure}[htp]
\begin{center}
\includegraphics[height=.17\textheight, angle =0]{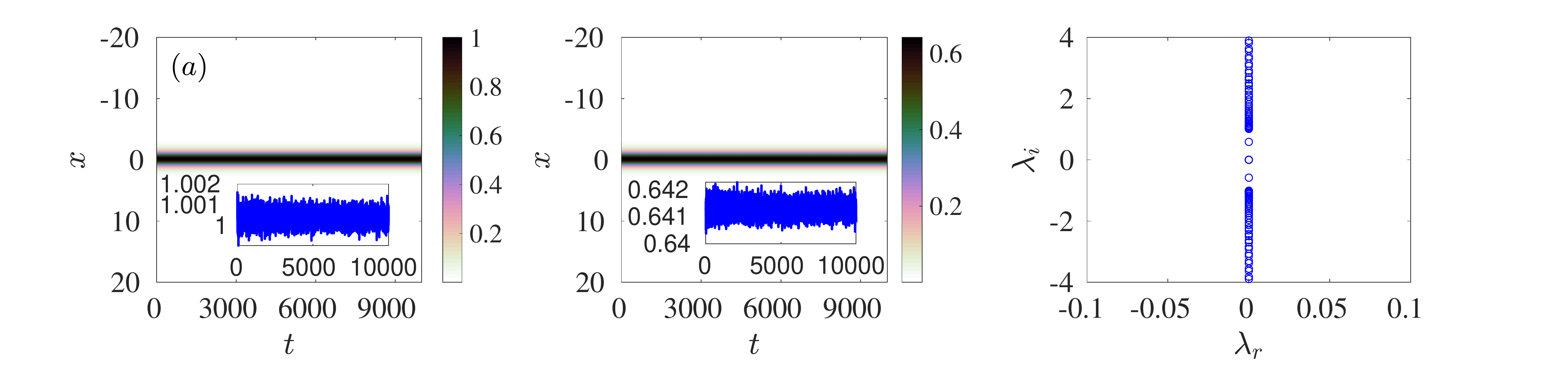}
\includegraphics[height=.17\textheight, angle =0]{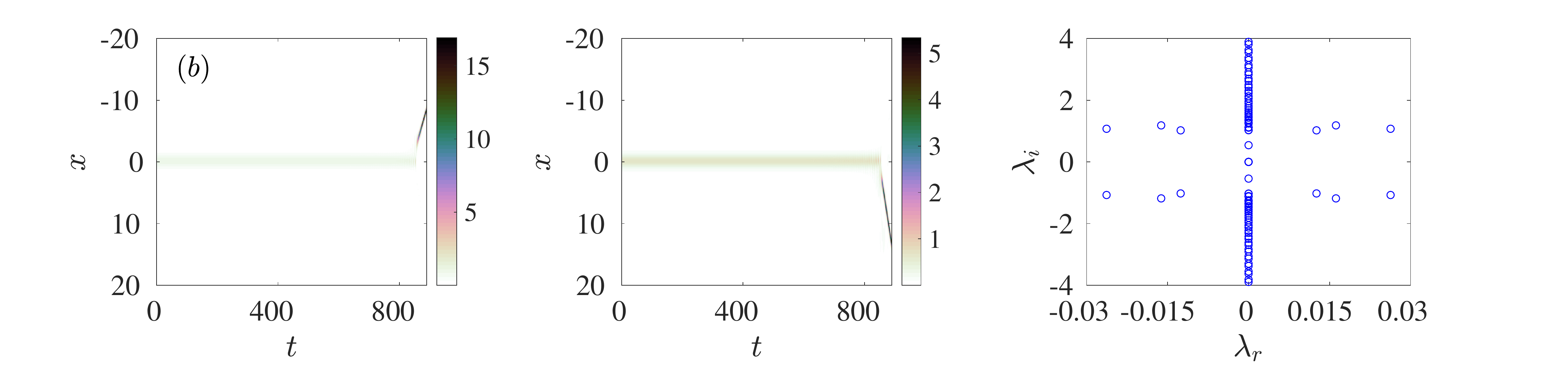}
\includegraphics[height=.17\textheight, angle =0]{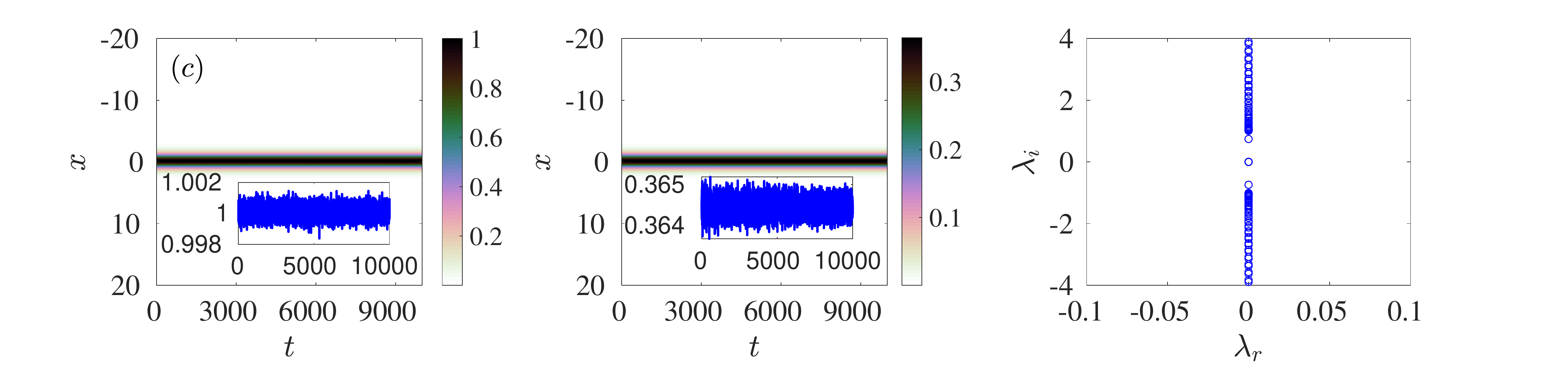}
\includegraphics[height=.17\textheight, angle =0]{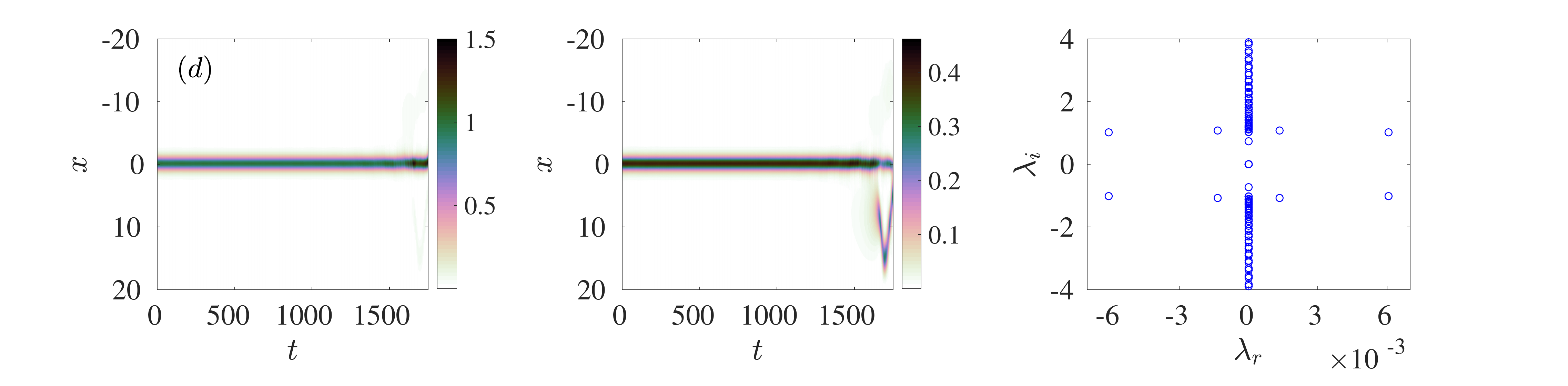}
\includegraphics[height=.17\textheight, angle =0]{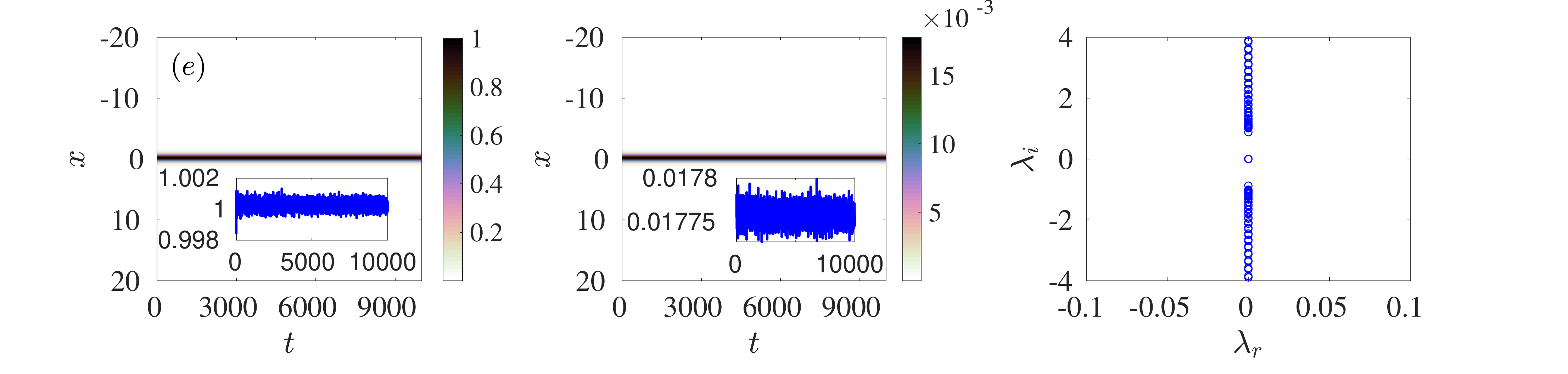}
\end{center}
\caption{
Same as Fig.~\ref{fig2} but for (a)-(b) $b=0.6$, (c)-(d) $b=0.8$ 
and (e) $b=1.0$. The results in (a) and (b) correspond to values 
of $d$ of (a) $d=0.1$ and (b) $d=0.5$ whereas those in (c) and (d) 
correspond to (c) $d=0.2$ and (d) $d=0.4$, respectively. Panel (e) 
corresponds to a value of $b=1.0$ with $d=0.4$.
}
\label{fig3}
\end{figure}

Fig.~\ref{fig1} presents our results on the stability of the steady-state 
solutions found via Newton's method as functions of the parameter $d$ and 
for different values of $b=0.2$, $b=0.4$, $b=0.6$, $b=0.8$, and $b=1.0$. 
Those panels correspond to results obtained from the FEAST algorithm and 
the range in $d$ considered therein is $[0,0.8]$. In particular, the top 
panel in this clustered figure showcases the growth rates of the most unstable
mode, i.e., $\max{(\lambda_{r})}$ as functions of $d$ and for different values 
in $b$ (see, the legend therein). It can be discerned from this panel that 
there exists parameter intervals of stability of the pertinent solutions. 
Indeed, the branches with $b=0.2$, $b=0.4$, $b=0.6$, $b=0.8$ and $b=1.0$ are
stable in parameter intervals of $d$ of $\approx [0,0.02]$, $[0,0.07]$, 
$[0,0.16]$, $[0,0.3]$, and $[0,0.67]$, respectively. A striking feature about
these results is that the parameter interval in $d$ in which solutions are 
(spectrally) stable becomes wider as soon as $b$ becomes larger, i.e., the 
coefficient appearing in $V_{0}(x)$. This suggests that one can controllably form a 
family of stable solitonic modes in the coupled NLS system by increasing the 
parameter $b$ which allows the existence of such (stable) solutions over a wide 
range in $d$. We summarize our presentation on the stability analysis results with
the panels (a)-(e) corresponding to the full spectrum of the solutions showcasing 
the imaginary $\lambda_{i}$ (left) and real $\lambda_{r}$ (right) parts of the eigenvalues
for the cases with $b=0.2$, $b=0.4$, $b=0.6$, $b=0.8$, and $b=1.0$, respectively. It 
should be noted that the instabilities we observe correspond to imaginary eigenvalues 
which bifurcate off the imaginary axis, thus resulting in {\it oscillatorily} unstable 
solutions characterized by a complex eigenvalue quartet. However, for the branch of 
$b=0.2$, a purely imaginary pair of eigenvalues passes through the origin at $d\approx 0.42$ 
creating a purely real unstable mode which becomes dominant past a value of $d\approx 0.43$. 
This suggests the emergence of a bifurcating branch out of this collision. Although this 
situation (which is also apparent for the other cases in $b$ we considered in this work 
but for larger values of $d$ than $0.8$) is quite interesting from the dynamical systems 
point of view, we are not pursuing bifurcations in this work.

We now turn our focus into our results on the dynamical evolution of steady-state solutions 
we obtained for different values of $b$ and $d$. In particular, our findings are summarized
in Figs.~\ref{fig2} and~\ref{fig3} showcasing the evolution of the densities $|\psi_{1}(x,t)|^{2}$ 
and $|\psi_{2}(x,t)|^{2}$ as well as the associated spectral plane $(\lambda_{r},\lambda_{i})$ 
of the steady-state solution identified in the left, middle and right columns, respectively. 
We advance the coupled NLS system [cf. Eq.~\ef{final}] forward in time by employing a standard
four-stage Runge-Kutta (RK4) method with fixed time-step size $\Delta t= 10^{-4}$. As per the
stable (according to our linear stability analysis) steady-state solutions, we add a random noise
of small amplitude ($\varepsilon\sim 10^{-3}$ in these cases) on top of the localized region 
of the pertinent steady-state profile and we use it as an initial condition in RK4. We then consider
time intervals of integration of $[0,10000]$ (or $[0,7000]$). In particular, the results on the 
dynamical evolution presented in Figs.~\ref{fig2}(a) (for $b=0.2$ and $d=0.02$), \ref{fig2}(d) 
(for $b=0.4$ and $d=0.05$), \ref{fig3}(a) (for $b=0.6$ and $d=0.1$), \ref{fig3}(c) (for $b=0.8$ 
and $d=0.2$), and \ref{fig3}(e) (for $b=1.0$ and $d=0.4$) correspond to stable solutions as it 
can be discerned from the spatio-temporal evolution of the respective densities over a very large 
time interval (see, the associated spectra). In addition, those panels offer the temporal distribution 
of the densities $|\psi_{1}(x=0,t)|^{2}$ and $|\psi_{2}(x=0,t)|^{2}$ suggesting the robustness of
the (perturbed) solutions and validating our linear stability analysis results. 

On the other hand, and as for the unstable steady-state solutions, we initialize the dynamics by 
perturbing the solutions along the most unstable eigendirection (with $\varepsilon\sim10^{-3}$ 
or $\varepsilon\sim10^{-2}$ depending upon the magnitude of $\lambda_{r}$). The results in 
Figs~\ref{fig2}(b), \ref{fig2}(c), \ref{fig2}(e) as well as~\ref{fig3}(b) and~\ref{fig3}(d) 
correspond to unstable steady-state solutions, and the instability is manifested in the dynamics
as this is evident in the spatio-temporal evolution of the densities shown in those panels.
For example, Fig.~\ref{fig2}(b) corresponds to $b=0.2$ and $d=0.1$ where the solutions are 
classified as oscillatorily unstable. In the left and middle panels of Fig.~\ref{fig2}(b), 
the solution starts performing oscillations (due to the oscillatory instability) until each 
component forms a narrower (in its width) bright pulse that propagates to the left (for the 
first component) and right (for the second component) and hits the boundaries (results are not
shown past that time). It should be noted that a similar phenomenology is observed in Figs.~\ref{fig2}(e)
($b=0.4$ and $d=0.25$) as well as in Figs.~\ref{fig3}(b) ($b=0.6$ and $d=0.5$) and \ref{fig3}(d) 
($b=0.8$ and $d=0.4$). Finally, the results shown in Fig.~\ref{fig2}(c) (with $b=0.2$ and $d=0.45$)
correspond to an example case scenario where the dominant unstable mode is characterized by a pair 
of purely real eigenvalues (on top of oscillatory unstable ones). It can be discerned from that 
figure that the density of the first component progressively becomes smaller at $t\approx 200$ 
whereas the second component (see, the middle panel) develops a bright pulse of higher amplitude. 
However, and past that time, both components start performing oscillations of gradually increasing 
amplitude until they are amplified substantially at $t\approx 450$ resulting in the breakdown of 
the pertinent waveforms.

\section{\label{s:Conclusions}Conclusions}
In this paper we have found exact solutions to the problem of two coupled 
NLSEs~\cite{cNLSreview} in the presence of a complex confining potential 
which has $\PT$ symmetry and is derivable from a superpotential 
$\bbW(x) = r \,\sigma_0 \, \tanh(x) + \rmi \, s \, \sigma_3 \, \sech(x) $.   
Such systems have started to be investigated experimentally in optical 
lattice environments~\cite{PhysRevA.99.013823}. Using numerical methods
we have mapped out the regimes of stability as well as studied the behavior 
of these solutions when they are subjected to small perturbations. We compared
the numerical solutions in the latter case with a variational approximation 
based on introducing  8 or 12 time-dependent CCs which are related to various 
low-order moments of the NLSEs. The CC approach allowed us to determine 
analytically approximate small oscillation frequencies. We compared the results
of the CC approach with the numerical simulations in two cases; one where the 
solutions are stable and one where the solutions are unstable. The 8 CC approximation
assumed that the average position and width of the two components of the NLSEs 
followed the same trajectory in time, whereas the 12 CC approximation allowed for 
these variables and their canonical conjugates to be different. Both CC approaches
quantitatively agreed with the numerically determined time evolution of the wave 
function in the first case, but only qualitatively agreed with the numerical solution
in the unstable regime. In the unstable case regime, the average position of the 
two components as well as the average width of the two components diverged from each
other, so that only the 12-CC approximation was able to track the behavior of the 
time evolution qualitatively. Then, we turned our focus to the coupled NLSEs and 
systematically studied the existence, stability and dynamical evolution of solitary
waves. Upon identifying branches of steady-state solutions via fixed-point methods, 
a bifurcation analysis was carried out over a two-parameter space where parametric 
intervals of stability were identified. Our spectral stability analysis results 
suggest that we can controllably form a wide range in the parameter $d$ by 
increasing the value of the parameter $b$ whereupon stable solitary waves can be
supported. This corresponds to the case where the real part $V_{0}$ of the potential 
$V(x)$ becomes larger. Finally, the stability results we report in this work were 
tested against direct numerical simulations where typical scenarios of blow-up were
involved for the unstable soliton solutions. The results and methods employed in this
work could be naturally applied and extended to other coupled, multi-component NLSEs
in order to explore the underlying configuration space of solutions. Such efforts are 
currently under consideration and will be reported in future publications.

\ack
FC, EGC and JFD would like to thank the Santa Fe Institute and the Center 
for Nonlinear Studies at Los Alamos National Laboratory for their hospitality.
EGC extends his deepest gratitude to Jesus Cuevas--Maraver (University of
Seville) for fruitful discussions about spectral collocation methods. AK 
is grateful to Indian National Science Academy (INSA) for awarding him 
INSA Senior Scientist position at Savitribai Phule Pune University, Pune, 
India. The work of AS was supported by the U.S.\ Department of Energy. 
%
%
\appendix

%
%
\section{\label{s:Integrals}Useful integrals and definitions}

We note that
\begin{subeqnarray}\label{e:I-1}
   \frac{d}{dz} \sech(z)
   &=
   - \sech(z) \, \tanh(z) 
   \label{e:I-1a} \,, \\
    \frac{d}{dz} \tanh(z)  
   &=
   \sech^2(z) \>.
   \label{e:I-1b}
\end{subeqnarray}
Also,
\begin{subeqnarray}\label{e:I-1.5}
   \phi(x) 
   &= 
   \alpha \tan^{-1}[\, \tanh(x/2) \,] \>,
   \label{e:I-1.5-a} \\
   \frac{d\phi(x)}{dx}
   &=
   \frac{\alpha}{2} \sech(x) \>.
   \label{e:I-1.5-b}
\end{subeqnarray}
Some useful integrals are the following:
\begin{subeqnarray}\label{e:I-2}
   \tint dz \sech^2(z)
   &=
   2 \>, 
   \label{e:I-2a} \\
   \tint dz\,\sech^3(z)
   &=
   \frac{\pi}{2} \>, 
   \label{e:I-2b} \\
   \tint dz\, \sech^4(z)
   &=
   \frac{4}{3} \>, 
   \label{e:I-2c} \\
   \tint dz \,z^2 \sech^2(z)
   &=
   \frac{\pi^2}{6} \>, 
   \label{e:I-2d} \\
   \tint dz\, \sech^2(z) \tanh^2(z)
   &=
   \frac{2}{3} \>,
   \label{e:I-2e} \\
   \tint dz \,z \sech^4(z) \tanh(z)
   &=
   \frac{1}{3} \>. 
   \label{e:I-2f}
\end{subeqnarray}
We define:
\begin{subeqnarray}\label{e:I-3}
   I_1(\beta,q)
   &:= 
   \tint dy \sech^2( \beta y ) \sech(y+q)
   \label{e:I-3a} \>, \\
   I_2(\beta,q)
   &:=
   \tint dy y \sech^2( \beta y ) \sech(y+q)
   \label{e:I-3b} \>, \\
   I_3(\beta,q)
   &:=
   \tint dy \sech^2( \beta y ) \sech^2(y+q)
   \label{e:I-3c} \>. 
\end{subeqnarray}
Also, we define:
\begin{subeqnarray}\label{e:I-4}
   f_1(\beta,q)
   &:=
   \tint dy \sech^2(\beta y) \, \sech(y + q)  \, \tanh(y + q) \>,
   \label{e:I-4a} \\
   f_2(\beta,q)
   &:=
   \tint dy y \, \sech^2(\beta y) \, \sech(y + q) \, \tanh(y + q) \>,
   \label{e:I-4b} \\
   f_3(\beta,q)
   &:=
   \tint dy y^2 \, \sech^2(\beta y) \, \sech(y + q) \, \tanh(y + q) \>.
   \label{e:I-4c}
\end{subeqnarray}
Partial derivatives are given by: 
\begin{subeqnarray}\label{e:I-5}
   \fl
   I_{1,q}(\beta,q)
   &=
   - \tint dy\,
   \sech^2(\beta y) \, \sech(y + q) \, \tanh(y + q)
   =
   - f_1(\beta,q) \>,
   \label{e:I-5a} \\
   \fl
   I_{1,\beta}(\beta,q)
   &=
   - 2 \tint dy \,y \,
   \sech^2(\beta y) \, \tanh(\beta y) \, \sech(y + q)
   =
   - 2 f_{10}(\beta,q) \>,
   \label{e:I-5b} \\
   \fl
   I_{2,q}(\beta,q)
   &=
   - \tint dy \, y \,
   \sech^2(\beta y) \, \sech(y + q) \, \tanh(y + q)
   =
   - f_2(\beta,q) \>,
   \label{e:I-7a} \\
   \fl
   I_{2,\beta}(\beta,q)
   &=
   - 2 \tint dy \,y^2 \,
   \sech^2(\beta y) \, \tanh(\beta y) \, \sech(y + q)
   =
   - 2 f_9(\beta,q) \>,
   \label{e:I-7b} \\
   \fl
   I_{3,q}(\beta,q)
   &=
   - 2 \tint dy\,
   \sech^2(\beta y) \, \sech^2(y + q) \, \tanh(y + q)
   =
   - 2 f_6(\beta,q) \>,
   \label{e:I-6a} \\
   \fl
   I_{3,\beta}(\beta,q)
   &=
   - 2 \tint dy\, y \,
   \sech^2(\beta y) \, \tanh(\beta y) \, \sech^2(y + q)
   =
   - 2 f_7(\beta,q) \>.
   \label{e:I-6b}
\end{subeqnarray}

%
%
\subsection{\label{ss:mixed}Mixing integrals}

The mixing integral is defined by
\begin{equation}\label{Cdef}
   C(\, \beta_1,q_1,\beta_2,q_2 \,)
   :=
   \tint dx \sech^2[\beta_1(x-q_1)] \sech^2[\beta_2(x-q_2)] \>.
\end{equation}
Derivatives of this integral are given by
\begin{eqnarray}
   \fl
   &C_{q_1}(\, \beta_1,q_1,\beta_2,q_2 \,)
   \\
   \fl
   & \hspace{1em}
   =
   2 \beta_1 \tint dx \,
   \sech^2[\beta_1(x-q_1)] \tanh[\beta_1(x-q_1)] \sech^2[\beta_2(x-q_2)] \>,
   \notag \\
   \fl
   &C_{q_2}(\, \beta_1,q_1,\beta_2,q_2 \,)
   \\
   \fl
   & \hspace{1em}
   =
   2 \beta_2 \tint dx \,
   \sech^2[\beta_1(x-q_1)]  \sech^2[\beta_2(x-q_2)] \tanh[\beta_2(x-q_2)] \>,
   \notag \\
   \fl
   &C_{\beta_1}(\, \beta_1,q_1,\beta_2,q_2 \,)
   \\
   \fl
   & \hspace{1em}
   =
   - 2 \tint dx \,(x - q_1) 
   \sech^2[\beta_1(x-q_1)] \tanh[\beta_1(x-q_1)] \sech^2[\beta_2(x-q_2)] \>,
   \notag \\
   \fl
   &C_{\beta_2}(\, \beta_1,q_1,\beta_2,q_2 \,)
   \\
   \fl
   & \hspace{1em}
   =
   - 2 \tint dx \,(x - q_2) 
   \sech^2[\beta_1(x-q_1)]  \sech^2[\beta_2(x-q_2)] \tanh[\beta_2(x-q_2)] \>.
   \notag
\end{eqnarray}

%
%
\subsection{\label{ss:expand}Expansion of the integrals}

To first order:
\begin{subeqnarray}\label{e:E-1}
   f_1(1+\delta\beta,\delta q)
   &=
   \frac{\pi}{4} \, \delta q \>,
   \\
   f_2(1+\delta\beta,\delta q)
   &=
   \frac{\pi}{6} + \Bigl( \frac{\pi}{3} - \frac{\pi^3}{16} \Bigr ) \, \delta \beta \>,
   \\
   f_3(1+\delta\beta,\delta q)
   &=
   - \Bigl( \frac{2 \, \pi}{3} - \frac{\pi^3}{16} \Bigr ) \, \delta q \>,
   \\
   I_1(1+\delta\beta,\delta q)
   &=
   \frac{\pi}{2} - \frac{\pi}{3} \, \delta \beta \>,
   \\
   I_2(1+\delta\beta,\delta q)
   &=
   - \frac{\pi}{6} \, \delta q \>,
   \\
   I_3(1+\delta\beta,\delta q)
   &=
   \frac{4}{3} - \frac{2}{3} \, \delta \beta \>,
   \\   
   I_{1,q}(1+\delta\beta,\delta q)
   &=
   - \frac{\pi}{4} \, \delta q \>,
   \\
   I_{2,q}(1+\delta\beta,\delta q)
   &=
   - \frac{\pi}{6} - \Bigl( \frac{\pi}{3} - \frac{\pi^3}{16} \Bigr ) \, \delta \beta \>,
   \\
   I_{3,q}(1+\delta\beta,\delta q)
   &=
   - \frac{16}{15} \, \delta q \>, 
   \\
   I_{1,\beta}(1+\delta\beta,\delta q)
   &=
   - \frac{\pi}{3} + \Bigl ( \pi - \frac{\pi^3}{16} \Bigr ) \, \delta \beta \>,
   \\
   I_{2,\beta}(1+\delta\beta,\delta q)
   &=
   - \Bigl ( \frac{\pi}{3} - \frac{\pi^3}{16} \Bigr ) \, \delta q \>,
   \\
   I_{3,\beta}(1+\delta\beta,\delta q)
   &=
   - \frac{2}{3} + \Bigl ( \frac{4}{3} - \frac{4 \, \pi^2}{45} \Bigr ) \, \delta \beta \>.
\end{subeqnarray}
For the mixing integrals, we find
\begin{subeqnarray}
   \fl
   C(1+\delta\beta_1,\delta q_1,1+\delta\beta_2,\delta q_2)
   &=
   \frac{4}{3} - \frac{2}{3} \, (\, \delta\beta_1 + \delta\beta_2 ) \>,
   \\
   \fl
   C_{q_1}(1+\delta\beta_1,\delta q_1,1+\delta\beta_2,\delta q_2)
   &=
   - \frac{16}{15} \, (\, \delta q_1 - \delta q_2 \, ) \>, 
   \\
   \fl
   C_{q_2}(1+\delta\beta_1,\delta q_1,1+\delta\beta_2,\delta q_2)
   &=
   - \frac{16}{15} \, (\, \delta q_2 - \delta q_1 \, ) \>, 
   \\
   \fl
   C_{\beta_1}(1+\delta\beta_1,\delta q_1,1+\delta\beta_2,\delta q_2)
   &=
   - 
   \frac{2}{3} 
   + 
   \Bigl ( \frac{4}{3} - \frac{4 \, \pi^2}{45} \Bigr ) \, \delta\beta_1
   +
   \frac{4 \, \pi^2}{45} \, \delta \beta_2 \>, 
   \\
   \fl
   C_{\beta_1}(1+\delta\beta_1,\delta q_1,1+\delta\beta_2,\delta q_2)
   &=
   - 
   \frac{2}{3} 
   + 
   \Bigl ( \frac{4}{3} - \frac{4 \, \pi^2}{45} \Bigr ) \, \delta\beta_2
   +
   \frac{4 \, \pi^2}{45} \, \delta \beta_1 \>.   
\end{subeqnarray}

\section*{Bibliography}
%
%
%

%
\end{document}